\documentclass[a4paper,onecolumn,11pt,accepted=2026-07-16]{quantumarticle}
\pdfoutput=1
\usepackage[utf8]{inputenc}
\usepackage[english]{babel}
\usepackage[T1]{fontenc}
\usepackage{amsmath,amssymb,amsthm}
\usepackage{hyperref}
\usepackage{algorithm}
\usepackage{algpseudocode}
\usepackage[dvipsnames]{xcolor}
 \usepackage{enumitem}
  \usepackage{mathtools}
 \usepackage{tcolorbox}
 \usepackage{tabularx}
 \usepackage{graphicx}
 \usepackage{colortbl}
 \usepackage{array}
\usepackage{tikz}
\usepackage{lipsum}
\usepackage{caption}
\usepackage{subcaption}
\usepackage{physics}
\usepackage{graphicx}
\newtheorem{theorem}{Theorem}
 \newtheorem{lemma}{Lemma}
 \newtheorem{corollary}{Corollary}
 \newtheorem{proposition}{Proposition}
 \newtheorem{example}{Example}
 \newtheorem{remark}{Remark}
 \newtheorem{definition}{Definition}

 \allowdisplaybreaks
\begin{document}
\title{Entanglement-assisted Quasi-cyclic Quantum Low-density Parity-check Codes over Qubits}

\author{Pavan Kumar}
\affiliation{Department of Electronic Systems Engineering, Indian Institute of Science, Bengaluru 560012, India}
\email{pavankumar1@iisc.ac.in}
\orcid{0000-0002-2450-8269}
\author{Abhi Kumar Sharma}
\affiliation{Department of Electronic Systems Engineering, Indian Institute of Science, Bengaluru 560012, India}
\email{abhisharma@iisc.ac.in}
%\orcid{0000-0003-1985-4623}

\author{Karthik Bharadwaj}
\affiliation{Department of Electronic Systems Engineering, Indian Institute of Science, Bengaluru 560012, India}
\email{karthikbgm@iisc.ac.in}
\orcid{0009-0008-1683-4129}

\author{Shayan Srinivasa Garani}
\affiliation{Department of Electronic Systems Engineering, Indian Institute of Science, Bengaluru 560012, India}
\email{shayangs@iisc.ac.in}
%\homepage{http://quantum-journal.org}
\orcid{0000-0002-2459-1445}
% \thanks{You can use the \texttt{\textbackslash{}email}, \texttt{\textbackslash{}homepage}, and \texttt{\textbackslash{}thanks} commands to add additional information for the preceding \texttt{\textbackslash{}author}. If applicable, this can also be used to indicate that a work has previously been published in conference proceedings.}

\maketitle

\begin{abstract}
We construct several families of entanglement-assisted quasi-cyclic quantum LDPC (EA-QC-QLDPC) codes via structured tilings of permutation matrices. The entanglement-unassisted portion of the joint Tanner graph of the proposed EA-QC-QLDPC code derived from two distinct classical QC-LDPC  codes is free of 4-cycles. Notably, one of the proposed families constructed from two distinct classical codes requires only a \textit{single} shared Bell pair between the quantum transmitter and receiver, highlighting its resource efficiency. We also analytically determine the exact code rates for some of the proposed constructions. Furthermore, two of the proposed families of EA-QC-QLDPC codes are derived from a single classical code whose Tanner graphs possess girth greater than six, further enhancing their error-correcting performance.

We also propose an encoding scheme with improved complexity by exploiting the proposed code structure. The performance of the proposed codes is assessed under both random and burst error models under the depolarizing and Markovian noise actions. Simulation results reveal nearly one order of improvement in error-correction performance with the quaternary block-layered normalized  min-sum  (QBLNMS) decoder compared to the layered binary sum-product decoder over both depolarizing and Markovian channels. Using  the QBLNMS decoder over a quaternary alphabet, we demonstrate that correlated Pauli errors can be effectively handled within the decoding framework.

Furthermore, under the QBLNMS decoding, the proposed codes achieve \textit{significant} performance improvements compared to prior works and can effectively handle both random and burst errors. The code constructions are scalable across various coding rates and quantum payloads, crucial for practical quantum communication and computing systems.
\end{abstract}
\section{Introduction}\label{sec: introduction}
 Quantum information is susceptible to errors due to decoherence (noise) and faulty gates. As a result, preserving quantum information through quantum error correcting codes (QECCs) becomes inevitable for realizing functional quantum computing and communication systems. The introduction of the stabilizer codes by Gottesman \cite{gottesman1997stabilizer} and a subclass of stabilizer codes, namely the CSS codes, derived from classical codes based on the dual-containment property \cite{calderbank1997quantum, calderbank1998quantum} provide a powerful framework to design QECCs from well-known classical codes. Following this, QECCs have been extensively studied and further developed \cite{lidar2013quantum}. For a more comprehensive background on general quantum codes, we refer the reader to the article by Terhal \cite{terhal2015quantum}. 

Although quantum codes with scaling properties similar to classical codes do exist \cite{calderbank1996good, ashikhmin2001asymptotically}, they face several practical challenges. In particular, the parity checks used for error detection and correction typically involve a large number of physical qubits for each logical qubit \cite{suzuki2022quantum,zhao2022realization}. This becomes problematic, as high-weight parity checks are difficult to implement reliably without additional fault-tolerance mechanisms. Moreover, these checks are often not easily parallelizable, causing some qubits to remain idle during measurements and thereby accumulate errors. Another important consideration is sparsity, which is essential for the efficiency of many decoding algorithms. These challenges closely resemble those encountered in classical coding theory, where they were successfully addressed by low-density parity-check (LDPC) codes. LDPC codes, introduced by Gallager \cite{gallager1963low} in the 1960s, are linear codes known for their near-capacity performance and efficient decoding. They gained renewed attention in the 1990s with the development of iterative decoding algorithms \cite{richardson2008modern}. Structured variants, such as quasi-cyclic (QC) LDPC codes, offer excellent error-correction performance and are well suited for hardware implementation. As a result, QC-LDPC codes are widely used in standards like DVB-S2 \cite{chen2018performance}, Wi-Fi \cite{wu2023high}, wireless systems~\cite{Wireless}, and data storage devices \cite{Shayan_1,Shayan_chann}. Motivated by their structured design, efficient encoding, and strong performance, we now turn to review their quantum counterparts.

%%%%%%%%%%%%%%%%%%%%%%%%%%%%%%%%%%%%%%%%%%%%%%%%%%%%%%%%%%%%%%%%%%%%%%%%%%%%%
%%%%%%%%%%%%%%%%%%%%%%%%%%%%%%%%%%%%%%%%%%%%%%%%%%%%%%%%%%%%%%%%%%%%%%%%%%%%%%%%%%%%%%%%%%%%%%%%%%%%%%
\subsection{Quantum LDPC Codes}
Quantum LDPC codes are built upon sparse bipartite graph structures, similar to their classical counterparts, but with the added complexity of handling quantum errors. Recent research has concentrated on developing such codes viable for quantum hardware implementation. For further insights into various types of quantum LDPC codes, the readers can refer to the review by Breuckmann and Eberhard \cite{breuckmann2021quantum}.

In a landmark paper, Gottesman \cite{gottesman2013fault} showed that quantum LDPC codes with a constant encoding rate could reduce the overhead of fault-tolerant quantum computation to a constant. This is in stark contrast to other quantum fault-tolerance schemes, where longer computations require suppressing errors to an increasingly smaller threshold, necessitating larger codes and a growing number of physical qubits \cite{kitaev1997quantum, knill1998resilient, aharonov1997fault}.

Random classical LDPC codes can achieve a constant encoding rate and a minimum distance that scales linearly with the code length, with high probability \cite{richardson2008modern, gallager1962low}. In comparison, constructing quantum LDPC codes along with efficient decoding algorithms under practical quantum constraints is significantly challenging, and it remains an open problem whether quantum LDPC codes exist that rival the parameters of their classical counterparts \cite{mackay2004sparse,panteleev2021quantum}. Despite these challenges, recent progress has been made towards improving the performance of QLDPC codes, with several families of code constructions that significantly outperform surface codes and color codes in terms of their asymptotic parameters~\cite{pecorari2025high}.

Constructing quantum LDPC codes requires classical codes to satisfy the dual containment property \cite{calderbank1998quantum,  ketkar2006nonbinary}, leading to $4$-cycles in the joint Tanner graph of the quantum code. Consequently, the quantum QC-LDPC code constructed by Hagiwara et al.~\cite{hagiwara2007quantum}, Pantaleev and Kalachev ~\cite{panteleev2021quantum} and Miao et al.~\cite{miao2024joint} have short cycles in the joint Tanner graph since the code constructions are based on the CSS framework. The reader should note that when the \(X\) and \(Z\) errors are statistically independent, their decoding can be performed separately using their respective Tanner graph representations, i.e., using the $H_z$ and $H_x$ stabilizers, respectively. The design strategy for such codes requires well-known approaches towards the optimization of classical LDPC codes, such as choosing large girth, avoiding harmful trapping sets, etc., to correct independent $X$ and $Z$ errors \cite{raveendran2017stochastic}. However, in the presence of correlated errors, the decoder must operate on the joint Tanner graph. Consequently, the CSS framework becomes inefficient, as the joint Tanner graph contains \(4\)-cycles, resulting in high error floors. This problem can be circumvented using the entanglement-assisted construction \cite{brun2006correcting}, requiring pre-shared error-free maximally entangled bits (ebits)\footnote{After the seminal work of Brun et al. on EA quantum codes over the binary field, there has been a significant progress toward generalization of EA QECCs over arbitrary finite fields (see, for instance, \cite{chen2017entanglement,guenda2018constructions, liu2018application,qian2018mds,galindo2019entanglement,luo2016non,nadkarni2021encoding,nadkarni2021entanglement,hsieh2011high}) along with efficient encoding and decoding circuits.} between the sender and the receiver. Pre-shared ebits are involved in measuring the stabilizer generators to obtain the syndromes for the EA-LDPC codes since the support of the stabilizer generators \footnote{The support of any operator on a quantum code is a set of indices of qubits where the operator does not apply trivial operator (Pauli $I$).} of the EA-quantum code involves all the qubits, i.e., the qubits received from the transmitter as well as halves of the pre-shared ebits with the receiver. Since the halves of the pre-shared ebits at the receiver are not involved in the decoding process, the corresponding nodes can be removed. Consequently, this enables the avoidance of \(4\)-cycles in the entanglement-unassisted portion of the joint Tanner graph of the EA quantum LDPC code.

Several constructions of EA-QLDPC codes have been reported in the literature~\cite{hsieh2011high,hsieh2009entanglement}; however, none of them address the removal of \(4\)-cycles from the entanglement-unassisted portion of the joint Tanner graph, which is crucial in the presence of correlated \(X\) and \(Z\) errors. It is important to note that eliminating \(4\)-cycles in the entanglement-unassisted portion of the joint Tanner graph is feasible only when the code is constructed from two distinct classical codes. In contrast, when an EA-QLDPC code is derived from a single classical LDPC code, \(4\)-cycles in the unassisted portion of the joint Tanner graph are unavoidable.  Hsieh et al. \cite{hsieh2009entanglement} provided an example of an EA-QC-QLDPC code using a single classical QC-LDPC code, where the Tanner graph of underlying classical code is devoid of $4$-cycles. It is desirable to work with a family of EA-QC-QLDPC codes devoid of short cycles with known properties, such as the precise rank of the parity check matrix, etc., along with flexible code rate and block lengths for quantum communication applications. It is also well-known that a large girth in classical LDPC codes leads to improved decoding performance, and this advantage naturally extends to their quantum counterparts. 

% Motivated by the work in~\cite{zhang2012new}, where QC-LDPC codes with Tanner graph girth greater than \(6\) were constructed using Sidon sequences and a parity-check matrix of column weight three, one can generalize this approach by exploring alternative sequences and higher column weights to design QC-LDPC codes with large girth, and subsequently employ these constructions to obtain quantum LDPC codes. 
Therefore, motivated by the works in~\cite{panteleev2021quantum,hagiwara2007quantum,miao2024joint,hsieh2009entanglement}, we propose several families of EA-QC-QLDPC codes with large girth, aimed at improving performance while remaining amenable to hardware implementation. In particular, when EA-QC-QLDPC codes are constructed from two distinct classical QC-LDPC codes, we focus on eliminating \(4\)-cycles in the entanglement-unassisted portion of the joint Tanner graph to mitigate correlated errors.
\subsection{Encoding and Decoding}
Efficient encoding and decoding of EA-QC-QLDPC codes is as crucial as their construction for practical deployment. Encoding techniques for stabilizer and entanglement-assisted quantum codes have been previously proposed in the literature. In particular, the encoder presented in~\cite{nadkarni2021encoding} relies on Gram--Schmidt orthonormalization and does not exploit the underlying structural properties of the quantum code. As a result, the computational complexity of such encoders can become prohibitive for large block lengths. Motivated by these limitations, we focus on the design of efficient encoding for EA-QC-QLDPC codes with structured stabilizer generators, achieving improved computational complexity compared to general stabilizer-based encoders. Furthermore, we demonstrate that the proposed encoding architecture is well-suited for QLDPC codes and yields additional complexity reductions when the underlying classical code exhibits a quasi-cyclic structure.

The overall decoding architecture comprises a syndrome computation unit done classically and a quantum decoding circuit to extract the quantum information, where the quantum decoding circuit is just the inverse of the quantum encoding circuit. We propose a quaternary block-layered normalized min-sum decoder for handling correlated quantum errors in the syndrome computation engine when operating over the joint Tanner graph of the EA-QC-QLDPC code.  The quaternary decoder yields significant performance gains when quantum errors are correlated.
\subsection{Our Contributions}
Our contributions in this paper are the following:
\begin{enumerate}[leftmargin=0pt, itemsep=0pt]
\item [] (a) We propose several families of EA-QC-QLDPC codes by employing single or two classical QC-LDPC  codes as follows: (i) Some of the families are  proposed by using a single QC-LDPC classical code such that the  Tanner graphs of the classical QC-LDPC codes have girths~$> 4$ or $>6$. We also analytically derive the code parameters of the proposed quantum code. (ii) Further, we propose EA-QC-QLDPC code families derived from two different classical QC-LDPC codes such that the entanglement-unassisted portion of the joint Tanner graph is devoid of  $4$-cycles.  Notably, in one of the proposed families of such quantum codes, we require just a \textit{single Bell pair} shared across the quantum transceiver. In the case of a binary decoder, errors are handled by treating X and Z errors independently. In contrast, a quaternary decoder treats the Y error as a single entity and operates on the joint Tanner graph to handle it effectively, achieving significant performance improvements. Consequently, in case of quaternary decoder, one can infer that EA quantum codes whose unassisted portion of the joint Tanner graph is free of 4-cycles will perform better than those whose unassisted portions contain 4-cycles. 
    % \item [] (a) \textcolor{blue}{ We propose several families of EA-QC-LDPC codes that work for a wide class of errors, i.e., when the errors are statistically independent on the X and Z portions, as well as errors that are correlated. (i) When  X and Z errors are independent, we employ the normalized layered binary decoder that operate on the individual Tanner graphs. We propose EA-QC-QLDPC codes for this family using a single QC-LDPC classical code such that the  Tanner graphs of the classical QC-LDPC codes have girths~$> 4$ or $>6$. We also analytically derive the code parameters of the quantum code for both these cases. (ii) Further, we propose EA-QC-QLDPC code families derived from two different classical QC-LDPC codes such that the entanglement-unassisted portion of the overall Tanner graph (joint Tanner graph) is devoid of  $4$-cycles. When X and Z errors are correlated, we use the normalized layered quaternary decoder that operates over the joint Tanner graph for handling correlated errors, yielding significant performance gains. Notably, in one of the proposed families of such quantum codes, we require just a \textit{single Bell pair} shared across the quantum transceiver.} 
      \item[] (b) We develop efficient encoding schemes for the proposed $[[n,k;c]]_2$ code families with gate complexity $O\left(n(\rho_1+\rho_2)-\rho_1^2\right)$, where $\rho_1$ and $\rho_2$ are the number of stabilizer generators for $H_x$ and $H_z$, in the CSS construction, respectively; and $n,k,c$ are the number of code qubits, information qubits and shared EPR pairs, respectively.
    % \item[] (c) We also investigate three transversal logical operators, namely, the Hadamard-SWAP, S-CZ, and Hadamard-S-CZ gates on the EA-QC code that utilizes one Bell pair for first family of codes with an eye to protect the logical state.
    \item[] (c) 
    Finally, we present simulation results for various classes of EA-QC-QLDPC codes over both the random depolarizing channel and the random depolarizing channel with Markovian burst errors. The performance of the proposed EA-QC-QLDPC codes is evaluated against decoder variants such as the layered binary, quaternary, normalized quaternary, and block-layered normalized quaternary decoders. Among these, the normalized block-layered quaternary decoder is assessed to have superior performance compared to others. The proposed EA-QC-QLDPC codes are also shown to correct a constant number of burst errors. 

We further compare the proposed codes with existing constructions. In particular, the $[[42,4]]_{2}$ quantum CSS code in \cite{hagiwara2007quantum} is compared with the proposed EA-QC-QLDPC code having parameters $[[42,10;6]]$, demonstrating an improvement of more than two orders of magnitude. This improvement arises because the $[[42,4]]_{2}$ code is a CSS code and contains $4$-cycles in its joint Tanner graph, which adversely affects the decoding performance under the quaternary decoder. In contrast, the proposed code $[[42,10;6]]$ is designed such that its unassisted portion of the joint Tanner graph is free of $4$-cycles. Since the quaternary decoder operates on the joint Tanner graph, the elimination of such short cycles in the proposed construction results in improved decoding performance. Similarly, the EA quantum code $[[128,58;18]]$ in \cite{hsieh2009entanglement} is compared with the proposed EA-QC-QLDPC code $[[121,70;51]]$, showing an improvement of more than one order of magnitude at a depolarizing noise probability of $10^{-2}$.

Furthermore, a proposed EA-QC-QLDPC code $[[390,132;128]]_{2}$ with girth $>6$ and another proposed code $[[529,176;353]]_{2}$ with girth $>4$ are compared to highlight the performance gains due to larger girth, despite the shorter length of the $[[390,132;128]]_{2}$ code relative to the $[[529,176;353]]_{2}$ code.

Finally, simulation results demonstrate that increasing the column weight of EA-QC-QLDPC codes leads to improved error-correcting performance and enhanced resilience to burst errors.
\end{enumerate}

The paper is organized as follows: Section \ref{sec: preliminaries} contains some basic definitions and preliminary results useful for establishing our main results. In Section \ref{Sec.2}, we revisit classical QC-LDPC codes and provide new code designs based on tiling permutation matrices with prime and composite orders. Several code properties are derived analytically. In Section \ref{Sec.3}, we construct three families of EA-QC-QLDPC codes. One of the families requires a single pre-shared EPR pair. In Section \ref{sec.5}, two families of EA-QC-QLDPC codes are proposed using a single classical QC-LDPC code. For each family, the Tanner graph of the underlying classical QC-LDPC code has girth greater than 6. In Section~\ref{Sec:Encoding}, an efficient encoder is proposed for the EA-QC-QLDPC codes. In Section~\ref{Sec.4}, we assess the performance of the proposed codes over the random depolarizing channel as well as the random depolarizing channel with Markovian bursts under different configurations of decoders through simulations, followed by conclusions in Section~\ref{Sec.5}.

\section{Preliminaries}\label{sec: preliminaries}
For a better presentation of the paper, we have divided this section into two subsections and provided some required definitions and preliminary results, to be useful later.
\subsection{Criterion for the existence of a $2k$-cycle}
In this subsection, we review the existence condition for $2k$-cycles presented in \cite{fossorier2004quasicyclic}.

For a positive integer $n$, let $\mathbb{Z}_{n}=\{0,1,2,\ldots, n-1\}$ be the ring of integers and $P$
% \begin{equation}\label{rightcirculant}
% P=\begin{bmatrix}
% 0 & 1 & 0 & \cdots & 0 \\
% 0 & 0 & 1 & \cdots & 0 \\
% \vdots & \vdots & \vdots & \ddots & \vdots\\
% 0 & 0 & 0 & \cdots & 1 \\
% 1 & 0 & 0 & \cdots & 0
% \end{bmatrix}
% \end{equation}
be a right circulant permutation matrix of order $n$. We define a degree matrix $M$ and the corresponding parity-check matrix $H$  by the following equation:
\begin{equation}\label{modelmatrix}%cycleparity
 	M=\begin{bmatrix}
 		a_{0,0}&a_{0,1}&\cdots &a_{0,t}\\
 		a_{1,0}&a_{1,1}&\cdots &a_{1,t}\\
 		\vdots&\vdots&\ddots&\vdots\\
 		a_{m,0}&a_{m,1}&\cdots&a_{m,t}	
 	\end{bmatrix}, \quad
H=\begin{bmatrix}
P^{a_{0,0}} & P^{a_{0,1}}& \cdots  & P^{a_{0,t}} \\
P^{a_{1,0}} & P^{a_{1,1}} & \cdots  & P^{a_{1,t}} \\
\vdots & \vdots & \ddots & \vdots\\
P^{a_{m,0}} & P^{a_{m,1}} & \cdots &  P^{a_{m,t}} 
\end{bmatrix},
\end{equation}
where $a_{i,j}\in\mathbb{Z}_{n}$ for all $1\leq i\leq m, 1\leq j\leq t$. In the parity-check matrix $H$, defined in \eqref{modelmatrix}, {\small $[P^{a_{i,0}}, P^{a_{i,1}}, \ldots, P^{a_{i,t}}]$} is known as the $i^{\text{th}}$ block-row,  whereas
     {\small $[(P^{a_{0,j}})^{T}, (P^{a_{1,j}})^{T}, \ldots, (P^{a_{m,j}})^{T}]^{T}$} stands for the $j^{\text{th}}$ block-column.
      
 A closed path of length $2k$ in any parity-check matrix of the form in \eqref{modelmatrix} is a sequence of block-row and block-column index pairs $(i_0, j_0), (i_0, j_1), (i_1, j_1), (i_1, j_2), \ldots, (i_{k-1}, j_{k-1}),$ $ (i_{k-1}, j_0)$, with $i_\ell \neq i_{\ell+1}, j_\ell \neq j_{\ell+1}$, for $\ell = 0, 1, \ldots, k-2$, and  $i_{k-1} \neq i_0, j_{k-1} \neq j_0 $.

The significance of closed paths arises from the following simple but important result.
\begin{theorem}\emph{\cite[Theorem 1]{fossorier2004quasicyclic}}\label{2k-cycletheorem}
A cycle of length $2k$ exists in the Tanner graph of a QC-LDPC code with parity-check matrix $H$, defined in \eqref{modelmatrix}, if and only if there exists a closed path $(i_0, j_0), (i_0, j_1), (i_1, j_1), (i_1, j_2), \ldots, (i_{k-1}, j_{k-1}), (i_{k-1}, j_0)$ in $H$ such that
\begin{equation}\label{cyclcondition.1}
P^{a_{i_{0},j_{0}}} (P^{a_{i_{0},j_{1}}})^{-1} P^{a_{i_{1},j_{1}}} (P^{a_{i_{1},j_{2}}})^{-1} \cdots P^{a_{i_{k-1},j_{k-1}}} (P^{a_{i_{k-1},j_{0}}})^{-1}=I~(\text{identity matrix}).
\end{equation}
\end{theorem}
\begin{proof} In \cite{fossorier2004quasicyclic}, the proof of the theorem is not provided. Hence, it is essential to include a complete proof here. We present a detailed proof in Appendix~\ref{2k-cycletheoremproof} by incorporating certain insights from~\cite{lieb2022number}. 
\end{proof}
%%%%%%%%%%%%%%%%%%%%%%%%%%%%%%%%%%%%%%%%%%%%%%%%%%%%%%%%%%%%%%%%%%%%%%%%%%%%%%%%%%%%%
Alternatively, the condition given by \eqref{cyclcondition.1} for the existence of a cycle of length $2k$ in the Tanner graph of a QC-LDPC code with parity-check matrix $H$ can be expressed as:
\begin{equation}\label{cyclcondition.2}
    (a_{i_{0},j_{0}}-a_{i_{0},j_{1}})+(a_{i_{1},j_{1}}-a_{i_{1},j_{2}})+\cdots+(a_{i_{k-1},j_{k-1}}-a_{i_{k-1},j_{0}})=0\pmod{n},
\end{equation}
where $n$ is the size of the right circulant permutation matrix $P$.
% \begin{corollary}\label{4-cyclecoro}
%    There does not exist a cycle of length 4 in the Tanner graph of a QC-LDPC code with parity-check matrix $H$, defined by \eqref{modelmatrix}, if and only if
%    \begin{equation*}
%     (a_{i_{0},j_{0}}-a_{i_{0},j_{1}})+(a_{i_{1},j_{1}}-a_{i_{1},j_{0}})\neq 0~\mathrm{mod }~(n) \textrm{ for all }   0\leq i_{0}<i_{1}\leq m,0\leq j_{0}<j_{1}\leq n.
% \end{equation*}
% \end{corollary}
% \begin{corollary}\label{6-cyclecoro}
%    There does not exist a cycle of length 6 in the Tanner graph of a QC-LDPC  code with parity-check matrix $H$, defined by \eqref{modelmatrix}, if and only if
%    \begin{equation*}
%     (a_{i_{0},j_{0}}-a_{i_{0},j_{1}})+(a_{i_{1},j_{1}}-a_{i_{1},j_{2}})+(a_{i_{2},j_{2}}-a_{i_{2},j_{0}})\neq 0\pmod{n},
% \end{equation*}
% where $i_\ell \neq i_{\ell+1}, j_\ell \neq j_{\ell+1}$, for $\ell = 0, 1$, and  $i_{2} \neq i_0, j_{2} \neq j_0 $.
% \end{corollary}
%%%%%%%%%%%%%%%%%%%%%%%%%%%%%%%%%%%%%%%%%%%%%%%%%%%%%%%%%%%%%%%%%%%%%%%%%%%%

Consider the parity-check matrix $H$ and the associated degree matrix $M$ defined by Eq.~\eqref{modelmatrix}, respectively. Let $i,j,k$ be any three row indices and $p,q,r$ be any three column indices from the matrix $M$. As a consequence of Eq.~\eqref{cyclcondition.2}, all possible 6-cycles along with their conditions are shown in Figure \ref{fig:all-6-cycles}.
\begin{figure}[!h]
    \centering
 {\small$\begin{array}{ccc} 
 \underset{(a_{k,p}-a_{i,p})=(a_{j,q}-a_{i,q})+(a_{k,r}-a_{j,r})}{\begin{tikzpicture}[node distance={14mm}, thick, main/.style]
 				\node[main] (1) {$a_{i,p}$}; 
 				\node[main] (2) [right of=1]  {$a_{i,q}$}; 
 				\node[main] (3) [right of=2]  {}; 
 				\node[main] (4) [below of=1] {}; 
 				\node[main] (5) [below of=2]{$a_{j,q}$}; 
 				\node[main] (6) [below of=3] {$a_{j,r}$}; 
 				\node[main] (7) [below of=4] {$a_{k,p}$}; 
 				\node[main] (8) [below of=5]{}; 
 				\node[main] (9) [below of=6] {$a_{k,r}$};
 				\draw [draw=blue,thick](1) -- (2); 
 				\draw [draw=blue,thick](2) -- (5); 
 				\draw [draw=blue,thick](5) -- (6); 
 				\draw [draw=blue,thick](6) -- (9); 
 				\draw [draw=blue,thick](9) -- (7); 
 				\draw [draw=blue,thick](7) -- (1);
 		\end{tikzpicture}} & \underset{(a_{k,p}-a_{i,p})=(a_{k,q}-a_{j,q})+(a_{j,r}-a_{i,r})}{\begin{tikzpicture}[node distance={14mm}, thick, main/.style] 
 				\node[main] (1) {$a_{i,p}$}; 
 				\node[main] (2) [right of=1]  {}; 
 				\node[main] (3) [right of=2]  {$a_{i,r}$}; 
 				\node[main] (4) [below of=1] {}; 
 				\node[main] (5) [below of=2]{$a_{j,q}$}; 
 				\node[main] (6) [below of=3] {$a_{j,r}$}; 
 				\node[main] (7) [below of=4] {$a_{k,p}$}; 
 				\node[main] (8) [below of=5]{$a_{k,q}$}; 
 				\node[main] (9) [below of=6] {};
 				\draw [draw=red,thick](1) -- (3); 
 				\draw [draw=red,thick](3) -- (6); 
 				\draw [draw=red,thick](6) -- (5); 
 				\draw [draw=red,thick](5) -- (8); 
 				\draw [draw=red,thick] (8) -- (7); 
 				\draw [draw=red,thick](7) -- (1);
 		\end{tikzpicture}} & \underset{(a_{k,r}-a_{i,r})=(a_{j,q}-a_{i,q})+(a_{k,p}-a_{j,p})}{\begin{tikzpicture}[node distance={14mm}, thick, main/.style] 
 				\node[main] (1) {}; 
 				\node[main] (2) [right of=1]  {$a_{i,q}$}; 
 				\node[main] (3) [right of=2]  {$a_{i,r}$}; 
 				\node[main] (4) [below of=1] {$a_{j,p}$}; 
 				\node[main] (5) [below of=2]{$a_{j,q}$}; 
 				\node[main] (6) [below of=3] {}; 
 				\node[main] (7) [below of=4] {$a_{k,p}$}; 
 				\node[main] (8) [below of=5]{}; 
 				\node[main] (9) [below of=6] {$a_{k,r}$};
 				\draw [draw=green,thick] (2) -- (3); 
 				\draw [draw=green,thick](3) -- (9); 
 				\draw [draw=green,thick](9) -- (7); 
 				\draw [draw=green,thick](7) -- (4); 
 				\draw [draw=green,thick](4) -- (5); 
 				\draw [draw=green,thick](5) -- (2);
 		\end{tikzpicture}}\\
        \\
        (\textrm{Type-I})&(\textrm{Type-II})&(\textrm{Type-III})\\
   \\
   \\
   \underset{(a_{k,r}-a_{i,r})=(a_{k,q}-a_{j,q})+(a_{j,p}-a_{i,p})}{\begin{tikzpicture}[node distance={14mm}, thick, main/.style] 
 				\node[main] (1) {$a_{i,p}$}; 
 				\node[main] (2) [right of=1]  {}; 
 				\node[main] (3) [right of=2]  {$a_{i,r}$}; 
 				\node[main] (4) [below of=1] {$a_{j,p}$}; 
 				\node[main] (5) [below of=2]{$a_{j,q}$}; 
 				\node[main] (6) [below of=3] {}; 
 				\node[main] (7) [below of=4] {}; 
 				\node[main] (8) [below of=5]{$a_{k,q}$}; 
 				\node[main] (9) [below of=6] {$a_{k,r}$};
 				\draw [draw=violet,thick](1) -- (3); 
 				\draw[draw=violet,thick] (3) -- (9); 
 				\draw [draw=violet,thick](9) -- (8); 
 				\draw [draw=violet,thick](8) -- (5); 
 				\draw [draw=violet,thick](5) -- (4); 
 				\draw [draw=violet,thick](4) -- (1);
 		\end{tikzpicture}} & \underset{(a_{k,q}-a_{i,q})=(a_{j,r}-a_{i,r})+(a_{k,p}-a_{j,p})}{\begin{tikzpicture}[node distance={14mm}, thick, main/.style] 
 				\node[main] (1) {}; 
 				\node[main] (2) [right of=1]  {$a_{i,q}$}; 
 				\node[main] (3) [right of=2]  {$a_{i,r}$}; 
 				\node[main] (4) [below of=1] {$a_{j,p}$}; 
 				\node[main] (5) [below of=2]{}; 
 				\node[main] (6) [below of=3] {$a_{j,r}$}; 
 				\node[main] (7) [below of=4] {$a_{k,p}$}; 
 				\node[main] (8) [below of=5]{$a_{k,q}$}; 
 				\node[main] (9) [below of=6] {};
 				\draw (2) -- (3); 
 				\draw (3) -- (6); 
 				\draw (6) -- (4); 
 				\draw (4) -- (7); 
 				\draw (7) -- (8); 
 				\draw (8) -- (2);
 		\end{tikzpicture}} & \underset{(a_{k,q}-a_{i,q})=(a_{k,r}-a_{j,r})+(a_{j,p}-a_{i,p})}{\begin{tikzpicture}[node distance={14mm}, thick, main/.style] 
 				\node[main] (1) {$a_{i,p}$}; 
 				\node[main] (2) [right of=1]  {$a_{i,q}$}; 
 				\node[main] (3) [right of=2]  {}; 
 				\node[main] (4) [below of=1] {$a_{j,p}$}; 
 				\node[main] (5) [below of=2]{}; 
 				\node[main] (6) [below of=3] {$a_{j,r}$}; 
 				\node[main] (7) [below of=4] {}; 
 				\node[main] (8) [below of=5]{$a_{k,q}$}; 
 				\node[main] (9) [below of=6] {$a_{k,r}$};
 				\draw[draw=orange,thick](1) -- (2); 
 				\draw[draw=orange,thick](2) -- (8); 
 				\draw [draw=orange,thick](8) -- (9); 
 				\draw [draw=orange,thick](9) -- (6); 
 				\draw[draw=orange,thick] (6) -- (4); 
 				\draw [draw=orange,thick](4) -- (1);
 		\end{tikzpicture}}\\
        \\
         (\textrm{Type-IV})&(\textrm{Type-V})&(\textrm{Type-VI})
   \end{array}$}
    \caption{All possible cycles of length 6 with their corresponding conditions.}
   \label{fig:all-6-cycles} 
\end{figure}
%%%%%%%%%%%%%%%%%%%%%%%%%%%%%%%%%%%%%%%
% \begin{definition}
%     A vector $(v_{1},v_{2},v_{3},\ldots,v_{m})$ over $\mathbb{Z}_{n}$ is said to be \textbf{multiplicity free} if all entries in the vector are different.
% \end{definition}
% \begin{example}
%    The vector $(1, 2, 3, 6)$ is multiplicity free over $\mathbb{Z}_{7}$ whereas $(1,1,5,6)$ is not.
% \end{example} 
\begin{proposition}\label{4-cyclepropo}
 Let $r_{i}=(a_{i,0},a_{i,1},a_{i,2},\ldots,a_{i,t})$ and $r_{j}=(a_{j,0},a_{j,1},a_{j,2},\ldots,a_{j,t})$ be any two rows from the degree matrix $M$, defined by~\eqref{modelmatrix}, of $H$. The Tanner graph of a QC-LDPC code with parity-check matrix $H$ contains no 4-cycles if and only if the difference $r_i - r_j$ has distinct elements under $\pmod{n}$, for every pair $(i,j)$ satisfying $0 \leq i < j \leq m$.
\end{proposition}
\begin{proof}
   The proof follows directly from Eq.~\eqref{cyclcondition.2}.
\end{proof}
%%%%%%%%%%%%%%%%%%%%%%%%%%%%%%%%

%%%%%%%%%%%%%%%%%%%%%%%%%%%%%%%%%%%%%%%%%%%%%%%%%%%%%%%%%%%%%%%%%%%%%%%%%%%%%%%%%%%%%%%%%%%%%%%%%%%%%%%%%%%%%%%%%%%%%%%%%%%%%%%%%%%%%%%%%%%%%%%%%
\subsection{Entanglement-assisted quantum codes (EAQCs)}
EAQCs form a subclass of QECCs that use pre-shared maximally entangled bits (ebits) between the sender and receiver to enhance coding efficiency. This entanglement enables for a more effective utilization of quantum resources, facilitating the transmission of quantum information through noisy channels. EA-QECCs possess multiple advantages. Below, we outline the two primary ones of particular interest in this work.
\begin{enumerate}[leftmargin=0pt]
    \item[] \textbf{Removal of Short Cycles in the Tanner Graph:} In EA-QECCs based on the CSS construction, short cycles can be eliminated in the entanglement-unassisted portion of the joint Tanner graph; see Figure~\ref{tannergraph} for an illustration of this unassisted portion.   
    \item[] \textbf{Utilization of Arbitrary Classical Codes:} Any one or two classical codes can be utilized to construct EA-QECCs using the CSS construction method, as outlined, for qubit case, in the following theorem.
\end{enumerate}
%%%%%%%%%%%%%%%%%%%%%%%%%%%%%%%%%%%%%%%%%%%%%%%%%%%%%%%%%%%%%%%%%%%%%%%%%%%%%%%%%%%%%
\begin{theorem}\label{maintheorem}
    \emph{\cite[Corollary 1]{wilde2008optimal}}
\label{cor:CSS}Let $[n,k_{1},d_{1}]_{2}$ and $[n,k_{2},d_{2}]_{2}$ be two binary linear codes with their parity-check matrices $H_{1}$ and $H_{2}$, respectively. Then there exists an EA quantum code with parameters $[[  n,k_{1}+k_{2}-n+c,\min\left(
d_{1},d_{2}\right)  ;c]]_2$, and
requires $c$ ebits, where
\begin{equation*}
c=\mathrm{rank}\left(  H_{1}H_{2}^{T}\right)  .
\end{equation*}    
\end{theorem}

\section{Classical QC-LDPC Codes: Tiling Permutation Matrices of Prime and Composite orders}\label{Sec.2}
This section is divided into two subsections. In the first subsection, we construct a family of parity-check matrices by tiling permutation matrices of prime order, and we derive the exact rank of the parity-check matrix to compute the code rate. In the second subsection, we introduce a different family of parity-check matrices formed by tiling permutation matrices of composite order, and we obtain an upper bound on the rank of the parity-check matrix, which enables us to compute a lower bound on the code rate.
\subsection{Classical QC-LDPC codes by tiling permutation matrices of prime order}\label{sec.2.1}
Let $\mathbb{F}_{p}=\{0,1,\ldots, p-1\}$ be the finite field of odd prime order~$p$. We define a matrix 
\begin{equation}
	M=\begin{bmatrix}
		b_{0,0} &b_{0,1}&\cdots&	b_{0, p-1}\\
		b_{1,0} &b_{1,1}&\cdots&	b_{1,p-1}\\
		\vdots &\vdots &\ddots&\vdots\\
		b_{p-1,0} &b_{p-1,1}&\cdots&	b_{p-1,p-1}
	\end{bmatrix},
\end{equation}
	where $b_{i,j}\in \mathbb{F}_{p} $ for all $0\leq i,j\leq (p-1)$ and the entries in $M$  are populated based on the Algorithm \ref{algoM}.
 \begin{algorithm}[!h]
	\caption{Constructing the Matrix $M$ in the Class \( C_{M} \)}
 \label{algoM}
	\begin{algorithmic}[1]
            \Procedure{Construct Matrix M}{}
		\State \textbf{Input:} Odd prime  $p$ and $\mathbb{F}_{p}=\{0,1,\ldots,p-1\}$.
		\State \textbf{Output:} Matrix $M \in C_{M}$.
		\State Initialize matrix $M = [b_{i,j}]$, where $0\leq i,j\leq (p-1)$, with all entries set to zero. 
		\For{each $j \in \{1, 2, \ldots, p-1\}$}
		\State Randomly choose $b_{1,j} \in \mathbb{F}_p \setminus \{0\}$ without repetition.
		\EndFor
		\State Choose $k_{1} \in \mathbb{F}_p \setminus \{0, 1\}$
		\For{each $j \in \{0, 1, \ldots, p-1\}$}
		\State Set $b_{2,j} = (k_{1}b_{1,j}) \pmod{p}$.
		\EndFor
		\For{each $i \in \{3, 4, \ldots, p-1\}$}
		\State Choose $k_{i-1} \in \mathbb{F}_{p} \setminus \{0, 1, k_{1}, \ldots, k_{i-2}\}$.
		\For{each $j \in \{0, 1, \ldots, p-1\}$}
		\State Set $b_{i,j} = (k_{i-1}b_{1,j}) \pmod{p}$.
		\EndFor
		\EndFor
		\State \Return $M$
		\EndProcedure
	\end{algorithmic}
\end{algorithm} 
Let $C_{M}$ be the collection of all possible distinct $p\times p$ matrices constructed using the Algorithm \ref{algoM}. For each $M\in C_{M}$, we define a block matrix~$\mathbf{A}_{M}$
as follows:
\begin{equation}\label{A}
	\mathbf{A}_{M}=
		\begin{bmatrix}
P^{b_{0,0}} & P^{b_{0,1}} & \cdots & P^{b_{0,p-1}} \\
P^{b_{1,0}} & P^{b_{1,1}} & \cdots & P^{b_{1,p-1}} \\
\vdots & \vdots & \ddots & \vdots \\
P^{b_{p-1,0}} & P^{b_{p-1,1}} & \cdots & P^{b_{p-1,p-1}}
\end{bmatrix},
	\end{equation}
where $P$ is the right circulant permutation matrix of order~$p$ and $M$ is known as the degree matrix of $\mathbf{A}_{M}$. Define a class $\mathcal{C}_{A}$ of block matrices associated with the class $C_{M}$ as follows:
\begin{equation}\label{classCA}
 \mathcal{C}_{A}=\{\mathbf{A}_{M} \mid M \in C_{M}\}.
\end{equation}
Let $r_{i}$ and $r_{j}$ be two distinct arbitrary rows from the matrix $M$. 
% (defined by \eqref{M}).
Then there exist distinct $k_{i},k_{j}\in \mathbb{F}_{p}$ such that $r_{i}=(k_{i}x_{0},k_{i}x_{1},\ldots,k_{i}x_{(p-1)})$ and $r_{j}=(k_{j}x_{0},k_{j}x_{1},\ldots,k_{j}x_{(p-1)})$, where $x_{q}\in \mathbb{F}_{p}$, for all $0\leq q\leq(p-1)$ and $x_{\lambda}\neq x_{\mu}$, for $\lambda\neq \mu$. It is easy to check that the vector $r_{i}-r_{j}$ has non-repeated elements under modulo~$p$. Therefore, by Proposition~\ref{4-cyclepropo}, the Tanner graph of the classical QC-LDPC code having $\mathbf{A}_{M}$ as a parity-check matrix is devoid of cycles of length 4.

To determine the code rate of the classical QC-LDPC code corresponding to a parity check matrix $\mathbf{A}_{M}$ in the class $\mathcal{C}_{A}$, first we compute the rank of the following particular block matrix $\mathbf{H}$ from the class $\mathcal{C}_{A}$
\begin{equation}\label{H}
		\mathbf{H}=\begin{bmatrix}
h_{0,0} & h_{0,1} & \cdots & h_{0,p-1} \\
h_{1,0} & h_{1,1} & \cdots & h_{1,p-1} \\
\vdots & \vdots & \ddots & \vdots \\
h_{p-1,0} & h_{p-1,1} & \cdots & h_{p-1,p-1}
\end{bmatrix},
	\end{equation}
    where $h_{i,j}=P^{\text{mod}(ij, p)}$ 
and then we comment on $\text{rank}(\mathbf{A}_{M})$. We require the following lemma to compute the $\text{rank}(\mathbf{H})$.
\begin{lemma}\label{lemma1}\emph{\cite{EA_ITW}} For $0\leq i,k\leq (p-1)$, let $R_{i}^{(k)}$ represents $i^{\text{th}}$ row in the $k^{\text{th}}$ block-row  of $\mathbf{H}$, and let $\langle(v_{1},\ldots,v_{p^{2}}),(u_{1},\ldots,u_{p^{2}})\rangle=v_{1}u_{1}+\cdots+v_{p^{2}}u_{p^{2}}$, where $v_{i}$ and $u_{j}$ are the binary scalars for all $i$ and $j$. Then, we have
	\begin{center}
		$\langle R_{i}^{(k)},R_{j}^{(l)}\rangle=\begin{cases}
			p,&\text{if } i=j\text{ and }k=l,\\
			0,&\text{if } i\neq j\text{ and }k=l,\\
			1,&\text{otherwise}.
		\end{cases}$
	\end{center}
	% \begin{equation*}
		% 	\langle R_{i}^{(k)},R_{j}^{(l)}\rangle=\begin{cases}
			% 		p,&\text{if } i=j\text{ and }k=l;\\
			% 		0,&\text{if } i\neq j\text{ and }k=l;\\
			% 		1,&\text{otherwise},
			% 	\end{cases}
		% \end{equation*}
\end{lemma}
% \begin{proof} First, with $k=l$, it is easy to check that $\langle R_{i}^{(k)},R_{j}^{(l)}\rangle=
%         p \text{ if }i=j$ and $\langle R_{i}^{(k)},R_{j}^{(l)}\rangle=
%         0,\text{ otherwise.}$
% % \begin{equation*}
% % \langle R_{i}^{(k)},R_{j}^{(l)}\rangle=
% %     \begin{cases}
% %         p,&\text{ if }i=j,\\
% %         0,&\text{ otherwise.}
% %     \end{cases}
% % \end{equation*}
% Next, we consider the case $k\neq l$. One can easily see that the position of entry 1 in the $j^{\text{th}}$  block-column \footnote{$C^{(j)}=[
% 	h_{0,j}^{\text{T}},h_{1,j}^{\text{T}},\ldots,h_{p-1,j}^{\text{T}}]^{\text{T}}$  denotes $j^{\text{th}}$  block column of $\mathbf{H}$} of $R_{r}^{(s)}$ is $jp+\text{mod}(r+\text{mod}(js,p),p)$.	Let $R_{x}^{(k)}$ and $R_{y}^{(l)}$ be arbitrary given rows from two arbitrary block-rows $k$  and $l$, respectively. If we choose $j=\frac{(y-x)}{(k-l)}\in \mathbb{F}_{p}$, then it is easy to check that
% 	\begin{equation*}
% 		 jp+\text{mod}(x+\text{mod}(kj,p),p)
% 		=jp+\text{mod}(y+\text{mod}(lj,p),p),
% 		\end{equation*}	
% which concludes that $R_{x}^{(k)}$ and $R_{y}^{(l)}$ have the same position of entry 1 in $j^{\text{th}}$ block-column. We cannot find more than one such $j$, since by construction, the Tanner graph of $\mathbf{H}$ is devoid of cycles of length 4. So, for each pair $(R_{x}^{(k)}, R_{y}^{(l)})$  there exists a unique $j$ in $\mathbb{F}_{p}$ such that $\langle R_{i}^{(k)}, R_{j}^{(l)}\rangle=1$, proving the result.
% \end{proof}
\begin{remark}\label{re1}
	The above lemma also holds for any block matrix from the class $\mathcal{C}_{A}$.
\end{remark}
\begin{theorem}\emph{\cite{EA_ITW}}\label{theorem1}
	Let $H$ be a submatrix  of $\mathbf{H}$ consisting distinct $k$ $(1\leq k\leq p)$ block-rows. Then, $\mathrm{rank}(H)=p+(k-1)(p-1)$.
\end{theorem}
\begin{corollary}\label{cor1}
	Let $B$ be any submatrix of $\mathbf{A}_{M}$ 
	% (defined by \eqref{H}) 
	with distinct $\lambda$ $(1\leq \lambda\leq p)$ block-rows. Then, $\mathrm{rank}(B)=p+(\lambda-1)(p-1)$.
\end{corollary}
% \begin{proof} Let $B$ be any submatrix of $\mathbf{A}_{M}$ consisting distinct $\lambda$ number of block-rows, and let 
% 	 \begin{equation}
% 		D=\begin{bmatrix}
% 			d_{0,0} &d_{0,1}&\cdots&	d_{0,p-1}\\
% 			d_{1,0} &d_{1,1}&\cdots&	d_{1,p-1}\\
% 			\vdots &\vdots &\ddots&\vdots\\
% 			d_{\lambda-1,0} &d_{\lambda-1,1}&\cdots&	d_{\lambda-1,p-1}
% 		\end{bmatrix}
% 	\end{equation}
%     	be the corresponding degree matrix of $B$. By employing elementary row and column operations on the degree matrix $D$, we can get the degree matrix associated with some submatrix of $\mathbf{H}$ in \eqref{H} consisting of distinct $\lambda$ block-rows. This completes the proof, since the $\mathrm{rank}(\mathbf{H})$ is given in Theorem \ref{theorem1}.
% \end{proof}

\begin{example}
	Let $E$ be any submatrix of $\mathbf{A}_{M}$
	% (defined by \eqref{H}) 
	consisting distinct $\lambda$ $(1\leq \lambda\leq p)$ block rows.
	Then, by Corollary \ref{cor1}, there exists a classical QC-LDPC code $\mathcal{C}$ with parameters $[p^{2},(p-1)(p-\lambda+1)]_{2}$ such that the Tanner graph of $\mathcal{C}$ has girth $>4$.
\end{example}

\subsection{Classical QC-LDPC codes by tiling permutation matrices of composite order}\label{sec.2.2}
Let $\mathbb{Z}_n$ denote a finite ring with $n$ elements, where $n$ is any positive integer. Given two positive integers $q$ and $r$ such that $q < r$, we aim to find the smallest composite number $n$ such that the following matrix $G$ can be defined:

\begin{equation}\label{G}
    G =\left[\begin{array}{c|ccc}
         a_{0,0} & a_{0,1} & \cdots & a_{0,r-1} \\
         \hline
        a_{1,0} & a_{1,1} & \cdots & a_{1,r-1} \\
        \vdots  & \vdots  & \ddots & \vdots \\
        a_{q-1,0} & a_{q-1,1} & \cdots & a_{q-1,r-1}
    \end{array} \right],
\end{equation}
where $a_{i,j}=0$ (for $i=0$ or $j=0$), $a_{i,j} \in \mathbb{Z}_n\setminus \{0\}$ for all $1 \leq i \leq q-1$ and $1 \leq j \leq r-1$,  and the nonzero entries are populated according to Algorithm \ref{algoG}. We now define $C_G$ as the set of all possible $G$ matrices that can be constructed using Algorithm \ref{algoG} for the same $n$.
\begin{algorithm}[!h]
	\caption{Constructing the Matrix $G$ in the Class $C_{G}$}
    \label{algoG}
	\begin{algorithmic}[1]
		\Procedure{Construct Matrix G}{}
		\State \textbf{Input:} Composite integer $n$, $\mathbb{Z}_{n}=\{0,1,\ldots,n-1\}$ and integers \( q, r \) such that $q < r < n$.
		\State \textbf{Output:} Matrix $G \in C_{G}$.	
		\State Initialize matrix $G' = [a_{i,j}]$, where $1 \leq i \leq q-1 $ and $1 \leq j \leq r-1$, with all entries set to zero.		
		\For{each $i \in \{1, 2, \ldots, q-1\}$}
		\State Initialize $\text{itr}_{\text{max}}=\frac{(n-1)!}{(n-r)!}$.
		\State Create a set $S = \{1, \ldots, i\}$. \Comment{Include all the indices of the filled rows}.
		
		\If{itr $<\text{itr}_{\text{max}}$}		
		\State Randomly generate a vector $R_{i}$ of length $(r-1)$ over $\mathbb{Z}_{n} \setminus \{0\}$ such that the entries of $R_{i}$ are distinct, $R_{i} \neq R_{j}$ for all $j<i$, and $R_{i}$ is not in the set of previously generated vectors.
        \State Replace the $i^{\text{th}}$ row in $G'$ by the vector $R_{i}$.
		\For{each $j \in S\setminus\{i\}$}
		\If{The vector $\text{mod}((R_{i}-R_{j}), n)$ contains repeated entries from $\mathbb{Z}_{n}$}
		\State \textbf{Error:} $\text{itr}=\text{itr}+1$ 
		\Return to step 8
		\EndIf
		\EndFor
		\Else
		\State Choose next $n$ and go to step 2.		
		\EndIf
		\EndFor		
		\State \Return \( G \)
		\EndProcedure
	\end{algorithmic}
\end{algorithm}

Let $P$ be the right circulant permutation matrix of order $n$. We define a class $\mathcal{C}_{K}$ of block matrices $\mathbf{K}_{G}=[P^{a_{i,j}}]$ associated with $G\in C_{G}$.
% \begin{equation}\label{K}
	% 	\mathbf{K}_{G}=[P^{a_{ij}}]. 		
	% \end{equation}
In the subsequent theorem, we have provided an upper bound on the rank of the matrix $\mathbf{K}_{G}$.
\begin{theorem}\emph{\cite{EA_ITW}}\label{theorem2}
	Let $\mathbf{K}_{G}$ be a matrix from the class $\mathcal{C}_{K}$
	%defined by \eqref{K},
	and let $S$ be a set such that  $i\in S$ if and only if $\mathrm{gcd}(a_{i,0},a_{i,1},\ldots,a_{i,r-1})\geq2$ and $\mathrm{gcd}(a_{i,0},a_{i,1},\ldots,a_{i,r-1})$ divides $n$, where $(a_{i,0},a_{i,1},\ldots,a_{i,r-1})$ is the $i^{\mathrm{th}}$ $(0\leq i\leq (q-1))$ row in the degree matrix of $\mathbf{K}_{G}$. Then we  have 
	\begin{equation*}
		\mathrm{rank}(\mathbf{K}_{G})\leq 1+q(n-1)+|S|-\sum_{i\in S }\mathrm{gcd}(a_{i,0},\ldots,a_{i,r-1}).
	\end{equation*} 
\end{theorem}
 
\begin{proof} The proof of the theorem is presented in \cite{EA_ITW}; however, for completeness and clarity, we include a detailed version here.
Let $\text{gcd}\left(a_{i,0}, a_{i,1},\ldots, a_{i,r-1}\right)=d\geq 2$ for some $i^{\mathrm{th}}$ row. Now, consider the $j^{\mathrm{th}}$ block-column of $i^{\mathrm{th}}$ block-row of the matrix $\mathbf{K}_{G}$. Since $d$ is a divisor of $n$ and $a_{i,j}$, we assume that $n=\ell d$, $a_{i,j}=md$ for some positive integers $\ell$ and $m$, where $m<\ell$. We partition the identity  matrix $I_{n}$ and $P^{a_{i,j}}$ into $\ell$ parts, each part with $d$ rows as in Figure~\ref{partition}. 
\begin{figure}[!h]
	\centering   
	$
{\small\left.\left[\begin{array}{cccc}	
	\textcolor{blue}{\begin{matrix}
		1 & 0 & 0 & \cdots & 0 \\
		0 & 1& 0& \cdots & 0\\
		\vdots & \vdots & \vdots & \ddots & \vdots \\
		0 & 0 & 0 & \cdots & 1
		\end{matrix}} &
	\left.\textcolor{red}{\begin{matrix}
	0 & 0 & 0 & \cdots & 0 \\
	0 & 0 & 0 & \cdots & 0 \\
	\vdots & \vdots & \vdots & \ddots & \vdots \\
	0 & 0 & 0 & \cdots & 0
	\end{matrix}}\right\} \text{$d$ rows} &
	\cdots &
	\textcolor{red}{\begin{matrix}
	0 & 0 & 0 & \cdots & 0 \\
	0 & 0 & 0 & \cdots & 0 \\
	\vdots & \vdots & \vdots & \ddots & \vdots \\
	0 & 0 & 0 & \cdots & 0
	\end{matrix}}\\
	\hline
	\textcolor{red}{\begin{matrix}
	0 & 0 & 0 & \cdots & 0 \\
	0 & 0 & 0 & \cdots & 0 \\
	\vdots & \vdots & \vdots & \ddots & \vdots \\
	0 & 0 & 0 & \cdots & 0
	\end{matrix}}&
	\left.\textcolor{blue}{
		\begin{matrix}
		1 & 0 & 0 & \cdots & 0 \\
		0 & 1 & 0 & \cdots & 0 \\
		\vdots & \vdots & \vdots & \ddots & \vdots \\
		0 & 0 & 0 & \cdots & 1
		\end{matrix}
	}\right\} \text{$d$ rows} &
	\cdots & 
	\textcolor{red}{\begin{matrix}
	0 & 0 & 0 & \cdots & 0 \\
	0 & 0 & 0 & \cdots & 0 \\
	\vdots & \vdots & \vdots & \ddots & \vdots \\
	0 & 0 & 0 & \cdots & 0
	\end{matrix}} \\
	\hline
	\vdots & \vdots & \vdots & \vdots \\
	\hline
	\textcolor{red}{\begin{matrix}
	0 & 0 & 0 & \cdots & 0 \\
	0 & 0 & 0 & \cdots & 0 \\
	\vdots & \vdots & \vdots & \ddots & \vdots \\
	0 & 0 & 0 & \cdots & 0
	\end{matrix}} & \left. \textcolor{red}{\begin{matrix}
	0 & 0 & 0 & \cdots & 0 \\
	0 & 0 & 0 & \cdots & 0 \\
	\vdots & \vdots & \vdots & \ddots & \vdots \\
	0 & 0 & 0 & \cdots & 0
	\end{matrix}}\right \} \text{$d$ rows}
	 &
	\cdots &
	 \textcolor{blue}{
		\begin{matrix}
		1 & 0 & 0 & \cdots & 0 \\
		0 & 1 & 0 & \cdots & 0 \\
		\vdots & \vdots & \vdots & \ddots & \vdots \\
		0 & 0 & 0 & \cdots & 1
		\end{matrix}
	}
	\end{array}\right]\right\}}
 \text{$\ell$ parts}$
	\caption{Partition of the identity matrix into $\ell$ parts where each part has $d$ number of rows.}
	\label{partition}
\end{figure}

Let $1\leq t\leq d$, then the positions of entry 1 in all the $t^{\mathrm{th}}$ rows from each part of the partition of the identity matrix are given by the set $Y_{1}=\{t,t+d,t+2d,\ldots,t+(\ell-1)d\}.$
Similarly, the positions of entry 1 in all the $t^{\mathrm{th}}$ rows from each part of the partition of the matrix $P^{a_{i,j}}$ are given by the following  set:
\begin{equation*}
    Y_{2}=\{\text{mod}(a_{i,j}+t,n),\text{mod}(a_{i,j}+d+t,n), \text{mod}(a_{i,j}+2d+t,n)\ldots,\text{mod}(a_{i,j}+(\ell-1)d+t,n)\}.
\end{equation*}
By substituting $n=\ell d$ and $a_{i,j}=md$ in the set $Y_{2}$, we have
	\begin{equation*}\label{eq10}
			Y_{2}=\left\{\text{mod}(md+t,\ell d),\text{mod}((m+1)d+t,\ell d),\ldots,\text{mod}((m+(\ell-1))d+t,\ell d)\right\}.
		\end{equation*}
	For any $x\neq y$, where $0\leq x,y\leq (\ell-1)$, we have
    \begin{align*}
        (m+x)d+t&=(m+y)d+t\pmod{\ell d}\\ 
	\iff (m+x)&=(m+y)\pmod{\ell}\quad(\text{since } \text{gcd}(d,\ell d)=d)\\
    \iff x&=y\pmod{\ell}.
    \end{align*}	
Therefore, all the elements are distinct in the set $Y_{2}$.
By the division algorithm for $\ell d$ and $(m+x)d+t$, there exist unique integers $q^\prime$ and $r^\prime$ such that
	\begin{equation*}\label{eq11}
			(m+x)d+t=(\ell d)q'+r',\text{ where } 0\leq r'<\ell d.
		\end{equation*}
Consequently, we have $ (r'-t)=kd \pmod{\ell d}, \text{ for some } k, \text{ where } 0\leq k\leq(\ell-1),$
	which concludes that $(m+x)d+t=kd+t$ under modulo $(\ell d)$ for some $k$, where $0\leq k\leq(\ell-1)$.	Hence, one can conclude that the set  $Y_{2}$  can equivalently be written as follows:
	\begin{equation*}
			Y_{2}=\{t,t+d,t+2d,\ldots,t+(\ell-1)d\} \pmod{n},
	\end{equation*}
    with the exception that $0 \equiv \ell d \pmod{n}$.
So, the vector sum of all the $t^{\mathrm{th}}$ rows from each part of the partition of the identity matrix is equal to the vector sum of all the $t^{\mathrm{th}}$ rows from each part of the partition of $P^{a_{i,j}}$. By following similar arguments, we can easily show that this holds for every block-column. Hence, for each $t$, we have $2\ell$  linearly dependent rows from the first block row and the $i^{\mathrm{th}}$ block-row. So, we can delete $d=\mathrm{gcd}(a_{i,0},\ldots,a_{i,r-1})$ rows from $i^{\mathrm{th}}$ block-row. By assumption, we have |S| number of such block-rows, consequently, we can delete total $\sum_{i\in S }\mathrm{gcd}(a_{i,0},\ldots,a_{i,r-1})$ number of rows from the all such block-rows. If the $\mathrm{gcd}(a_{i,0},a_{i,1},\ldots,a_{i,r-1})=1$ or $\mathrm{gcd}(a_{i,0},a_{i,1},\ldots,a_{i,r-1})$ does not divide $n$, then, by fixing the first block-row, we can delete any row from such a block-row since the vector sum of all the rows in each block-row is an all-ones vector. This proves the result.          
\end{proof}
\begin{example} Let 
	\begin{equation*}
		H=\begin{bmatrix}
			I&I&I&I&I\\
			I&P&P^{2}&P^{3}&P^{4}\\
			I&P^{2}&P^{4}&P^{6}&P^{8}\\
		\end{bmatrix}
	\end{equation*}
be a parity-check matrix, for $n=10$.
	Computationally, we get $\mathrm{rank}(H)=27$ that meets the upper bound given in Theorem \ref{theorem2} with the corresponding   $[50,23]_{2}$  QC-LDPC code.
	% So, there exists a classical QC LDPC code $\mathcal{C}$ with the parameters $[50,23]$ such that the Tanner graph of $\mathcal{C}$ has girth $>4$.
\end{example}
%%%%%%%%%%%%%%%%%%%%%%%%%%%%%%%%%%%%%%%%%%%%%%%%%%%%%%%%%%%%%%%%%%%%%%%%%%%%%%%%%%%%%%%%%%%%%%%%%%%%%%%%%%%%%%%%%%%%%%%%%%%%%%%%%%%%%%%%%
\section{EA-QC-QLDPC Code with Tanner Graph of Girth Greater than 4}\label{Sec.3}
In this section, we discuss the construction of EA-QC-QLDPC codes. Based on Theorem \ref{maintheorem}, we construct two families of EA-QC-QLDPC codes, depending on the classical QC-LDPC codes used, i.e., one in which a single classical QC-LDPC code is used, and another where two distinct classical QC-LDPC codes are used. Furthermore, when the EA-QC-QLDPC code is constructed using two different classical QC-LDPC codes, the entanglement-unassisted portion of joint Tanner graph has girth greater than 4.

 Let 
{\small\begin{equation}\label{H1}
H_{x}=\begin{bmatrix}
		A_{0,0} &A_{0,1}&\cdots&A_{0,p-1}\\
		A_{1,0} &A_{1,1}&\cdots&A_{1, p-1}\\
		\vdots &\vdots &\ddots&\vdots\\
		A_{\ell_{1}-1,0} &A_{\ell_{1}-1,1}&\cdots&	A_{\ell_{1}-1,p-1}
	\end{bmatrix},\quad  H_{z}=\begin{bmatrix}
		B_{0,0} &B_{0,1}&\cdots&B_{0,p-1}\\
		B_{1,0} &B_{1,1}&\cdots&	B_{1,p-1}\\
		\vdots &\vdots &\ddots&\vdots\\
		B_{\ell_{2}-1,0} &B_{\ell_{2}-1,1}&\cdots&	B_{\ell_{2}-1,p-1}
	\end{bmatrix},
\end{equation}}
% \begin{align}
% 	H_{x}&=\begin{bmatrix}
% 		A_{0,0} &A_{0,1}&\cdots&A_{0,p-1}\\
% 		A_{1,0} &A_{1,1}&\cdots&A_{1, p-1}\\
% 		\vdots &\vdots &\ddots&\vdots\\
% 		A_{\ell_{1}-1,0} &A_{\ell_{1}-1,1}&\cdots&	A_{\ell_{1}-1,p-1}
% 	\end{bmatrix}, \label{H1} \\
% 	H_{z}&=\begin{bmatrix}
% 		B_{0,0} &B_{0,1}&\cdots&B_{0,p-1}\\
% 		B_{1,0} &B_{1,1}&\cdots&	B_{1,p-1}\\
% 		\vdots &\vdots &\ddots&\vdots\\
% 		B_{\ell_{2}-1,0} &B_{\ell_{2}-1,1}&\cdots&	B_{\ell_{2}-1,p-1}\label{H2}
% 	\end{bmatrix},
% \end{align}
be two submatrices of the matrix $\mathbf{A}_{M}$ (refer to Eq.~\eqref{A}).
Assume that $1\leq \ell_{1},\ell_{2}\leq (p-1)$ and $ \ell_{1}+\ell_{2}\leq p$. In addition, consider that there is no block-row common to $H_{x}$ and $H_{z}$. Further, let all the block-rows in each matrix be distinct. In the following theorem we construct a family of EA-QC-QLDPC codes based on $H_{x}$ and $H_{z}$ in Eq.~\eqref{H1}.
\begin{theorem}\emph{\cite{EA_ITW}}\label{theorem3}
	Let $C_{1}$ and $C_{2}$ be two classical QC-LDPC codes, and let $H_{x}$ and $H_{z}$ be their parity-check matrices, respectively, defined in Eq.~\eqref{H1}. Then there exists an EA-QC-QLDPC code with the parameters $[[p^{2}, p^{2}-2p-(p-1)(\ell_{1}+\ell_{2}-2)+1;1]]_2$. Furthermore, the unassisted portion of the joint Tanner graph of the EA-QC-QLDPC code has girth~$>4$.
\end{theorem}
\begin{proof} The proof of the theorem is omitted in \cite{EA_ITW}. For completeness and clarity, we provide a detailed proof here.

Let $H_{x}$ and $H_{z}$, defined in Eq.~\eqref{H1}, be the parity-check matrices of the classical QC-LDPC codes  $C_{1}$ and $C_{2}$, respectively. Then, by Theorem~\ref{maintheorem}, there exists an EA-QC-QLDPC code with parameters $[[n, k_{1}+k_{2}-n+c; c]]_{2}$, where $n$ denotes the length of the codes $C_{1}$ and $C_{2}$, $k_{i}$ denotes the dimension of the QC-LDPC code $C_{i}$, and $c$ is the Minimum number of maximally entangled bits (ebits) required.  The length of the proposed EA quantum code is $p^{2}$ which is evident, since each matrix $A_{i,j}$ or $B_{i,j}$ within the matrices $H_x$ and $H_z$ is a right circulant permutation matrix of size $p$, and there are $p$ such matrices in each block-row. The Minimum number of maximally entangled bits required for the constructed EA-QC-QLDPC code is $1$, which is given by
\begin{equation*}
    \mathrm{rank}(H_{x}H_{z}^{\mathrm{T}})=\mathrm{rank}\left(J_{p\ell_{1}\times p\ell_{2}}\right), \text{ by Remark~\ref{re1}}, 
\end{equation*}
where $J_{p\ell_{1}\times p\ell_{2}}$ is an all-ones matrix. Therefore, $\mathrm{rank}(H_{x}H_{z}^{\mathrm{T}})=1$.
Moreover, $H_x$ and $H_z$ are submatrices of the matrix $\mathbf{A}_M$ defined in Eq.~\eqref{A}.
 Therefore, by Corollary~\ref{cor1}, we have 
\begin{equation}\label{eq:rankofhxandhz}
    \mathrm{rank}(H_{x})=p+(\ell_{1}-1)(p-1) \text{ and } \mathrm{rank}(H_{z})=p+(\ell_{2}-1)(p-1).
\end{equation}
Eq.~\eqref{eq:rankofhxandhz}, together with the minimum number of required ebits and the code length, directly yields the dimension of the proposed EA-QC-QLDPC code constructed from $H_{x}$ and $H_{z}$. By the selection of $H_x$ and $H_z$, the unassisted portion of the joint Tanner graph of the EA-QC-QLDPC code has girth greater than 4; hence, the result follows.
\end{proof}
%%%%%%%%%%%%%%%%%%%%%%%%%%%%%%%%%%%%%%%%%%%%%%%%%%%%%%%%%%%%%%%%%%%%%%%%%
\begin{example}
	For $p=3$, let  $H_{x}=\begin{bmatrix}
		I& P &P^2
	\end{bmatrix}$ and $ H_{z}=\begin{bmatrix}
		I& P^{2} &P
	\end{bmatrix}$
	be two parity-check matrices of the classical QC-LDPC codes $\mathcal{C}_{1}$ and $\mathcal{C}_{2}$, respectively. Then, by Theorem~\ref{theorem3}, there exists an EA-QC-QLDPC code with the parameters $[[9, 4, 2; 1]]_2$. Moreover, the associated Tanner graph is shown in Figure~\ref{tannergraph}.
	\begin{figure}
		\centering		\includegraphics[width=\columnwidth,height=2.5in]{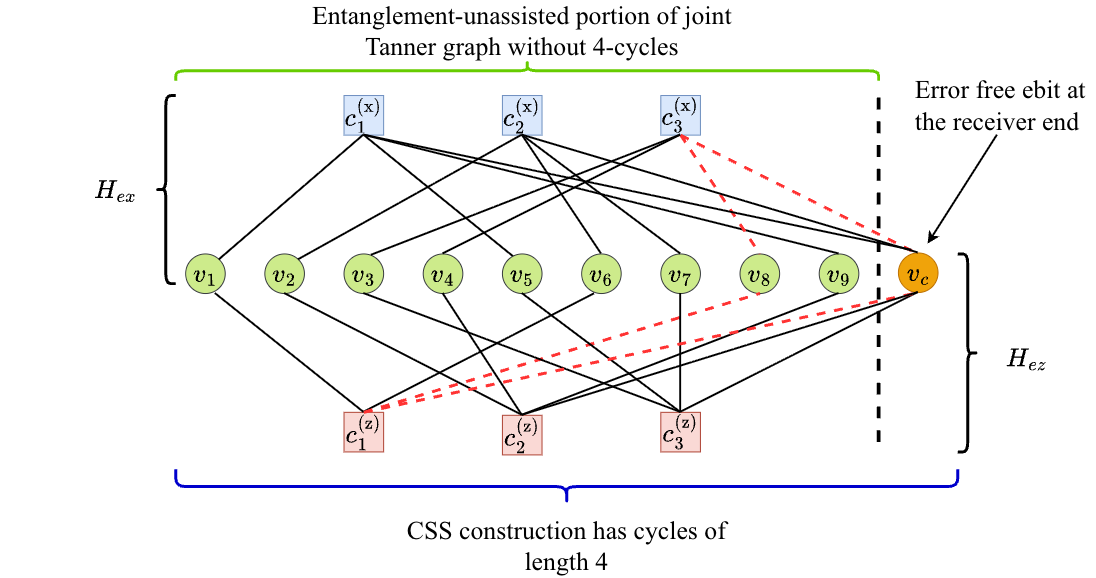}
		\caption{Tanner graph for the code $[[9,4,2;1]]_2$ contains nine variable nodes $v_1,\ldots,v_9$ at the transmitter end and variable node $v_c$ at the receiver end corresponding to the pre-shared entangled bits. Since the node $v_c$ is assumed to be error-free, it does not participate in the decoding. For entanglement-unassisted case, the Tanner graph has short cycles of length 4, as shown by red dotted lines.}
		\label{tannergraph}   
	\end{figure}
\end{example}
%%%%%%%%%%%%%%%%%%%%%%%%%%%%%%%%%%%%%%%%%%%%%%%%%%%%%%%%%%%%%%%%%%%%%%%%%%%%%%%%%%%%%%%%%%%%%%%%%%%%%%%%%%%%%%%%%%%%%%%%%%%%%%%%%%%%%%%%%%%%%%%%%%%%%%%%%%%%%%%%%%%%%%%%%%%%%%%%%%%%%%%%%%%%%%%%%%%%%%%%%%%%%%%%%%%%%%%%%%%%%%%%%%%%%%%%%%%%%%%%%%%%%%%%%%%%%%%%%%%%%%%%%%%%%%%%%%%%%%%%%%%%%%%%%%%%%%%%%%%%%%%%%%%%%%%%%%%%%%%%%%%%%%%%%%%%%%%%%%%%%%%%%%%%%%%%%%%%%%%%%%%%%%%%%%%%%%%%%%%%%%%%%%%%%%%%%%%%%%%%%%%%%%%%%%%%%%%%%%%%%%%%%%%%%%%%%%%%%%%%%%%%%%%%%%%%%%%%%%%%%%%%%%%%%%%%%%%%%%%%%%%%%%%%%%%%%%%%%%%%%%%%%%%%%%%%%%%%%%%%%%%%%%%%%%%%%%%%%%%%%%%%%%%%%%%%%%%%%%%%%%%%%%%%%%%%%%%%%%%%%%%%%%%%%%%%%%%%%%%%%%%%%%%%%%%%%%%%%%%%%%%%%%%%%%%%%%%%%%%%%%%%%%%%%%%%%%%%%%%%%%%%%%%%%%%%%%%%%%%%%%%%%%%%%%%%%%%%%%%%%%%%%%%%%%%%%%%%%%%%%%%%%%%%%%%%%%%%%%%%%%%%%%%%%%%%%%%%%%%%%%%%%%%%%%%%%%%%%%%%%%%%%%%%%%%%%%%%%%%%%%%%%%%%%%%%%%%%%%%%%%%%%%%%%%%%%%%%%%%%%%%%%%%%%%%%%%%%
\begin{example}  For $p=7$, let
			 \begin{equation*}
			     H_{x}=\begin{bmatrix}
				     I& I &I &I &I & I& I\\
				      I& P &P^2 &P^3 &P^4 &P^{5}&P^{6}\\
				      I& P^{2} &P^{4} &P^{6} &P &P^{3}&P^{5}
				 \end{bmatrix} \text{ and } 
			      H_{z}=\begin{bmatrix}
				    I& P^{4} &P &P^{5} &P^{2} &P^{6}&P^{3}\\
				      I& P^{5} &P^{3} &P &P^{6} &P^{4}&P^{2}\\
				       I& P^{6} &P^{5} &P^{4} &P^{3} &P^{2}&P
				 \end{bmatrix}
			 \end{equation*}   	
be two parity-check matrices of the classical QC-LDPC codes $\mathcal{C}_{1}$ and $\mathcal{C}_{2}$, respectively. Then, by Theorem \ref{theorem3}, there exists an EA-QC-QLDPC code with the parameters $[[49,12,6;1]]_2$. The minimum stabilizer generators of the code are as follows:
\begin{enumerate}[label=, leftmargin=0pt, labelwidth=0pt, itemsep=0pt]
    \item \textbf{$X$-stabilizers:}  $(I^{\otimes i}\otimes X\otimes I^{\otimes(6-i)})^{\otimes 7}\otimes X$ is the stabilizer generator corresponding to $i^{\mathrm{th}}$ ($0\leq i\leq 6$) row in the first block-row of $H_{x}$. Further, $\otimes_{w=0}^{6}(I^{\otimes(w+i)}\otimes X\otimes I^{\otimes(6-w-i)})\otimes X$ and $\otimes_{w=0}^{6}(I^{\otimes(2w+i)}\otimes X\otimes I^{\otimes(6-2w-i)})\otimes X$ are the stabilizer generators corresponding to the $i^{\mathrm{th}}$ $(0\leq i\leq 5)$  row in the second and third block-rows of $H_{x}$, respectively.
    \item \textbf{$Z$-stabilizers:} $\otimes_{w=0}^{6}(I^{\otimes(4w+i)}\otimes Z\otimes I^{\otimes(6-4w-i)})\otimes Z$ is the stabilizer generator corresponding to $i^{\mathrm{th}}$ ($0\leq i\leq 6$) row in the first block-row of $H_{z}$.  In addition, $\otimes_{w=0}^{6}(I^{\otimes(5w+i)}\otimes Z\otimes I^{\otimes(6-5w-i)})\otimes Z$ and $\otimes_{w=0}^{6}(I^{\otimes(6w+i)}\otimes Z\otimes I^{\otimes(6-6w-i)})\otimes Z$ are the stabilizer generators corresponding $i^{\mathrm{th}}$ $(0\leq i\leq 5)$  row in the second and third block-rows of $H_{z}$, respectively.
\end{enumerate}
The extended parity-check matrices of $H_{x}$ and $H_{z}$ are given as follows:
	\begin{equation*}
			  {\small  H_{ex}= \left[\begin{array}{ccccccc|c}
				   I& I &I &I &I & I& I& \mathbf{\bar{1}}\\
				     I& P &P^2 &P^3 &P^4 &P^{5}&P^{6}& \mathbf{\bar{1}}\\
				     I& P^{2} &P^{4} &P^{6} &P &P^{3}&P^{5}& \mathbf{\bar{1}}
				  \end{array} \right],
			 H_{ez}=\left[\begin{array}{ccccccc|c}
				   I& P^{4} &P &P^{5} &P^{2} &P^{6}&P^{3}& \mathbf{\bar{1}}\\
				     I& P^{5} &P^{3} &P &P^{6} &P^{4}&P^{2}& \mathbf{\bar{1}}\\
				      I& P^{6} &P^{5} &P^{4} &P^{3} &P^{2}&P& \mathbf{\bar{1}}
				\end{array}\right],}
			\end{equation*}
            where $\mathbf{\bar{1}}=[1,1,1,1,1,1,1]^{\mathrm{T}}$.
\end{example}
%%%%%%%%%%%%%%%%%%%%%%%%%%%%%%%%%%%%%%%%%%%%%%%%%%%%%%%%%%%%%%%%%%%%%%%%%%%%%%%%%%%%%%%%%%%%%%%%%%%%%%%%
\begin{lemma}\label{I+J}
Let $p > 3$ be a prime, and let $M=J_{p}+I_{p},$
	% 	\begin{equation*}
	% M=\left[\begin{array}{cccccc}
	% 		0 & 1& 1 &\cdots&1& 1\\
	% 		1 & 0 & 1&\cdots&1& 1\\
	% 		\vdots & \vdots &\vdots&\ddots&\vdots& \vdots\\
	% 		1 & 1& 1 &\cdots&0& 1\\
	% 		1 & 1& 1 &\cdots&1& 0
	% 	\end{array}\right]_{p\times p}=J_{p}+I_{p},
	% 	\end{equation*}
where $J_{p}$ is the $p\times p$ all-ones matrix and $I_{p}$ is the $p\times p$ identity matrix. Then, $\mathrm{rank}(M) =(p-1)$.
\end{lemma}
\begin{proof} For all $3\leq i\leq p$, replacing row $R_{i}$ by $R_{i}+R_{2}$ in the given matrix $M$ results in the following matrix $M_{1}$:
\begin{equation*}
M \sim M_{1}= \left[\begin{array}{c|c}
	\begin{array}{cc}
		0&1\\
		1 &0
	\end{array}&\begin{array}{ccccc}
		1&1&\cdots&1&1\\
		1&1&\cdots&1&1
	\end{array}\\
	\hline
	\begin{array}{cc}
		0&1\\
		0&1\\
		\vdots&\vdots\\
		0&1\\
		0&1
	\end{array}&\begin{array}{ccccc}
		1&0&\cdots&0&0\\
		0&1&\cdots&0&0\\
		\vdots&\vdots&\ddots&\vdots&\vdots\\
		0&0&\cdots&1&0\\
		0&0&\cdots&0&1
	\end{array}
\end{array}\right].
\end{equation*}
Next, replacing $R_{1}$ by $R_{1} + \sum\limits_{i=3}^{p} R_{i}$ followed by interchanging $R_{2}$ and $R_{1}$ ($R_{2}\leftrightarrow R_{1}$) in the above matrix $M_{1}$ results in the following matrix $M_{2}$:
	\begin{equation*}
	M_{1}\sim M_{2}=\left[\begin{array}{c|c}
		\begin{array}{cc}
			1 &0\\
			0&	0	
		\end{array}&\begin{array}{ccccc}
			1&1&\cdots&1&1\\
			0&	0&\cdots&	0&	0	
		\end{array}\\
		\hline
		\begin{array}{cc}
			0&1\\
			0&1\\
			\vdots&\vdots\\
			0&1\\
			0&1
		\end{array}&\begin{array}{ccccc}
			1&0&\cdots&0&0\\
			0&1&\cdots&0&0\\
			\vdots&\vdots&\ddots&\vdots&\vdots\\
			0&0&\cdots&1&0\\
			0&0&\cdots&0&1
		\end{array}
	\end{array}\right].
	\end{equation*}
Finally, for all $3\leq i\leq(p-1)$, replacing $R_{i}$ by $R_{i}+R_{p}$ and $R_{2}\leftrightarrow R_{p}$ in the matrix $M_{2}$ results in the following matrix $M_{3}$:
	\begin{equation*}
	M_{2}\sim M_{3}=\left[\begin{array}{c|c}
		\begin{array}{cc}
			1 &0\\
			0&1		
		\end{array}&\begin{array}{ccccc}
			1&1&\cdots&1&1\\
			0&	0&\cdots&	0&	1\\
			
		\end{array}\\
		\hline
		\begin{array}{cc}
			0&0\\
			0&0\\
			\vdots&\vdots\\
			0&0\\
			0&0
		\end{array}&\begin{array}{ccccc}
			1&0&\cdots&0&1\\
			0&1&\cdots&0&1\\
			\vdots&\vdots&\ddots&\vdots&\vdots\\
			0&0&\cdots&1&1\\
			0&0&\cdots&0&0
		\end{array}
	\end{array}\right].
	\end{equation*}
Therefore, $\operatorname{rank}(M) = p - 1$, as required.
\end{proof} 
We employ the previously established lemma in the proof of the following theorem.
%%%%%%%%%%%%%%%%%%%%%%%%%%%%%%%%%%%%%%%%%%%%%%%%%%%%%%%%%%%%%%%%%%%%%%%%
\begin{theorem}\label{theorem3.1}
	Let $C_{1}$ and $C_{2}$ be two classical QC-LDPC codes, and let $H_{x}$ and $H_{z}$ be their parity-check matrices, respectively, defined as follows:   
    \begin{equation*}\label{H1.1}
H_{x}=\begin{bmatrix}
		P&P^{2}&\cdots&P^{(p-1)}\\
		P^{2}&P^{4}&\cdots&P^{2(p-1)}\\
		\vdots &\vdots &\ddots&\vdots\\		P^{\ell_{1}}&P^{2\ell_{1}}&\cdots&P^{(p-1)\ell_{1}}
	\end{bmatrix},\quad H_{z}=\begin{bmatrix}	P^{(\ell_{1}+1)}&P^{2(\ell_{1}+1)}&\cdots&P^{(p-1)(\ell_{1}+1)}\\
P^{(\ell_{1}+2)}&P^{2(\ell_{1}+2)}&\cdots&P^{(p-1)(\ell_{1}+2)}\\
\vdots &\vdots &\ddots&\vdots\\		P^{(\ell_{1}+\ell_{2})}&P^{2(\ell_{1}+\ell_{2})}&\cdots&P^{(p-1)(\ell_{1}+\ell_{2})}\\
	\end{bmatrix},
\end{equation*}     
where $p$ is an odd prime, $2\leq \ell_{1},\ell_{2}\leq (p-2)$, $ \ell_{1}+\ell_{2}\leq (p-1)$ and $P$ is the right circulant permutation matrix of order $p$. Then there exists an EA-QC-QLDPC code with the parameters $[[p^{2}-p, p^{2}-3p-(p-1)(\ell_{1}+\ell_{2}-2)+(p-1);(p-1)]]_2$. Furthermore, unassisted portion of the joint Tanner graph of the EA-QC-QLDPC code has girth $>4$.
\end{theorem}
\begin{proof} Let $H_{x}$ and $H_{z}$, defined in the statement of the theorem, be the parity-check matrices of the classical QC-LDPC codes  $C_{1}$ and $C_{2}$, respectively. Then, by Theorem~\ref{maintheorem}, there exists an EA-QC-QLDPC code with parameters $[[n, k_{1}+k_{2}-n+c; c]]_{2}$, where $n$ denotes the length of the codes $C_{1}$ and $C_{2}$, $k_{i}$ denotes the dimension of the QC-LDPC code $C_{i}$, and $c$ is the minimum number of maximally entangled bits (ebits) required. 

The length of the proposed EA quantum code is $p^{2}-p$, as each entry of the matrices $H_x$ and $H_z$ is a right circulant permutation matrix of size $p$, and each block-row contains $p-1$ such matrices.

 Next, we find the minimum number of required maximally entangled bits.
Let $i^{\mathrm{th}}$ block-row of the matrix $H_{x}$ be $(P^{k_{i}},P^{2k_{i}},\ldots,P^{(p-1)k_{i}})$ and $j^{\mathrm{th}}$ block-row of the matrix  $H_{z}$ be $(P^{m_{j}},P^{2m_{j}},\ldots,P^{(p-1)m_{j}})$, where $k_{i}, m_{j}\in \mathbb{F}_{p}^{*}$, $1\leq i\leq \ell_{1}$ and $(\ell_{1}+1)\leq j\leq \ell_{2}$. Then the $\left(i,j\right)$-th entry of the matrix  $H_{x}H_{z}^{T}$ is $P^{(k_{i}-m_{j})}+P^{2(k_{i}-m_{j})}+\cdots+P^{(p-1)(k_{i}-m_{j})}$. Given that $k_{i} \neq m_{j}$ for all $i$ and $j$, we obtain
\begin{equation*}
		  H_{x}H_{z}^{T}= \begin{bmatrix}
           J_{p}+I_{p}&J_{p}+I_{p}&\cdots&J_{p}+I_{p}&J_{p}+I_{p}\\		    J_{p}+I_{p}&J_{p}+I_{p}&\cdots&J_{p}+I_{p}&J_{p}+I_{p}\\		      \vdots&\vdots&\ddots&\vdots&\vdots\\		       J_{p}+I_{p}&J_{p}+I_{p}&\cdots&J_{p}+I_{p}&J_{p}+I_{p}\\    J_{p}+I_{p}&J_{p}+I_{p}&\cdots&J_{p}+I_{p}&J_{p}+I_{p}
			\end{bmatrix}_{(\ell_{1} p)\times (\ell_{2} p)},
            \end{equation*}
where $J_{p}$ is the $p\times p$ all-ones matrix and $I_{p}$ is the $p\times p$ identity matrix. By Lemma \ref{I+J}, one can easily conclude that $\mathrm{rank}(H_{x}H_{z}^{T})$ is $(p-1)$.

It remains to establish the rank of $H_{x}$ and $H_{z}$. It is straightforward to see that
\begin{align*}
  \mathrm{rank} \left( \begin{bmatrix}
		P&P^{2}&\cdots&P^{(p-1)}\\
		P^{2}&P^{4}&\cdots&P^{2(p-1)}\\
		\vdots &\vdots &\ddots&\vdots\\		P^{\ell_{1}}&P^{2\ell_{1}}&\cdots&P^{(p-1)\ell_{1}}
	\end{bmatrix}\right)=\mathrm{rank} \left( \begin{bmatrix}
		\textbf{0}&P&P^{2}&\cdots&P^{(p-1)}\\
		\textbf{0}&P^{2}&P^{4}&\cdots&P^{2(p-1)}\\
		\vdots &\vdots &\ddots&\vdots\\		\textbf{0}&P^{\ell_{1}}&P^{2\ell_{1}}&\cdots&P^{(p-1)\ell_{1}}
	\end{bmatrix}\right).
\end{align*}
Now, after adding all the block-columns to the first block-column in the matrix on the right-hand side, we obtain
\begin{align*}
  \ \mathrm{rank} \left( \begin{bmatrix}
		P&P^{2}&\cdots&P^{(p-1)}\\
		P^{2}&P^{4}&\cdots&P^{2(p-1)}\\
		\vdots &\vdots &\ddots&\vdots\\		P^{\ell_{1}}&P^{2\ell_{1}}&\cdots&P^{(p-1)\ell_{1}}
	\end{bmatrix}\right)
    =\mathrm{rank} \left( \begin{bmatrix}
		J_{p}+I_{p}&P&P^{2}&\cdots&P^{(p-1)}\\
		J_{p}+I_{p}&P^{2}&P^{4}&\cdots&P^{2(p-1)}\\
		\vdots &\vdots &\ddots&\vdots\\		J_{p}+I_{p}&P^{\ell_{1}}&P^{2\ell_{1}}&\cdots&P^{(p-1)\ell_{1}}
	\end{bmatrix}\right),
\end{align*}
where $J_{p}$ is the $p\times p$ all-ones matrix and $I_{p}$ is the $p\times p$ identity matrix. Next, by adding all the columns of any block-column—except the first one—to each column of the first block-column in the matrix on the right-hand side, we get
\begin{align*}
   \mathrm{rank} \left( \begin{bmatrix}
		P&P^{2}&\cdots&P^{(p-1)}\\
		P^{2}&P^{4}&\cdots&P^{2(p-1)}\\
		\vdots &\vdots &\ddots&\vdots\\		P^{\ell_{1}}&P^{2\ell_{1}}&\cdots&P^{(p-1)\ell_{1}}
	\end{bmatrix}\right)&=\mathrm{rank} \left( \begin{bmatrix}
		I_{p}&P&P^{2}&\cdots&P^{(p-1)}\\
		I_{p}&P^{2}&P^{4}&\cdots&P^{2(p-1)}\\
		\vdots &\vdots &\ddots&\vdots\\		I_{p}&P^{\ell_{1}}&P^{2\ell_{1}}&\cdots&P^{(p-1)\ell_{1}}
	\end{bmatrix}\right)\\
    &=p+(p-1)(\ell_{1}-1) \text{ by Theorem}~\ref{theorem1}.
\end{align*}
 Therefore, $\mathrm{rank}(H_{x}) = p + (p-1)(\ell_{1}-1).$ Similarly, we can show that $\mathrm{rank}(H_{z}) = p + (p-1)(\ell_{2}-1)$. By using the ranks of $H_{x}$ and $H_{z}$, together with the minimum number of required ebits, the dimension of the constructed EA-QC-LDPC code can be readily determined.  By the choice of $H_{x}$ and $H_{z}$, the unassisted portion of the joint Tanner graph of the EA-QC-QLDPC code has girth $>4$; hence, proved.
\end{proof}
\begin{remark}\label{Remark2}
 The extended parity-check matrices of $H_{x}$ and $H_{z}$ in the previous theorem can be written
as {\footnotesize \begin{equation*}\label{H1.1}
H_{ex}=\left[\begin{array}{cccc|c}
   P&P^{2}&\cdots&P^{(p-1)}&\mathbf{A}\\
		P^{2}&P^{4}&\cdots&P^{2(p-1)}&\mathbf{A}\\
		\vdots &\vdots &\ddots&\vdots&\vdots\\		P^{\ell_{1}}&P^{2\ell_{1}}&\cdots&P^{(p-1)\ell_{1}}&\mathbf{A}
\end{array}\right],~H_{ez}=
\left[\begin{array}{cccc|c}
    P^{(\ell_{1}+1)}&P^{2(\ell_{1}+1)}&\cdots&P^{(p-1)(\ell_{1}+1)}&\mathbf{B}\\
P^{(\ell_{1}+2)}&P^{2(\ell_{1}+2)}&\cdots&P^{(p-1)(\ell_{1}+2)}&\mathbf{B}\\
\vdots &\vdots &\ddots&\vdots&\vdots\\		P^{(\ell_{1}+\ell_{2})}&P^{2(\ell_{1}+\ell_{2})}&\cdots&P^{(p-1)(\ell_{1}+\ell_{2})}&\mathbf{B}
\end{array}\right],
\end{equation*}} 
           $\text{where } 
\mathbf{A} =
\begin{bmatrix}
I_{p-1} \\
\mathbf{\bar{1}}
\end{bmatrix},~
\mathbf{B} =
\begin{bmatrix}
I_{p-1} + J_{p-1} \\
\mathbf{\bar{1}}
\end{bmatrix},$
with \(I_{p-1}\) denoting the \((p-1)\times(p-1)\) identity matrix, \(J_{p-1}\) the all-ones matrix of the same size, and \(\mathbf{\bar{1}} = \underbrace{[1,1,\ldots,1]}_{(p-1)\text{-tuple}}\).
\end{remark}
\begin{example} \label{ex:hagiwara_comparison}
For $p=7$, let
			 \begin{equation*}
			     H_{x}=\begin{bmatrix}
				       P &P^2 &P^3 &P^4 &P^{5}&P^{6}\\
				       P^{2} &P^{4} &P^{6} &P &P^{3}&P^{5}\\
                        P^{3} &P^{6} &P^{2} &P^{5} &P&P^{4}
				 \end{bmatrix} \text{ and } 
			      H_{z}=\begin{bmatrix}
				     P^{4} &P &P^{5} &P^{2} &P^{6}&P^{3}\\
				       P^{5} &P^{3} &P &P^{6} &P^{4}&P^{2}\\
				        P^{6} &P^{5} &P^{4} &P^{3} &P^{2}&P
				 \end{bmatrix}
			 \end{equation*}   	
be two parity-check matrices of the classical QC-LDPC codes $\mathcal{C}_{1}$ and $\mathcal{C}_{2}$, respectively. Then, by Theorem \ref{theorem3.1}, there exists an EA-QC-QLDPC code with the parameters $[[42,10;6]]_2$. 

The extended parity-check matrices of $H_{x}$ and $H_{z}$ are given as follows:
	\begin{equation*}
			  {\small   H_{ex}= \left[\begin{array}{cccccc|c}
				     P &P^2 &P^3 &P^4 &P^{5}&P^{6}& \mathbf{A}\\
				     P^{2} &P^{4} &P^{6} &P &P^{3}&P^{5}& \mathbf{A}\\
                      P^{3} &P^6 &P^2 &P^5 &P&P^{4}& \mathbf{A}\\
				  \end{array} \right],
			 H_{ez}=\left[\begin{array}{cccccc|c}
				   P^{4} &P &P^{5} &P^{2} &P^{6}&P^{3}& \mathbf{B}\\
				     P^{5} &P^{3} &P &P^{6} &P^{4}&P^{2}& \mathbf{B}\\
				      P^{6} &P^{5} &P^{4} &P^{3} &P^{2}&P& \mathbf{B}
				\end{array}\right],}
			\end{equation*}
$\text{where } 
\mathbf{A} =
\begin{bmatrix}
I_{6} \\
\mathbf{\bar{1}}
\end{bmatrix},~
\mathbf{B} =
\begin{bmatrix}
I_{6} + J_{6} \\
\mathbf{\bar{1}}
\end{bmatrix},$ where \(I_{6}\) denotes the \(6 \times 6\) identity matrix, \(J_{6}\) is the all-ones matrix of the same size, and \(\mathbf{\bar{1}} = [1,1,1,1,1,1]\).
\end{example}
In the next theorem, we construct EA-QC-QLDPC code using a single classical code.
\begin{theorem}\emph{\cite{EA_ITW}}\label{theorem4}
	Let $\mathcal{C}$ be a classical QC-LDPC code with the parity-check matrix $H$ that contains $1\leq \ell\leq p$ distinct number of block-rows from the matrix $\mathbf{A}_{M}$. Then, there exists an EA-QC-QLDPC code with the parameters $[[p^{2},(p-1)(p-\ell+1);p+(\ell-1)(p-1)]]_2$. Moreover, the Tanner graph of the code $\mathcal{C}$ has girth $> 4$.    
\end{theorem}
\begin{proof}
	%Let $\Tilde{I}$ be a $p\times p$ matrix with zero entries on the diagonal and 1 entry elsewhere, i.e.,
	% \begin{equation*}
		% \Tilde{I}=\begin{bmatrix}
			%     0&1&\cdots&1\\
			%      1&0&\cdots&1\\
			%       \vdots&\ vdots&\ddots&\vdots\\
			%        1&1&\cdots&0
			% \end{bmatrix}.   
		% \end{equation*}
	% Then the matrix $\Tilde{I}$ can be reduced to the matrix \begin{equation}\label{eq14}
		%    \Tilde{I}=\left[\begin{array}{c|ccc|c}
			%         0 & 0 &\cdots & 0 & 0\\
			%         1 & 1 &\cdots & 1 & 0\\
			%         \hline
			%         \textbf{0} & & I_{p-2}& & \textbf{1}          
			%     \end{array}\right]
		% \end{equation}
	%  by applying the following steps:
	% \begin{enumerate}
		%     \item Add the second row of the matrix to all the rows below it;
		%     \item Add the last $(p-2)$ rows of the matrix to the first row.
		%     \item Swap the second column of the matrix subsequently with the columns next to it until it reaches the very end.    
		% \end{enumerate}
The code length is straightforward to determine. Moreover, by applying Corollary~\ref{cor1}, we obtain $\mathrm{rank}(H)$.

Next, we obtain the minimum number of required entangled bits as follows: Let $i^{\mathrm{th}}$ block-row of $H$ be $(P^{k_{i}x_{0}},P^{k_{i}x_{1}},\ldots,P^{k_{i}x_{(p-1)}})$, where $k_{i}\in \mathbb{F}_{p}$ for all $0\leq i\leq (\ell-1)$. Then the $\left(i,j\right)$-th entry of the matrix  $HH^{T}$ is $P^{(k_{i}-k_{j})x_{0}}+P^{(k_{i}-k_{j})x_{1}}+\cdots+P^{(k_{i}-k_{j})x_{(p-1)}}$. Therefore,
\begin{equation*}
		   HH^{T}= \begin{bmatrix}
			    I_{p}&J_{p}&\cdots&J_{p}&J_{p}\\
			     J_{p}&I_{p}&\cdots&J_{p}&J_{p}\\
			      \vdots&\vdots&\ddots&\vdots&\vdots\\
                   J_{p}&J_{p}&\cdots&J_{p}&I_{p}
			\end{bmatrix}_{(\ell p)\times (\ell p)},
            \end{equation*}
where $J_{p}$ is the $p\times p$ all-ones matrix and $I_{p}$ is the $p\times p$ identity matrix. For each $0\leq u\leq (p-1)$ and $1\leq v\leq (l-1)$, we replace $R_{u}^{(v)}$  by $(R_{u}^{(v)}+\sum_{u=0}^{(p-1)}R_{u}^{(0)})$ in the matrix $HH^{T}$ to obtain 
	\begin{equation*}
			\begin{bmatrix} I_{p}&J_{p}&\cdots&J_{p}&J_{p}\\  \mathbf{0}&I_{p}+J_{p}&\cdots&\mathbf{0}&\mathbf{0}\\ 
				\vdots&\vdots&\ddots&\vdots&\vdots\\		\mathbf{0}&\mathbf{0}&\cdots&I_{p}+J_{p}&\mathbf{0}\\                \mathbf{0}&\mathbf{0}&\cdots&\mathbf{0}&I_{p}+J_{p}
			\end{bmatrix}_{(\ell p)\times (\ell p)},  
	\end{equation*}
	where $\mathbf{0}$ represents the $p\times p$ zero matrix. By Lemma \ref{I+J}, one can easily conclude that $\mathrm{rank}(HH^{T})$ is $p+(\ell -1)(p-1)$. Using the rank of $H$ together with the required minimum number of entangled bits, we can determine the code dimension. By construction, the Tanner graph of the code $\mathcal{C}$ has girth $>4$; hence, proved.
\end{proof}
\begin{remark}
In the previous result, if the required catalytic rate is positive, then one may choose $\ell$ such that $2\ell \leq p$.
\end{remark}
%%%%%%%%%%%%%%%%%%%%%%%%%%%%%%%%%%%%%%%%%%%%%%%%%%%%%%%%%%%%%%%%%%%%%%%%%%%%%%%%%
\section{EA-QC-QLDPC Code with Tanner Graph of Girth Greater than 6}\label{sec.5}
In this section, we present the construction of EA-QC-QLDPC codes of girth > 6. This section has been divided into two subsections, namely Subsection \ref{clm-3} and Subsection \ref{clm-4}. First, in Subsection \ref{clm-3}, we introduce an EA-QC-QLDPC code with a parity-check matrix that has a uniform column weight of 3. Subsequently, in Subsection \ref{clm-4}, we present two families of EA-QC-QLDPC codes with a parity-check matrix having a uniform column weight of~4.   
\subsection{EA-QC-QLDPC codes with a parity-check matrix of column weight 3
}\label{clm-3}
  In the following theorem, we have proposed a class of EA-QC-QLDPC codes with girth~$>6$.
\begin{theorem}
For two integers $\ell\geq 6$ and $w\geq2$, let $P$ be a right circulant permutation matrix of order $w^{\ell}+1$. Let
 \begin{equation*}
  M=\begin{bmatrix}
       0 & 0&\cdots&0\\
        w^{1} & w^{2}&\cdots&w^{\ell}\\   
        -w^{1} & -w^{2}&\cdots&-w^{\ell}
  \end{bmatrix}
 \end{equation*} be a degree matrix associated with the parity-check matrix $H$, where $-w^{x}$ stands for the additive inverse of $w^{x}$ in $\mathbb{Z}_{(w^{\ell}+1)}$. Then there exists  an EA-QC-QLDPC code with the parameters $[[(w^{\ell}+1)\ell,2k-(w^{\ell}+1)\ell+c;c]]_2$, where $k\geq (\ell-3)(w^{\ell}+1)+2$ and $c\leq 3(w^{\ell}+1)-2$. In addition, the Tanner graph of $H$ is of girth greater than 6.
  \end{theorem}
  \begin{proof}
The length is straightforward to determine, and the lower bound on the dimension $k$ can be easily obtained using Theorem \ref{theorem2}. Now, we will show that the Tanner graph of $H$ is free from cycles of length 4. The existence of 4-cycle requires two columns and two rows from the matrix $M$. Each nonzero row in $M$ contains distinct entries under modulo $(w^{\ell} + 1)$; thus, by Proposition \ref{4-cyclepropo}, the presence of a 4-cycle is not possible for one zero row and one nonzero row. Consequently, a 4-cycle may exist between two nonzero rows.
By Proposition~\ref{4-cyclepropo}, a 4-cycle exists between the two nonzero rows in $M$ if and only if  
 \begin{align*}
     2w^{a}&=2w^{b}\quad\textrm{mod }(w^{\ell}+1),\text{ for some } 1\leq a, b\leq \ell, \text{ where } a\neq b\\
      \iff 2&=2w^{b-a}\quad\textrm{mod }(w^{\ell}+1)\quad\quad(\textrm{since }\textrm{gcd}(w^{a},w^{\ell}+1)=1),
 \end{align*}
 which leads to a contradiction. Hence, there is no $4$-cycle in the Tanner graph of $H$.
 
Next, we show that the Tanner graph of $H$ is devoid of
cycles of length 6. To prove that, consider any three columns from the matrix $M$ as follows
 \begin{equation}\label{0-portion_of_H}
 \begin{bmatrix}
       \cdots&0&\cdots & 0& \cdots&0&\cdots\\
        \cdots&w^{a}&\cdots& w^{b}&\cdots&w^{c}&\cdots\\   
        \cdots&-w^{a}&\cdots & -w^{b}&\cdots&-w^{c}&\cdots
 \end{bmatrix},
 \end{equation}
 where $1\leq a<b<c\leq \ell$. We examine the presence of all possible 6-cycles, as illustrated in Figure \ref{fig:all-6-cycles}, in the segment of the Tanner graph corresponding to a portion of the degree matrix defined by Eq.~\eqref{0-portion_of_H}.

\noindent \textbf{Existence of Type-I 6-cycle:} A Type-I cycle of length 6 exists in the Tanner graph of $H$ if and only if the following condition is satisfied:
 \begin{align*} 
    -w^{a}&= -w^{c}-w^{c}+w^{b}\quad\textrm{mod }(w^{\ell}+1)\\
    \implies 2w^{c}&=w^{b}+w^{a}\quad\textrm{mod }(w^{\ell}+1)\\
    \implies 2w^{c-1}&=w^{b-1}+w^{a-1}\quad\textrm{mod }(w^{\ell}+1)\quad\quad(\textrm{since }\textrm{gcd}(w,w^{\ell}+1)=1).
 \end{align*}
 This yields a contradiction, since the left-hand side exceeds the right-hand side.

 Similarly, the remaining $6$-cycles can be easily analyzed, and it is found that the parity-check matrix $H$ does not contain any $6$-cycles.

The minimum number of required ebits is given by $c=\text{rank}(HH^{T})$,
      where
      \begin{align*}
          HH^{T}&= \begin{bmatrix}
              I &I&\cdots&I\\
    P^{w} &P^{w^{2}}&\cdots&P^{w^{\ell}}\\
      P^{-w} &P^{-w^{2}}&\cdots&P^{-w^{\ell}}
          \end{bmatrix}  \begin{bmatrix}
              I&P^{-w}&P^{w}\\
    I&P^{-w^{2}}&P^{w^{2}}\\
     \vdots&\vdots&\vdots\\
      I&P^{-w^{\ell}}&P^{w^{\ell}} 
          \end{bmatrix} \\      
&=\begin{bmatrix}
     \ell I& \sum\limits_{i=1}^{\ell}P^{-w^{i}}&\sum\limits_{i=1}^{\ell}P^{w^{i}}\\
        \sum\limits_{i=1}^{\ell}P^{w^{i}}& \ell I&\sum\limits_{i=1}^{\ell}P^{2w^{i}}\\     \sum\limits_{i=1}^{\ell}P^{-w^{i}}&\sum\limits_{i=1}^{\ell}P^{-2w^{i}}&\ell I
\end{bmatrix}.
      \end{align*}
 It is easy to verify that the $\text{rank}(HH^{T}) \leq 3(w^{\ell} + 1) - 2$, since the vector sum of each block-row is an all-ones vector or all-zeros vector, depending on whether $\ell$ is even or odd. This completes the proof. \end{proof}

 \begin{remark}
In the preceding theorem, the parameter $\ell \geq 6$ is chosen so that the \emph{catalytic rate}
$\frac{2k-(w^{\ell}+1)\ell}{(w^{\ell}+1)\ell}$ is positive for the entanglement-assisted code $[[(w^{\ell}+1)\ell,\,2k-(w^{\ell}+1)\ell+c;\,c]]_2$. We do not discuss the condition guaranteeing a positive entanglement-assisted rate, since the upper bound for the parameter $c$ is given, which is not sufficient to establish this. The same holds for the theorems in the next subsection, i.e., some of the conditions on the code parameters based on positive catalytic rate.
\end{remark}
 \begin{example} For $\ell=6$ and $w=2$, we define the
 degree matrix $M$ and the associated parity-check matrix $H$ as follows:
 \begin{equation*}
     M=\begin{bmatrix}
        0 & 0 & 0& 0&0& 0\\
			2 & 4 &8& 16&32& 64\\
   			63 & 61 &57& 49&33& 1
    \end{bmatrix},~ H=\begin{bmatrix}
         I & I & I& I&I& I\\
	P^{2} & P^{4} &P^{8}& P^{16}&P^{32}& P^{64}\\
   	P^{63} & P^{61} &P^{57}& P^{49}&P^{33}& P
     \end{bmatrix},   
\end{equation*}
where $P$ is the right circulant permutation matrix of order 65. Then there exists an EA-QC-QLDPC code with the parameters $[[390,2k-390+c;c]]_2$, where $k\geq 197$ and $c\leq 193$.
 \end{example}  
 \begin{remark}
   Since $2+63=64+1$, the entries in M do not form a Sidon sequence \cite{zhang2012new}.
 \end{remark} %%%%%%%%%%%%%%%%%%%%%%%%%%%%%%%%%%%%%%%%%%%%%%%%%%%%%%%%%%%%%%%%%%%%%%%%%%%%%%%%%%%%%%%%%%%%%%%%%%%%%%%%%%%%%%%%%%%%%%%%%%%%%%%%%%%%%%%%%%%%%%%%%%%%%%%%%%%%%%%%%%%%%%%%%%%%%%%%%%%%%%%%%%%%%%%%%%%%
 \subsection{EA-QC-QLDPC codes with a parity-check matrix of column weight 4}\label{clm-4}
 In the following theorem, we propose a class of EA-QC-QLDPC codes, constructed using classical QC-LDPC codes which is characterized by a Tanner graph of girth > 6.   
\begin{theorem}\label{columnweight-4-even}
	Let $S=\{a_{1},a_{2},\ldots,a_{t}\}$, where $2\leq a_{1}<a_{2}\cdots<a_{t}$ and for any $p<q<r$, $a_{q}-a_{p}=a_{r}-a_{q}=1$ does not hold. Let $P$ be a right circulant permutation matrix of order $(w^{a_{t}+1}-1)$, where $a_{t}\geq 12$ is an integer and $w\geq2$ is an even integer, and let
	\begin{equation*}
		M=\begin{bmatrix}
		   0 & 0&\cdots&0\\
			w^{a_{1}} & w^{a_{2}}&\cdots&w^{a_{t}}\\   
			w^{a_{1}-1} & w^{a_{2}-1}&\cdots&w^{a_{t}-1}\\   
			w^{a_{1}-2} & w^{a_{2}-2}&\cdots&w^{a_{t}-2} 
		\end{bmatrix}
	\end{equation*} be a degree matrix associated with the parity-check matrix $H$. Then there exists an EA-QC-QLDPC code with parameters $[[n,2k-n+c;c]]_2$, where $n=(w^{a_{t}+1}-1)t$, $k\geq (w^{a_{t}+1}-1)(t-4)+3$ and $c\leq 4(w^{a_{t}+1}-1)-3$. In addition, the Tanner graph of the parity-check matrix $H$ is of girth greater than 6.
\end{theorem}
\begin{proof}
    The detailed analysis is provided in Appendix~\ref{appendix.a}.
\end{proof}
\begin{theorem}\label{columnweight-4-general}
    Let $S=\{a_{1},a_{2},a_{3},\ldots,a_{t}\}$, where $2\leq a_{1}<a_{2}<a_{3}\cdots<a_{t}$ and for any $1\leq p<q\leq t$, $a_{q}-a_{p}\geq 2$. For two integers $a_{t}\geq 16$ and $w\geq2$, let $P$ be a right circulant permutation matrix of order $(w^{a_{t}+1}-1)$, and let
	\begin{equation*}
		M=\begin{bmatrix}
		    0 & 0&\cdots&0\\
			w^{a_{1}} & w^{a_{2}}&\cdots&w^{a_{t}}\\   
			w^{a_{1}-1} & w^{a_{2}-1}&\cdots&w^{a_{t}-1}\\   
			w^{a_{1}-2} & w^{a_{2}-2}&\cdots&w^{a_{t}-2}
		\end{bmatrix}
	\end{equation*} be a degree matrix associated with the parity-check matrix $H$. Then there exists an entanglement assisted quantum QC-LDPC code with parameters $[[n,2k-n+c;c]]_2$, where $n=(w^{a_{t}+1}-1)t$, $k\geq (w^{a_{t}+1}-1)(t-4)+3$ and $c\leq 4(w^{a_{t}+1}-1)-3$. In addition, the Tanner graph of the parity-check matrix $H$ is of girth greater than 6.   
\end{theorem}
\begin{proof}
    See the Appendix \ref{appendix.b}.
\end{proof}
In the following section, we present an efficient encoding scheme for certain families of the proposed EA-QC-QLDPC codes by leveraging their structural properties.
\section{Encoding of the Proposed Quantum Codes}\label{Sec:Encoding}
In this section, we develop an efficient encoding scheme for the codes  in Theorems~\ref{theorem3} and~\ref{theorem3.1}. Before presenting the complete encoding procedure, we introduce several preliminary notions to facilitate understanding. We begin by defining the Pauli group $\mathcal{P}_n$ for an $n$-qubit system as the following subgroup of the unitary group $U(2^n)$ over $\mathbb{C}$:
\begin{equation*}
    \mathcal{P}_n = \left\{ i^{\ell} \bigotimes_{j=1}^n X^{x_j} Z^{z_j} : \ell \in \mathbb{Z}_4,\; (x_{1},x_{2},\ldots,x_{n}|z_{1},z_{2},\ldots,z_{n}) \in \mathbb{F}_2^{2n} \right\},
\end{equation*}
 where  $X=\ket{0}\bra{1}+\ket{1}\bra{0}$ and $Z=\ket{0}\bra{0}-\ket{1}\bra{1}.$
 
Next, we define the Clifford group on $n$ qubits, denoted by $\mathcal{C}_n$, as the normalizer of the Pauli group $\mathcal{P}_n$ within the unitary group $U(2^n)$ over $\mathbb{C}$, that is,
\begin{equation*}
    \mathcal{C}_n = \left\{ U \in U(2^n) : U \mathcal{P}_n U^{\dagger} = \mathcal{P}_n \right\}.
\end{equation*}
The group $\mathcal{C}_n$ is generated by the controlled-NOT (CNOT), Hadamard ($H$), and phase ($S$) gates, whose conjugation action maps Pauli operators to Pauli operators and preserves their commutation relations; this action is summarized in Table~\ref{Table:Operators}.

% A single qubit is represented by a normalized state vector $\ket{\psi} \in \mathbb{C}^2$ of the form $\ket{\psi}=\alpha\ket{0}+\beta\ket{1}$, where $\alpha,\beta\in\mathbb{C}$ and $|\alpha|^2+|\beta|^2=1$. Physical operations acting on $\ket{\psi}$ are described by norm-preserving linear transformations, which are implemented by unitary operators. Such operations can be expressed in terms of the single-qubit Pauli group $\mathcal{P}_1=\{ i^{\ell} X^{x} Z^{z} : \ell \in \mathbb{Z}_4,\; (x,z)\in\mathbb{F}_2^2 \}$, where  $X=\ket{0}\bra{1}+\ket{1}\bra{0}$ and $Z=\ket{0}\bra{0}-\ket{1}\bra{1}$. The element from the set $\mathcal{P}_1$ are referred to as  Pauli operators. 
\begin{table*}[t]
  \caption{Action of the generators of the Clifford group $\mathcal{C}_n$ on binary symplectic vectors corresponding to elements of the $n$-qubit Pauli group $\mathcal{P}_n$. A vector $[\,w\: x \mid y\: z\,]$ is isomorphic to the Pauli operator $X^{w}Z^{y} \otimes X^{x}Z^{z}$. For $\mathrm{CNOT}^{(i,j)}$, the $i^{\text{th}}$ qubit acts as the control and the $j^{\text{th}}$ qubit as the target.}
    \centering
    \begin{tabular}{|p{2.8cm}|p{2cm}|p{4.5cm}|p{4cm}|}
        \hline
        \multicolumn{4}{|c|}{\textbf{Generators of the Clifford Group}} \\
        \hline
        \textbf{Gate} & \textbf{Notation} & \textbf{Unitary Representation} & \textbf{Symplectic Vector Action} \\
        \hline
        Controlled-NOT (CNOT) 
        & $\mathrm{CNOT}^{(i,j)}$ 
        & $\ket{0}\bra{0}_{i}\!\otimes\! I_{j} + \ket{1}\bra{1}_{i}\!\otimes\! X_{j}$ 
        & $[\,w\; x \mid y\; z\,] \mapsto [\,w\; w{+}x \mid y{+}z\; z\,]$ \\
        \hline
        Hadamard 
        & $\mathrm{H}^{(i)}$ 
        & $\frac{1}{\sqrt{2}} \displaystyle\sum_{c,d \in \mathbb{F}_2} (-1)^{cd} \ket{c}\bra{d}_{i}$ 
        & $[\,w \mid x\,] \mapsto [\,x \mid w\,]$ \\
        \hline
        Phase 
        & $\mathrm{S}^{(i)}$ 
        & $\displaystyle\sum_{c \in \mathbb{F}_2} i^{c}\ket{c}\bra{c}_{i}$ 
        & $[\,w \mid x\,] \mapsto [\,w \mid x{+}w\,]$ \\
        \hline
%         SWAP
%         &  $\mathrm{SWAP}^{(i,j)}$
%         & $\sum\limits_{i,j \in \{0,1\}}
% \lvert i\rangle \lvert j\rangle
% \langle j\rvert \langle i\rvert.$ 
%         & $[\,w~ x \mid y ~z\,] \mapsto [\,x ~w \mid z~ y\,]$ \\
%         \hline
    \end{tabular}
    \label{Table:Operators}
\end{table*}
An $n$-qubit quantum state is represented by a normalized vector $\ket{\psi} \in (\mathbb{C}^2)^{\otimes n}$ and can be expressed in the computational basis $\{\ket{x}\}_{{x\in\mathbb{F}_2^n}}$ as
\[
\ket{\psi}=\sum_{x\in\mathbb{F}_2^n} \alpha_x \ket{x},
\]
where $\alpha_x \in \mathbb{C}$ and $\sum_{x\in\mathbb{F}_2^n} |\alpha_x|^2 = 1$. Physical operations acting on $\ket{\psi}$ are described by unitary operators in $U(2^n)$. Such operations can be conveniently characterized using the $n$-qubit Pauli group $\mathcal{P}_n$. 

\textit{The reason for describing any physical operation on the quantum state $\ket{\psi}$ using elements of the Pauli group is that every element of $U(2^n)$ can be expressed as a linear combination of elements of the Pauli group $\mathcal{P}_n$.} 

%  Before proceeding, we present the following lemma, which is key to understanding the insight behind the decoding procedure.
% \begin{lemma}\label{le:conjuagacy_relation}
% Let $Q$ be a quantum code with stabilizer group $S$, generated by $\{g_i\}_{i=1}^{\rho}$. A quantum state $\ket{\psi}$ is stabilized by a generator $g_{i}$ if and only if, for any $U \in \mathcal{C}_n$, the state $U\ket{\psi}$ is stabilized by the conjugated operator $U g_{i} U^{\dagger}$.
% \end{lemma}
% \begin{proof}
% Suppose the quantum state $\ket{\psi}$ is stabilized by $g_{i}$, i.e.,
% \begin{equation*}
%     g_{i}\ket{\psi} = \ket{\psi}.
% \end{equation*}
% Applying a unitary operator $U \in \mathcal{C}_n$ to both sides and inserting $UU^{\dagger} = I$, we obtain
% \begin{equation*}
%       Ug_{i}U^{\dagger}(U\ket{\psi}) = U\ket{\psi},
% \end{equation*}
% which shows that $U\ket{\psi}$ is stabilized by $Ug_{i}U^{\dagger}$.

% Conversely, assume that vector $U\ket{\psi}$ is stabilized by $Ug_{i}U^{\dagger}$, i.e.,
% \begin{equation*}
%       UgU^{\dagger}(U\ket{\psi}) = U\ket{\psi}.
% \end{equation*}
% Multiplying both sides on the left by $U^{\dagger}$ yields
% \begin{equation*}
%   g\ket{\psi} = \ket{\psi},   
% \end{equation*}
% which shows that $\ket{\psi}$ is stabilized by $g_{i}$. This completes the proof.
% \end{proof}
The action of a Clifford operator $U$ on an $n$-qubit state can be equivalently described in terms of its action on the corresponding stabilizer generators.  More precisely, if $\rho$ denotes the number of stabilizer generators, then for each $i \in \{1, \ldots, \rho\}$, a quantum state $\ket{\psi}$ is stabilized by a generator $g_{i}$ if and only if the state $U\ket{\psi}$ is stabilized by the conjugated operator $U g_{i} U^{\dagger}$. Since stabilizer generators correspond bijectively to the rows of the stabilizer matrix, conjugation by a Clifford operator induces corresponding column operations on the associated $X$- and $Z$-type stabilizer matrices as illustrated in Table~\ref{Table:Operators}.

To construct the encoder, we first design the corresponding decoder. The decoding procedure consists of a sequence of Clifford operations, listed in Table~\ref{Table:Operators}, which are applied systematically to transform the stabilizer generators of the code into a form that stabilizes the initial states. Once the decoder is obtained, the encoder $\mathcal{E}$ is naturally defined as its adjoint, i.e., $\mathcal{E} = \mathcal{D}^{\dagger}$.

With these preliminaries established, we are now prepared to present the complete encoding procedure for the codes constructed in Theorems~\ref{theorem3} and \ref{theorem3.1}. The same line of argument also applies when the quantum code is constructed using a single classical QC-LDPC code.

% Furthermore, the reader should note that, in developing the encoding procedure based on Theorem~\ref{theorem3.1}, we must consider at least two block-rows in the matrix $H_z$, one of which must contain the block-row $\begin{bmatrix}
%     P^{p-1} & P^{p-2} & \cdots & P
% \end{bmatrix},$
% for any prime $p \geq 5$. These assumptions are necessary for the proper construction of the encoding procedure. \textcolor{red}{Some more details and why.}

Let $H_x$ and $H_z$ denote the parity-check matrices of two distinct classical QC-LDPC codes, respectively, and the  parameters of  the  constructed entanglement-assisted code be $[[n,k;c]]$. Further, let $H_{ex}$ and $H_{ez}$ denote the extended parity-check matrices  satisfying the CSS condition.  We apply Gaussian elimination to $H_{ex}$ and $H_{ez}$ to obtain their \emph{row-reduced echelon forms (RREF)} and remove any zero rows from each matrix. After Gaussian elimination, let $H_{ex}$ have $\rho_{1}$ number of rows and $H_{ez}$ have $\rho_{2}$ number of rows. Further, 
without loss of generality, we assume that $\rho_1 \ge \rho_2$. The resulting matrices are given by the following equation:
\begin{equation}\label{Eq:Mat_Encod_GA_ELI}
    \left.
    \begin{array}{l}
    \tilde{H}_{ex} =
\Bigg[\,
\overset{n-\text{transmitter qubits}}{\overbrace{
I_{\rho_1} ~\Bigg |~ A^{(x)}_{\rho_{1}\times (n-\rho_{1}-c)} ~\Bigg |~
\overset{c \text{ halves of ebits}}{\overbrace{E^{(x)}_{\rho_{1}\times c}}}
}}
~ \Bigg |~ 
\overset{c\text{ halves of ebits}}{\overbrace{D^{(x)}_{\rho_{1}\times c}}}
\Bigg], \\[1.5em]
\tilde{H}_{ez} =
\Bigg[\,
\underset{(n-c)-\text{ancilla and information qubits}}{\underbrace{
J_{\rho_2 \times \rho_1} ~\Bigg |~ A^{(z)}_{\rho_{2}\times (n-\rho_{1}-c)}
}}~\Bigg |~ 
E^{(z)}_{\rho_{2}\times c}~ \Bigg | ~D^{(z)}_{\rho_{2}\times c}
\Bigg],    
  \end{array}
    \right\}
\end{equation}
 where the columns of $\tilde{H}_{ex}$ and $\tilde{H}_{ez}$ are partitioned to facilitate a clearer understanding of the encoding steps.
 
The submatrix $[I_{\rho_1} \mid A^{(x)}]$ represents the binary support of the $X$-type stabilizer generators acting on the ancilla and information qubits. Similarly, the submatrix $[J_{\rho_2 \times \rho_1} \mid A^{(z)}]$ represents the binary support of the $Z$-type stabilizer generators on the ancilla and information qubits. The matrices $E^{(x)}$ and $E^{(z)}$ specify the $X$- and $Z$-type stabilizer support on the sender’s halves of the $c$ shared ebits, respectively. Likewise, $D^{(x)}$ and $D^{(z)}$ describe the $X$- and $Z$-type stabilizer support on the receiver’s halves of the ebits, respectively.

For the construction of entanglement-assisted quantum codes, the transformed matrices in
\eqref{Eq:Mat_Encod_GA_ELI} must satisfy the dual-containment condition, i.e.,  $\tilde{H}_{ex} \tilde{H}_{ez}^{\text{T}} \equiv \mathbf{0} \pmod{2}.$
% \begin{equation*}
%    \tilde{H}_x H_z^{\text{T}} \equiv \mathbf{0} \pmod{2}. 
% \end{equation*}
This condition leads to the following equation:
\begin{align}
A^{(x)} A^{(z)^{\text{T}}}
\oplus
E^{(x)} E^{(z)^{\text{T}}}
\equiv
J^{\text{T}}
\oplus
D^{(x)} D^{(z)^{\text{T}}},
\label{Eq:DUAL_Containm_CSS}
\end{align}
where $\oplus$ denotes addition over the finite field $\mathbb{F}_2$. 

Before discussing the encoder and decoder for the EA-QC-QLDPC code, we first inform the reader that the encoder $\mathcal{E}$ takes as input the state
$\ket{0}^{\otimes a}\ket{\psi}\ket{\phi}^{\otimes c},$
where $\ket{\psi}$ denotes the state of the information qubits, $\ket{0}^{\otimes a}$ represents $a$ ancilla qubits initialized in the zero state, and $\ket{\phi}=\tfrac{1}{\sqrt{2}}(\ket{00}+\ket{11})$ is a maximally entangled Bell pair. We describe the decoding procedure in detail below; in developing the decoder $\mathcal{D}$, we draw upon the methodology proposed in~\cite{nadkarni2021encoding}.

\begin{enumerate}[leftmargin=0pt]
\item[]\textbf{(a)} For the EA quantum code $[[n,k;c]]_{2}$, post received valid state $\ket{\Psi}$, the ancilla qubits $\ket{0}^{\otimes a}$ are stabilized by Pauli-$Z$ operators, and the information qubits $\ket{\psi}$,   are stabilized by identity operators. Therefore, we transform the submatrix $\begin{bmatrix}
    I_{\rho_1} \mid A^{(x)}_{\rho_{1}\times (n-\rho_{1}-c)}
\end{bmatrix}$ to zero submatrix, i.e., $\begin{bmatrix}
    \mathbf{0}_{\rho_{1}} \mid \mathbf{0}_{\rho_{2}\times (n-\rho_{1}-c)}
\end{bmatrix}$. Moreover, since the information qubits are stabilized by the identity operator, we make the last $k=(n-\rho_{1}-\rho_{2}+c)$ columns of $A^{(z)}$ zero.
\item []\textbf{(b)} The submatrices $E^{(x)}$ and $E^{(z)}$ must be transformed into $D^{(x)}$ and $D^{(z)}$, respectively, since the Bell pair $\ket{\phi}=\tfrac{1}{\sqrt{2}}(\ket{00}+\ket{11})$ is stabilized by $X \otimes X$ and $Z \otimes Z$. Consequently, the corresponding columns of the augmented matrices  $[\,E^{(x)} \mid D^{(x)}\,]$ and $[\,E^{(z)} \mid D^{(z)}\,]$
must coincide pairwise.
\end{enumerate}
We now present the decoding procedure that implements the above transformations through the following steps.
\paragraph{STEP 1:}
The transformation of the matrix $\tilde{H}_{ex}$ in \eqref{Eq:Mat_Encod_GA_ELI} is carried out by applying row-wise $\mathrm{CNOT}$ operations. These operations are chosen such that the submatrix $A^{(x)}$ is reduced to the zero matrix, while the submatrix $E^{(x)}$ is transformed into $D^{(x)}$. The corresponding unitary operator that implements the desired transformation is given by
\begin{equation}\label{Eq:U_xgate} 
   U_1=
\prod_{i=1}^{\rho_1}\Bigg(
\underbrace{\Bigg( \prod_{k\in Q} \text{CNOT}^{(i, k)}\Bigg) }_{U_2^{(i)}}
\underbrace{\Bigg(\prod_{\ell\in P} \text{CNOT}^{(i, \ell)}\Bigg)}_{U_1^{(i)}}\Bigg),
\end{equation}
where 
\begin{align}
    P&=\left\{\rho_1+j:A^{(x)}_{i,j}\neq 0 \text{ for } j\in\{1,2,\ldots,n-c-\rho_1\}\right\},\label{Eq:U_x_target_1}\\
    Q&=\left\{n-c+j:E^{(x)}_{i,j} \oplus D^{(x)}_{i,j}\neq 0 \text{ for } j\in\{1,2,\ldots,c\}\right\},\label{Eq:U_x_target_2}
\end{align}
 denote the indices of the target columns in the $i^{\text{th}}$ rows of $\tilde{H}_{ex}$ when $A^{(x)}_{i,*}$ and
$E^{(x)}_{i,*} \oplus D^{(x)}_{i,*}$ are nonzero, respectively~\footnote{The symbol $M_{i,*}$ denotes an arbitrary entry of the $i^{\text{th}}$ row of the matrix $M$.}.

For the $i^{\text{th}}$ row of $\tilde{H}_{ex}$, the operator $U_1^{(i)}$ adds the $i^{\text{th}}$ control column of $I_{\rho_1}$ to the target columns of $A^{(x)}_{i,*}$ that contain nonzero entries, refer Eq.~\eqref{Eq:U_x_target_1}, thereby making the entries zero, over $\mathbb{F}_2$. Similarly, the operator $U_2^{(i)}$ replaces $E^{(x)}_{i,*}$ with
$D^{(x)}_{i,*}$, since the target columns are selected according to the nonzero indices of 
$E^{(x)}_{i,*} \oplus D^{(x)}_{i,*},$ refer Eq.~\eqref{Eq:U_x_target_2}, i.e., $E^{(x)}_{i,*}\rightarrow E^{(x)}_{i,*}\oplus E^{(x)}_{i,*} \oplus D^{(x)}_{i,*}=D^{(x)}_{i,*}$.

Therefore, the application of the operator $U_{1}$ in Eq.\eqref{Eq:U_xgate}, transforms the $X$-type stabilizer matrix $\tilde{H}_{ex}$ as follows:
\begin{equation*}
 \tilde{H}_{ex} =
\begin{bmatrix}
I_{\rho_1} \mid \mathbf{0}_{\rho_1 \times (n-\rho_1-c)} \mid D^{(x)}_{\rho_{1}\times c} \mid D^{(x)}_{\rho_{1}\times c}
\end{bmatrix}.   
\end{equation*}
As we know that the application of CNOT gate affects the columns in $\tilde{H}_{ex}$ and $\tilde{H}_{ez}$ parts, i.e., under the action of $U_1$, the target columns of $\tilde{H}_{ez}$ are added to its control columns, refer to Table~\ref{Table:Operators}. Therefore, the operator $U_{1}$ induces a corresponding transformation on the matrix $J$ in the $Z$-type stabilizer matrix. For the $i^{\text{th}}$ row of $\tilde{H}_{ex}$, the target columns of $U_1$ in \eqref{Eq:U_xgate} are determined by the nonzero entries of the $i^{\text{th}}$ row of the matrix $S =
[\,A^{(x)} \mid E^{(x)} \oplus D^{(x)}\,]$,
% \[
% S^{(\tilde{H}_x)}_i =
% [\,A^{(x)}_{i,*} \mid E^{(x)}_{i,*} \oplus D^{(x)}_{i,*}\,],
% \],
which are controlled by the columns of $I_{\rho_1}$. Since the columns of $J$ are indexed by the same control qubits, the $i^{\text{th}}$ column of $J$ is updated using the corresponding target columns of
$[\,A^{(z)} \mid E^{(z)}\,]$, resulting in the transformation
\begin{equation}\label{Eq:W_maps_to_B}
   J \mapsto
J \oplus
\big[\,A^{(z)} \mid E^{(z)}\big]
\big[\,A^{(x)} \mid E^{(x)} \oplus D^{(x)}\,\big]^{T}=J\oplus
A^{(z)} A^{(x)^{\text{T}}}
\oplus
E^{(z)} E^{(x)^{\text{T}}}
\oplus
E^{(z)} D^{(x)^{\text{T}}}.   
\end{equation}
Therefore, along with Eq.~\eqref{Eq:W_maps_to_B}, the complete action of the operator $U_1$ in Eq.~\eqref{Eq:U_xgate}, transforms the stabilizer matrices as follows:
\begin{align*}
    \tilde{H}_{ex} &=\begin{bmatrix}
        I_{\rho_1} \mid \mathbf{0}_{\rho_1 \times (n-\rho_1-c)} \mid D^{(x)}_{\rho_{1}\times c} \mid D^{(x)}_{\rho_{1}\times c}
    \end{bmatrix},\\
\tilde{H}_{ez}& =
\begin{bmatrix}
    B_{\rho_{2}\times\rho_{1}} \mid A^{(z)}_{\rho_2 \times (n-\rho_1-c)} \mid E^{(z)}_{\rho_{2}\times  c} \mid D^{(z)}_{\rho_{2}\times c}
\end{bmatrix},
\end{align*}
where
\begin{equation}\label{Eq:Mat_B_Trans}
   B_{\rho_{2}\times\rho_{1}}=J\oplus
A^{(z)} A^{(x)^{\text{T}}}
\oplus
E^{(z)} E^{(x)^{\text{T}}}
\oplus
E^{(z)} D^{(x)^{\text{T}}}. 
\end{equation}
 \paragraph{STEP 2:} By substituting $A^{(z)} A^{(x)^{\text{T}}}
\oplus
E^{(z)} E^{(x)^{\text{T}}}
\equiv
J\oplus
D^{(z)} D^{(x)^{\text{T}}}$ from Eq.~(\ref{Eq:DUAL_Containm_CSS}) into Eq.~\eqref{Eq:Mat_B_Trans}, the submatrix $B_{\rho_{2}\times\rho_{1}}$ in Eq.~\eqref{Eq:Mat_B_Trans} can be simplified to
\begin{equation} \label{Eq:Simp_Mat_B}
       B_{\rho_{2}\times\rho_{1}} = \left(D^{(z)} \oplus E^{(z)}\right)D^{(x)^\text{T}} .
\end{equation} 
For Theorems~\ref{theorem3} and \ref{theorem3.1}, the matrix $\bigl(D^{(x)}\bigr)^{T}$ has full row rank equal to $c$. Therefore, we have
\begin{equation}\label{eq:rank_of_B}
\operatorname{rank}\!\left(B_{\rho_{2}\times\rho_{1}}\right)
=
\operatorname{rank}\!\left(D^{(z)} \oplus E^{(z)}\right).  
\end{equation}
We, now, partition $D^{(x)^\text{T}}$ as $\begin{bmatrix}
    D^{(x)^\text{T}}_1|D^{(x)^\text{T}}_{2}
\end{bmatrix}$, where $D^{(x)^\text{T}}_1$ consists of the first $c$ linearly independent columns of $D^{(x)^\text{T}}$, and  $D^{(x)^\text{T}}_2$ contains the remaining columns $\rho_1-c=\lambda~(\text{say})$. Based on the partition of $D^{(x)^\text{T}}$, the matrix $B_{\rho_{2}\times\rho_{1}}$ can be written as:
\begin{equation}\label{eq:decompo_of_B0}
   B_{\rho_{2}\times\rho_{1}}=\begin{bmatrix}
      V_{\rho_{2}\times c} \mid W_{\rho_{2}\times (\rho_{1}-c)}
   \end{bmatrix}
 =\left[
 (D^{(z)} \oplus E^{(z)})D^{(x)^\text{T}}_1 \,\middle|\,
 (D^{(z)} \oplus E^{(z)})D^{(x)^\text{T}}_2
 \right].
\end{equation}
 Since the first $c$ columns of $D^{(x)^{\mathrm{T}}}$ form the identity matrix of order $c$, therefore, $D_{1}^{(x)^{\mathrm{T}}} = I_c$, we refer the reader to Remark~\ref{Remark2} following Theorem~\ref{theorem3.1} for clarification. For Theorem~\ref{theorem3}, this fact is immediate. Consequently, the above Eq.~\eqref{eq:decompo_of_B0} reduces to
 \begin{equation}\label{eq:decompo_of_B}
   B_{\rho_{2}\times\rho_{1}}=\begin{bmatrix}
      V_{\rho_{2}\times c} \mid W_{\rho_{2}\times (\rho_{1}-c)}
   \end{bmatrix}
 =\left[
 (D^{(z)} \oplus E^{(z)}) \,\middle|\,
 (D^{(z)} \oplus E^{(z)})D^{(x)^\text{T}}_2
 \right].
\end{equation}
 Hence, by Eqs.~\eqref{eq:rank_of_B} and \eqref{eq:decompo_of_B}, we obtain
\begin{equation*}
  \operatorname{rank}(V_{\rho_{2}\times c})=\operatorname{rank}\left(D^{(z)} \oplus E^{(z)}\right)=\operatorname{rank}(B_{\rho_{2}\times\rho_{1}}).  
\end{equation*}
Consequently, each column of the matrix $W_{\rho_{2}\times (\rho_{1}-c)}$ can be written as a linear combination of the columns of the matrix $V_{\rho_{2}\times c}$, i.e., $V_{\rho_{2}\times c}$ and $W_{\rho_{2}\times (\rho_{1}-c)}$ can be written as 
\begin{equation}\label{eq:relationb1b2}
 \underset{V}{\underbrace{ 
    \begin{bmatrix}        b_{1,1}&b_{1,2}&\cdots&b_{1,c}\\      b_{2,1}&b_{2,2}&\cdots&b_{2,c}\\
    \vdots&\vdots&\ddots&\vdots\\     b_{\rho_{2},1}&b_{\rho_{2},2}&\cdots&b_{\rho_{2},c}
    \end{bmatrix}}}
    \underset{X}{\underbrace{\begin{bmatrix}        x_{1,1}&x_{1,2}&\cdots&x_{1,\lambda}\\      x_{2,1}&x_{2,2}&\cdots&x_{2,\lambda}\\
    \vdots&\vdots&\ddots&\vdots\\     x_{c, 1}&x_{c,2}&\cdots&x_{c,\lambda}
    \end{bmatrix}}}
   =\underset{W}{\underbrace{\begin{bmatrix}        b'_{1,1}&b'_{1,2}&\cdots&b'_{1,\lambda}\\      b'_{2,1}&b'_{2,2}&\cdots&b'_{2,\lambda}\\
    \vdots&\vdots&\ddots&\vdots\\     b'_{\rho_{2},1}&b'_{\rho_{2},2}&\cdots&b'_{\rho_{2},\lambda}
    \end{bmatrix}}}, 
\end{equation}
% \begin{equation*}
%     B_{1}X_{1}=B_{2}
% \end{equation*}
for some non-zero matrix $X_{c\times \lambda}$, where each column of $X$ contains at least one nonzero entry. It is obvious from the Eq.~\eqref{eq:relationb1b2} that the $k^{\text{th}}$ column of $W$ can be written as follow:
\begin{equation*}
    x_{1,k}\begin{bmatrix}
b_{1,1}\\
b_{2,1}\\
\vdots\\
b_{\rho_2,1}
\end{bmatrix}
+
x_{2,k}\begin{bmatrix}
b_{1,2}\\
b_{2,2}\\
\vdots\\
b_{\rho_2,2}
\end{bmatrix}
+\cdots+
x_{c,k}\begin{bmatrix}
b_{1,c}\\
b_{2,c}\\
\vdots\\
b_{\rho_2,c}
\end{bmatrix}=\begin{bmatrix}
    b^{'}_{1,k}\\
     b^{'}_{2,k}\\
     \vdots\\
      b^{'}_{\rho_{2},k}
\end{bmatrix}.
\end{equation*}
Therefore, the $k^{\text{th}}$ column in $W$ can be made zero by applying $\prod_{\ell\in\mathcal{C}_{k}}\mathrm{CNOT}^{(c+k,~\ell)},$ where $\mathcal{C}_{k}\subseteq\{1,2,\ldots,c\}$ denotes the set of row indices of nonzero entries in the $k^{\text{th}}$ column of $X$.  The $\mathrm{CNOT}$ gates are applied with controls corresponding to the $k^{\text{th}}$ column of $W_{\rho_{2}\times (\rho_{1}-c)}$, while the target columns of $V_{\rho_{2}\times c}$ are determined by the nonzero entries of $k^{\text{th}}$ column of $X$, i.e, $X_{*,k}$.  Finally, all columns in $W_{\rho_{2}\times (\rho_{1}-c)}$ are made zero by applying
\begin{equation}
\label{Eq:Operator_U_B_z}
U_{2}=\prod_{k=1}^{\lambda}\left(\prod_{\ell\in\mathcal{C}_{k}}\mathrm{CNOT}^{(c+k,~\ell)}\right),
\end{equation}
where $\lambda=\rho_1-c$ and $\mathcal{C}_{k}\subseteq\{1,2,\ldots,c\}$ denotes the set of row indices of nonzero entries in the $k^{\text{th}}$ column of $X_{c\times(\rho_{1}-c)}$. As we know, according to Table~\ref{Table:Operators}, the CNOT gate affects both X- and Z-stabilizer matrices. As a result, $U_2$~adds $  [\mathbf{0}_{(\rho_1-c)\times c} \mid I_{\rho_1-c}]^{\text{T}} X^{\text{T}}$ to the first $c$ columns, which serve as the target columns of $U_{2}$.

Finally, the action of the operator $U_2$ in Eq.~\eqref{Eq:Operator_U_B_z} transforms the stabilizer matrices as
\begin{equation}
\label{Eq:STabilizer_matrix_N_FNE}
\left.
\begin{array}{l}
\tilde{H}_{ex}=
\left[\begin{array}{c|c|c|c|c}
I_{c} & \mathbf{0}_{c\times(\rho_1-c)} & \mathbf{0}_{c\times (n-\rho_{1}-c)} & D^{(x)}_1 & D^{(x)}_1 \\
\hline
X_{(\rho_{1}-c)\times c}^{\text{T}} & I_{\rho_1-c} & \mathbf{0}_{(\rho_{1}-c)\times (n-\rho_{1}-c)} & D^{(x)}_2 & D^{(x)}_2
\end{array}\right], \\[1.5em]
\tilde{H}_{ez}=
\left[\begin{array}{c|c|c|c|c}
   V_{\rho_{2}\times c} & \mathbf{0}_{\rho_2\times(\rho_1-c)} &
 A^{(z)}_{\rho_2 \times (n-\rho_1-c)} &
E^{(z)}_{\rho_{2}\times c} &
D^{(z)}_{\rho_{2}\times c}
\end{array}\right],
\end{array}
\right\}
\end{equation}
where $\mathbf{0}_{s\times t}$, for any positive integers $s$ and $t$, denotes the zero matrix of dimension $s \times t$, and this notation is used consistently throughout the encoding section.

  \paragraph{STEP 3:} We now determine the operator that maps the submatrix $E^{(z)}$ in Eq.~\eqref{Eq:STabilizer_matrix_N_FNE} to $D^{(z)}$. By simply adding $V_{\rho_{2}\times c}$ to $E^{(z)}$, we can transform $E^{(z)}$ to $D^{(z)}$ as 
\begin{equation*}
  E^{(z)} \oplus V_{\rho_{2}\times c}= E^{(z)} \oplus E^{(z)} \oplus D^{(z)}= D^{(z)}\quad(\text{by Eq.~\eqref{eq:decompo_of_B}}~ V_{\rho_{2}\times c}=  E^{(z)} \oplus D^{(z)}).
\end{equation*}
% \begin{align}
%     B_1= E^{(z)} \oplus D^{(z)},
%     \label{Eq:Line_Equ_B}
% \end{align}
Accordingly, the resulting unitary operator that transform $E^{(z)}$ to $D^{(z)}$ is given by
\begin{equation} \label{Eq:Oper_U_e_2_z}
    U_{3} = \prod_{j=1}^{c} \mathrm{CNOT}^{(n-c+j,~j)}.
\end{equation}
 The application of operator $U_{3}$ in Eq.~\eqref{Eq:Oper_U_e_2_z} transforms the stabilizer matrices $\tilde{H}_{ex}$ and $\tilde{H}_{ez}$ as follows:
\begin{equation}\label{Eq:After_U_e_2_S_M}
\left.
\begin{array}{l}
\tilde{H}_{ex}=
\left[\begin{array}{c|c|c|c|c}
    I_{c}\oplus D_{1}^{(x)}
 & \mathbf{0}_{c\times(\rho_1-c)} & \mathbf{0}_{c\times (n-\rho_{1}-c)} & D^{(x)}_1 & D^{(x)}_1 \\
\hline
  X^{T}\oplus D_{2}^{(x)} & I_{\rho_1-c} & \mathbf{0}_{(\rho_{1}-c)\times (n-\rho_{1}-c)} & D^{(x)}_2 & D^{(x)}_2
\end{array}\right], \\[1.5em]
\tilde{H}_{ez}=
\left[\begin{array}{c|c|c|c|c}
    V_{\rho_{2}\times c} & \mathbf{0}_{\rho_2\times(\rho_1-c)} & A^{(z)}_{\rho_{2}\times (n-\rho_{1}-c)} & D^{(z)}_{\rho_{2}\times c} &D^{(z)}_{\rho_{2}\times c}
\end{array}\right]
\end{array}
\right\}
\end{equation}
By Eqs.~\eqref{Eq:Simp_Mat_B} and \eqref{eq:decompo_of_B}, it is evident that  
\begin{align}
    D^{(x)}   V^{\mathrm{T}}_{\rho_{2}\times c}
    = D^{(x)} \bigl( E^{(z)^\mathrm{T}} \oplus D^{(z)^\mathrm{T}} \bigr)=B^{T}=\begin{bmatrix}
      V_{\rho_{2}\times c} \mid W_{\rho_{2}\times (\rho_{1}-c)}
   \end{bmatrix}^{T}.
    \label{Eq:D_X_cal}
\end{align}
By substituting $D^{(x)}=\left[\begin{array}{c}
      D_{1}^{(x)} \\
      \hline
     D_{2}^{(x)}
\end{array}\right]$ in Eq.~\eqref{Eq:D_X_cal}, we have
\begin{equation}
\left[\begin{array}{c}
D_1^{(x)}  V^{\mathrm{T}}_{\rho_{2}\times c} \\
\hline
D_2^{(x)} V^{\mathrm{T}}_{\rho_{2}\times c}
\end{array}\right]
=
\left[\begin{array}{c}
V^{\mathrm{T}}_{\rho_{2}\times c}\\
\hline
W^{\mathrm{T}}_{\rho_{2}\times (\rho_{1}-c)}
\end{array}\right].
\label{Eq:Ans_f_subM_zero}
\end{equation}
 Assuming that, in developing the encoding procedure based on Theorem~\ref{theorem3.1}, we consider the first block-row
$\begin{bmatrix}
P^{p-1} & P^{2(p-1)} & \cdots & P^{2}
\end{bmatrix}$ in $H_{z}$, for any odd prime $p$, it can be readily verified that $\operatorname{rank}\!\left(D^{(z)} \oplus E^{(z)}\right)=\operatorname{rank}(V_{\rho_{2}\times c})=c.$ For Theorem~\ref{theorem3}, this fact is immediate.
Now, by Eqs. \eqref{eq:relationb1b2} and \eqref{Eq:Ans_f_subM_zero}  and full row rank property of $V^{\mathrm{T}}$, we have
\begin{equation}\label{eq:eq27}
    D_2^{(x)} V^{\mathrm{T}} = W^{\mathrm{T}}_{\rho_{2}\times (\rho_{1}-c)} \implies D_2^{(x)}  V^{\mathrm{T}} = X^{T}V^{\mathrm{T}}\implies D_2^{(x)}  = X^{\mathrm{T}}.
\end{equation}
By substituting the value of $D_1^{(x)}=I_{c}$ and $D_2^{(x)}$ from Eq.~\eqref{eq:eq27} into $\tilde{H}_{ex}$ given in Eq.~\eqref{Eq:After_U_e_2_S_M}, we have
  \begin{equation}\label{Eq:After_U_e_2_S_M2}
\left.
\begin{array}{l}
\tilde{H}_{ex}=
\left[\begin{array}{c|c|c|c|c}
    \mathbf{0}_{c}
 & \mathbf{0}_{c\times(\rho_1-c)} & \mathbf{0}_{c\times (n-\rho_{1}-c)} & D^{(x)}_1 & D^{(x)}_1 \\
\hline
   \mathbf{0}_{(\rho_{1}-c)\times c} & I_{\rho_1-c} & \mathbf{0}_{(\rho_{1}-c)\times (n-\rho_{1}-c)} & D^{(x)}_2 & D^{(x)}_2
\end{array}\right], \\[1.5em]
\tilde{H}_{ez}=

\left[\begin{array}{c|c|c|c|c}
   V_{\rho_{2}\times c} & \mathbf{0}_{\rho_2\times(\rho_1-c)} & A^{(z)}_{\rho_{2}\times (n - \rho_{1}-c)}& D^{(z)}_{\rho_{2}\times c} & D^{(z)}_{\rho_{2}\times c}
\end{array}\right].
\end{array}
\right\}
\end{equation}
 \paragraph{STEP 4:}
 Next, we enforce that the last 
$k = n - \rho_{1} - \rho_{2} + c$
columns of the submatrix $A^{(z)}$ in $\tilde{H}_{ez}$ are zero.  Consider the submatrix $
L = \big[\, V_{\rho_{2}\times c}  \;\big|\; A^{(z)}_{\rho_{2}\times (n - \rho_{1}-c)} \;\big|\; D^{(z)}_{\rho_{2}\times c} \,\big]$
of $\tilde{H}_{ez}$. Since $\operatorname{rank}(\tilde{H}_{ez}) = \rho_{2}$, it follows that 
$\operatorname{rank}(L) = \rho_{2}.$  

Since all the $c$ columns of $D^{(z)}$ are linearly independent, they can be extended to a set of $\rho_{2}$ linearly independent columns in the matrix $L$ by suitably incorporating columns from $V$ and $A^{(z)}$. We then reorder the columns of $L$—by appropriately applying SWAP gates—so that it takes the form $\big[\, \tilde{V}_{\rho_{2}\times c}\;\big|\; \tilde{A}^{(z)}_{\rho_{2}\times (n - \rho_{1}-c)} \;\big|\; D^{(z)}_{\rho_{2}\times c} \,\big]. $

In this arrangement, all $c$ columns of $\tilde{V}$, the first $\rho_{2}-2c$ ($\rho_{2}-2c>0$ for the codes proposed in Theorems~\ref{theorem3} and \ref{theorem3.1}) columns of $\tilde{A}^{(z)}$, and all $c$ columns of $D^{(z)}$ together form a set of $\rho_{2}$ linearly independent columns.
 
Let $\tilde{A}^{(z)}_{1}$ denote the submatrix formed by the first $\rho_{2}-2c$ linearly independent columns of $\tilde{A}^{(z)}$, and let $\tilde{A}^{(z)}_{2}$ denote the submatrix consisting of remaining $k = n - \rho_{1} - \rho_{2} + c$ columns of $A^{(z)}$. Then, the last $k = n - \rho_{1} - \rho_{2} + c$ columns of $\tilde{A}^{(z)}$ can be written in the linear combination of the columns of the matrix $\big[\, \tilde{V}_{\rho_{2}\times c} \;\big|\; \tilde{A}_{1}^{(z)} \;\big|\; D^{(z)}_{\rho_{2}\times c} \,\big]$, i.e., 
 \begin{equation*}    \big[\, \tilde{V}_{\rho_{2}\times c} \;\big|\; \tilde{A}^{(z)}_{1} \;\big|\; D^{(z)}_{\rho_{2}\times c} \,\big]\tilde{X}_{\rho_{2}\times k}=\tilde{A}^{(z)}_{2}, 
 \end{equation*}
 for some matrix $\tilde{X}_{\rho_{2}\times k}$. For $1\leq j\leq k$, let $Y_{j}$ represent the set of row indices of nonzero entries in the $j^{\text{th}}$ column of  $\tilde{X}_{\rho_{2}\times k}$. Then we apply the following operator to make the all columns of $\tilde{A}^{(z)}_{2}$ zero:
\begin{equation}
U_{4}=\prod_{j=1}^{k}\left(\prod_{\ell\in Y_{j}}\text{CNOT}^{\big(\rho_{1}+\rho_{2}-2c+j,\;Y_j+\{m\}\big)}\right),
\label{Eq:Uza2Gate}
\end{equation}
where $\{\rho_{1}+\rho_{2}-2c+j\}_{j=1}^{k}$ are the positions of the last $k$-columns of $\tilde{A}^{(z)}$ in $\tilde{H}_{ez}$ and 
\begin{equation}
   m = \begin{cases}
0,\text{ if } 1\leq \ell\leq c,\\    
\rho_{1}-c,\text{ if } c+1 \leq \ell \leq \rho_{2}-c,\\
\rho_{1}-c+k,\text{ if } \rho_{2}-c+1 \leq \ell \leq \rho_{2}.
 \end{cases}
\end{equation}
Under the action of $U_{4}$, the stabilizer matrices transform as
 \begin{equation}\label{Mat:UZa2StabMat}
\left.
\begin{array}{l}
\tilde{H}_{ex}=
\left[\begin{array}{c|c|c|c|c}
    \mathbf{0}_{c}
 & \mathbf{0}_{c\times(\rho_1-c)} & \mathbf{0}_{c\times (n-\rho_{1}-c)} & D^{(x)}_1 & D^{(x)}_1 \\
\hline
   \mathbf{0}_{(\rho_{1}-c)\times c} & I_{\rho_1-c} & \mathbf{0}_{(\rho_{1}-c)\times (n-\rho_{1}-c)} & D^{(x)}_2 & D^{(x)}_2
\end{array}\right], \\[1.5em]
\tilde{H}_{ez}=
\Big[\overset{\text{ancilla qubits}}{\overbrace{\tilde{V}_{\rho_{2}\times c} \mid \mathbf{0}_{\rho_2\times(\rho_1-c)} \mid \tilde{A}_1^{(z)}}} \overset{\text{information qbits}}{\overbrace{\mathbf{0}_{\rho_2\times k} }}\mid D^{(z)}_{\rho_{2}\times c} \mid D^{(z)}_{\rho_{2}\times c}\Big].
\end{array}
\right\}
\end{equation} 
Consequently, the final stabilizer matrix $H_{S}=\left[\begin{array}{c|c}
    \tilde{H}_{ex}&\textbf{0}\\
    \hline
 \textbf{0}& \tilde{H}_{ez}
\end{array}\right]$ will look like\\
\begin{equation}
 \left[\begin{array}{c|c}
     \begin{array}{c|c|c|c|c}
    \mathbf{0}_{\rho_{1}\times c}
 &\mathrm{M} & \mathbf{0}_{\rho_{1}\times (n-\rho_{1}-c)} & D^{(x)} & D^{(x)} 
\end{array}& \mathbf{0}_{\rho_{1}\times (n+c)} \\
\hline
   \mathbf{0}_{\rho_{2}\times (n+c)}  & \tilde{V}_{\rho_{2}\times c} \mid \mathbf{0}_{\rho_2\times(\rho_1-c)} \mid \tilde{A}_1^{(z)}\,\mathbf{0}_{\rho_2\times k} \mid D^{(z)} \mid D^{(z)}
\end{array}\right], 
\end{equation}
where $M_{\rho_{1}\times(\rho_{1}-c)}=\left[\begin{array}{cc}
         \mathbf{0}_{c\times (\rho_{1}-c)}\\
     \hline
       I_{\rho_{1}-c}
 \end{array}\right]$ and $k=n-\rho_{1}-\rho_{2}+c$.
 \paragraph{STEP 5:} As mentioned earlier, for the transformed matrix to establish the initial state
$\ket{0}^{\otimes a}\ket{\psi}\ket{\phi}^{\otimes c},$
the binary support of the \(X\)-type stabilizers must be zero at the first \(n-2c\) positions. Therefore, to exchange the nonzero columns $\begin{bmatrix}
M^{\mathrm{T}}\;\mid
\mathbf{0}_{(\rho_{1}-c)\times \rho_{2}}
\end{bmatrix}^{\mathrm{T}}$
in $H_{S}$ with the zero columns, we apply the Hadamard gates to the qubits indexed from $c+1$ to $\rho_1$, as follows:
 \begin{equation*}
     U_5=\prod_{i=c+1}^{\rho_1} H^{(i)}.
 \end{equation*}
By combining the operators introduced in Steps~1--5, the overall encoding operator for the entanglement-assisted quantum code $Q$ is expressed as
\begin{equation*}
    \mathcal{E}=\left(U_5 U_4 U_3 U_2 U_1\right)^{\dagger}.
\end{equation*}
Steps~1-5 complete the encoding process for the codes constructed in Theorems~\ref{theorem3} and~\ref{theorem3.1}. 
We now show that the number of CNOT gates required in the decoding procedure is less than that used by the alternative methods presented in~\cite{nadkarni2021encoding}. The justification for the same is given below:

The maximum number of CNOT-gates required by the $U_1$-operator in Eq.~\eqref{Eq:U_xgate} is  $\rho_1(n-\rho_1)$. Similarly, the $U_2$, $U_3$ and $U_4$-operators require $c(\rho_1-c)$, $c$ and $k\times \rho_{2}$ maximum number of CNOT-gates in Eqs.~\eqref{Eq:Operator_U_B_z}, \eqref{Eq:Oper_U_e_2_z} and \eqref{Eq:Uza2Gate}, respectively. Therefore, the upper bound on the total number of CNOT-gates required for the encoder $\mathcal{E}$ is given by
    $$\rho_1(n-\rho_1)+c(\rho_1-c)+c+k\rho_{2}.$$
 Substituting $k=n-\rho_{1}-\rho_{2}+c$ into the above expression, we then have
    \begin{align}
   \label{Eq:enc_simp_complex} 
   n(\rho_{1}+\rho_{2})+(c-\rho_{2})(\rho_{1}+\rho_{2})-\rho_{1}^{2}-c^{2}+c<n(\rho_1+\rho_2)-\rho_{1}^{2}\quad (\text{since } c<\rho_{2}).
    \end{align}
    Therefore, it is quite obvious that the complexity of the CNOT gate $\mathcal{O}\left(n(\rho_1+\rho_2)-\rho_{1}^{2}\right)$ is less than
 the complexity of the gate in~\cite{nadkarni2021encoding}, which is reported as $\mathcal{O}\left(n(\rho_1+\rho_2)\right)$. 
  
\begin{figure}
    \centering
    \includegraphics[width=0.95\linewidth]{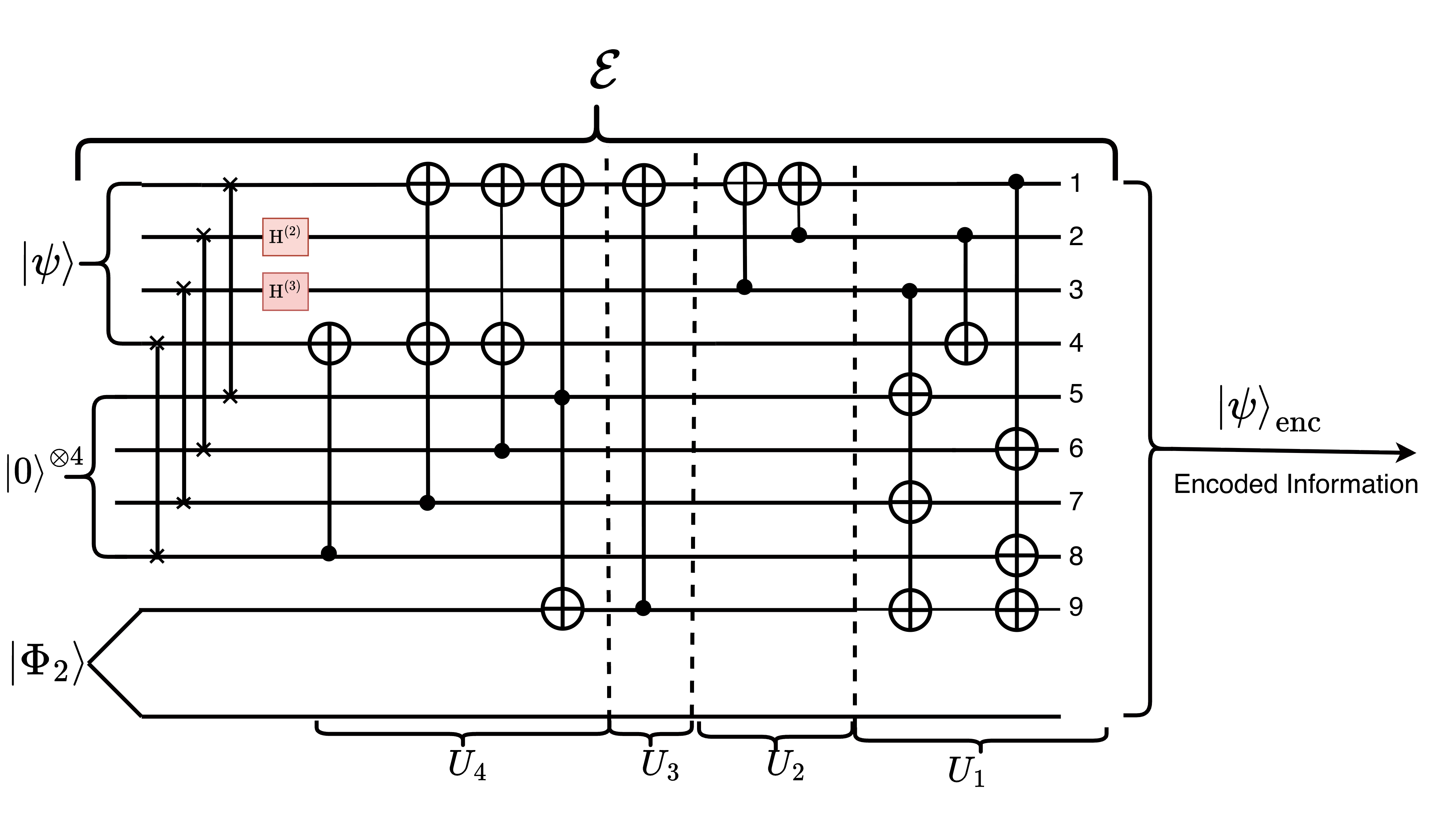}
   \caption{Encoding circuit of the EA-QC-QLDPC code  $[[9,4,2;1]]_2$ presented in the Example~\ref{example1}.}
    \label{fig:Examp_entang_rate}
\end{figure}

  The reader should also note that, although, the maximum number of CNOT-gates required by the $U_1$-operator is $\rho_1(n-\rho_1)$, given in Eq.~(\ref{Eq:enc_simp_complex}). This maximum bound is achieved only when all the entries of the $T=\left[A^{(x)}|E^{(x)}\right]$ are assumed to be 1. However, the exact number of CNOT-gates required by the $U_1$ operator is exactly equal to the number of nonzero elements in $A^{(x)}$ and $E^{(x)}+ D^{(x)}$, refer Eq.~\eqref{Eq:U_xgate}. Therefore, for quasi-cyclic structure, we can further reduce the gate complexity, which is significant.
  % Therefore, if the matrix $T$ is quasi-cyclic in nature then the complexity of the $U_x$-operator is reduced to $\frac{\rho_1(n-\rho_1)}{p}$ since invidual rows of $T$ have $\frac{(n-\rho_1)}{p}$ nonzero entries. For the quasi-cyclic construction, the complexity of the encoder scales as $\mathcal{O}\left(\rho_1\frac{(n-\rho_1)}{p}+n\rho_2\right)$
\begin{example}\label{example1}
   Let $C_1$ and $C_2$ be two $[9,6]_2$ classical QC-LDPC codes with  the following parity-check matrices:
\begin{align}
    H_x=\left[\begin{array}{ccc|ccc|ccc}
        1 & 0 & 0 & 0 & 0 & 1 &  0 & 1 & 0\\
       0 & 1 & 0 & 1 & 0 & 0 &  0 & 0 & 1\\
       0 & 0 & 1 & 0 & 1 & 0 &  1 & 0 & 0
    \end{array}\right],\nonumber\\
      H_z=\left[\begin{array}{ccc|ccc|ccc}
        1 & 0 & 0 & 0 & 1 & 0 &  0 & 0 & 1\\
       0 & 1 & 0 & 0 & 0 & 1 &  1 & 0 & 0\\
       0 & 0 & 1 & 1 & 0 & 0 &  0 & 1 & 0
    \end{array}\right],\nonumber
    \label{Examp:Stab_matrices}
\end{align} respectively. It follows from Theorem~\ref{theorem3} that there exists an EA-QC-QLDPC code with parameters $[[9,4,2;1]]_2$.

%  The minimum number of required maximally entangled qubits pairs can be calculated as follows:
% \begin{align*}
%     c=\mathrm{rank}(H_xH^{\text{T}}_z)=\mathrm{rank}\left(\left[\begin{array}{ccc}
%         1 & 1 & 1\\
%         1 & 1 & 1 \\
%         1 & 1 & 1
%     \end{array}\right]\right)=1.
% \end{align*}
By using a single entangled qubit pair, stabilizer generator matrices $H_{ex}$ and $H_{ez}$ of the quantum code $Q=[[9,4,2;1]]_2$ satisfying the dual-containing criteria can be generated, i.e., $H_{ex}H_{ez}^{^{\text{T}}}=\mathbf{0}\pmod{2}$. The matrices $H_{ex}$ and $H_{ez}$ are given in the following equation:

\begin{equation} \label{Eq:Stab_Mat_Gen_enc}
    \left.
    \begin{array}{l}
    H_{ex} =
\scalebox{1}[1.2]{\Bigg[}
\overset{9-\text{transmitter qubits}}{\overbrace{
\begin{array}{ccc}
1&0&0 \\
   0&1&0 \\
   0&0&1
\end{array} \scalebox{1}[1.2]{\Bigg |} \begin{array}{ccccc}
    0 & 0 & 1 & 0 & 1 \\
    1 & 0 & 0 & 0 & 0 \\
    0 & 1 & 0 & 1 & 0 
\end{array}  \scalebox{1}[1.2]{\Bigg |}
\overset{\text{half of 1 ebit}}{\overbrace{\begin{array}{c}
     0\\
      1 \\
      0
\end{array}}}
}}
  \scalebox{1}[1.2]{\Bigg |}  \scalebox{1}[1.2]{\Bigg |}
\overset{\text{half of 1 ebit}}{\overbrace{\begin{array}{c}
     1\\
      1 \\
      1
\end{array}}}
\scalebox{1}[1.2]{\Bigg]}, \\[1.5em]
H_{ez} =
\overset{8-\text{ancilla and information qubits}}{\overbrace{\left[\begin{array}{ccc|ccccc}
   1&0&0 &0 & 1 & 0 & 0 & 0 \\
   0&1&0  &  0 & 0 & 1 & 1 & 0\\ 
   0&0&1  &  1 & 0 & 0 & 0 & 1
\end{array} \right. }} \scalebox{1}[1.3]{\Bigg |}
\begin{array}{c}
     1\\
      0 \\
      0
\end{array}  \scalebox{1}[1.3]{\Bigg |} \scalebox{1}[1.3]{\Bigg |} \left. \begin{array}{c}
     1\\
      1 \\
      1
\end{array} \right]. 
  \end{array}
    \right\}
\end{equation}
Now, we are ready to design the quantum encoder for the EA-QC-QLDPC code $Q=[[9,4,2;1]]_2$ based on the techniques described in this paper.
On comparing the stabilizer matrices in Eq.~\eqref{Eq:Stab_Mat_Gen_enc} with the matrices in Eq. (\ref{Eq:Mat_Encod_GA_ELI}), we get 
\begin{equation*}
    J=I_3=\left[\begin{array}{ccc}
    1 & 0 & 0\\
    0 & 1 & 0\\
    0 & 0 & 1 
\end{array}\right],\quad A^{(x)}=\left[\begin{array}{ccccc}
    0 & 0 & 1 & 0 & 1 \\
    1 & 0 & 0 & 0 & 0 \\
    0 & 1 & 0 & 1 & 0 
\end{array}\right],\quad
A^{(z)}=\left[\begin{array}{ccccc}
    0 & 1 & 0 & 0 & 0 \\
    0 & 0 & 1 & 1 & 0\\
    1 & 0 & 0 & 0 & 1
\end{array}\right],\quad
\end{equation*}
\begin{equation*}
E^{(x)}=\left[\begin{array}{ccc}
     0 \\
     1 \\
     0
\end{array}\right],\quad
E^{(z)}=\left[\begin{array}{c}
     1\\
      0 \\
      0
\end{array}\right],\quad D^{(x)}=\left[\begin{array}{c}
        1\\1\\1
\end{array}\right] \text{ and } D^{(z)}=\left[\begin{array}{c}
1\\1\\1
\end{array}\right].
\end{equation*}
 The submatrices $\left[E^{(x)}|D^{(x)}\right]$ and $\left[E^{(z)}|D^{(z)}\right]$ have the $X$ and $Z$ type stabilizer support on the transmitter and receiver side entangled qubits, respectively.

In the next step, we iterate over the rows of $H_{ex}$ and apply $\text{CNOT}$ gates to make the $A^{(x)}$ submatrix zero, which implies that the support of X-type stabilizers on the ancillae qubits become Pauli-$I$ and transform $E^{(x)}$ submatrix to $D^{(x)}$ in $H_{ex}$ for column-wise matching of the X-type stabilizer support on the transmitter and the receiver end qubits, and these 
desired transformations can be done by the following operator:
\begin{align*}
U_1=&\left(\underset{\text{ Transform third row of }H_{ex}}{\underbrace{\left(\text{CNOT}^{(3,9)}\text{CNOT}^{(3,7)}\right)\left(\text{CNOT}^{(3,5)}\right)}}\right)\left(\underset{\text{ Transform second row of }H_{ex}}{\underbrace{\left(\text{CNOT}^{(2,4)}\right)}}\right)\\
&\left(\underset{\text{ Transform first row of }H_{ex}}{\underbrace{\left(\text{CNOT}^{(1,9)}\text{CNOT}^{(1,8)}\right)\left(\text{CNOT}^{(1,6)}\right)}}\right).
\end{align*}

 The above operator $U_1$ transform the stabilizer matrices as follows:
 \begin{equation} \label{Eq:Appen_N_F}
     \left.
     \begin{array}{l}
        H_{ex}=\left[\begin{array}{ccc|ccc|ccc||c}
        1 & 0 & 0 & 0 & 0 & 0 &  0 & 0 & 1 & 1\\
       0 & 1 & 0 & 0 & 0 & 0 &  0 & 0 & 1 & 1\\
       0 & 0 & 1 & 0 & 0 & 0 &  0 & 0 & 1 & 1
    \end{array}\right],\\
    \\
H_{ez}=\left[\begin{array}{ccc|ccc|ccc||c}
        0 & 0 & 0 &  0 & 1 & 0 &  0 & 0 & 1 & 1\\
       1 & 1 & 1 & 0 & 0 & 1 &  1 & 0 & 0 & 1\\
       1 & 1 & 1 & 1 & 0 & 0 &  0 &  1 & 0 & 1
    \end{array}\right].    
     \end{array}
     \right\}
 \end{equation}
From Eq. (\ref{Eq:Appen_N_F}), the submatrix $B$ is $\left[\begin{array}{ccc}
    0 & 0 & 0 \\
    1 & 1 & 1  \\
    1 & 1 & 1
\end{array}\right].$ It is straightforward that the column rank $c$ of the submatrix $B$ is $1$. Therefore, we can solve the following system of linear equations:
$$VX=\left[\begin{array}{cc}
    0 \\
    1  \\
    1 
\end{array}\right]X=\left[\begin{array}{cc}
     0 & 0 \\
     1 & 1\\
     1 & 1
\end{array}\right]=W.$$ It is trivial to check that the solution $X$ is equal to row matrix $\begin{bmatrix}
    1&1
\end{bmatrix}$. Using $X$, we apply $$U_2=\underset{\text{Make 3rd column of }B\text{ zero}}{\underbrace{\text{CNOT}^{(3,1)}}}\ \ \underset{\text{Make 2nd column of }B\text{ zero}}{\underbrace{\text{CNOT}^{(2,1)}}}$$ which makes the second and third columns of $B$ zero.  The transformed matrices $H_{ex}$ and $H_{ez}$ after applying the operators $U_{1}$ and $U_{2}$ are 
\begin{align}
    H_{ex}=&\left[\begin{array}{ccc|ccc|ccc||c}
        1 & 0 & 0 & 0 & 0 & 0 &  0 & 0 & 1 & 1\\
       1 & 1 & 0 & 0 & 0 & 0 &  0 & 0 & 1 & 1\\
      1 & 0 & 1 & 0 & 0 & 0 &  0 & 0 & 1 & 1
    \end{array}\right],\nonumber\\
      H_{ez}=&\left[\begin{array}{ccc|ccc|ccc||c}
        0 &  0 & 0 & 0 & 1 & 0 &  0 & 0 & 1 & 1\\
       1 & 0 & 0 & 0 & 0 & 1 &  1 & 0 & 0 & 1\\
       1 & 0 &  0 & 1 & 0 & 0 &  0 & 1 & 0 & 1
    \end{array}\right]\nonumber.
\end{align}
Next, we apply the CNOT gate 
$U_3 = \mathrm{CNOT}^{(9,1)},$
which maps $E^{(z)}$ to $D^{(z)}$. This operation transforms the stabilizer matrices as follows:
\begin{equation}
    \left.
    \begin{array}{l}
     H_{ex}=\left[\begin{array}{ccc|ccc|ccc||c}
         0 & 0 & 0 & 0 & 0 & 0 &  0 & 0 & 1 & 1\\
       0 & 1 & 0 & 0 & 0 & 0 &  0 & 0 & 1 & 1\\
        0 & 0 & 1 & 0 & 0 & 0 &  0 & 0 & 1 & 1
    \end{array}\right],\\
    \\
H_{ez}=\left[\begin{array}{ccc|ccc|ccc||c}
        0 & 0 & 0 & 0 & 1 & 0 &  0 & 0 &  1 & 1\\
       1 & 0 & 0 & 0 & 0 & 1 &  1 & 0 & 1 & 1\\
       1 & 0 & 0 & 1 & 0 & 0 &  0 & 1 & 1 & 1
    \end{array}\right].
    \end{array}
    \right\}
\end{equation}

Next, we apply the operator $U_{4}$ to make the last $k=4$ columns of $A^{(z)}$ zero. Now, we partition $A^{(z)}$ as follows:
\begin{equation*}
A^{(z)}=
\scalebox{1}[1.15]{\Bigg[}
\underbrace{\begin{array}{c}
0\\
1\\
1
\end{array}}_{A^{(z)}_{1}}
\scalebox{1}[1.15]{\Bigg |}
\underbrace{\begin{array}{cccc}
1&0&0&0\\
0&1&1&0\\
0&0&0&1
\end{array}}_{A^{(z)}_{2}}
\scalebox{1}[1.15]{\Bigg]}
\end{equation*}
% \begin{equation*}
%  A^{(z)}=
%      \underset{A^{(z)}_{1}}{\left[\underbrace{\begin{array}{c}
%          0\\
%           1\\
%           1 \end{array}}\right.}\Bigg| \underset{A^{(z)}_{2}}{\left.\underbrace{\begin{array}{cccc}
%           1&0&0&0  \\
%          0&1&1&0  \\
%            0&0&0&1  \\
%     \end{array}}\right]} 
% \end{equation*}

% \begin{equation*}
%     A^{(z)}=\underset{$A^{(z)}_{1}$}{\begin{array}{c}
%          0 \\
%          1\\
%          1
%     \end{array}}|\underset{$A_{2}^{(Z)}$}{\begin{array}{cccc}
%          1&0&0&0  \\
%          0&1&1&0  \\
%           0&0&0&1  \\
%     \end{array}}.
% \end{equation*}
For obtaining $U_{4}$, we solve the linear equation $$[V|A^{(z)}_1|D^{(z)}]\tilde{X}=A^{(z)}_2\implies \left[\begin{array}{c|c|c}
    0 & 0 & 1\\
    1 & 0 & 1\\
    1 & 1 & 1
\end{array}\right]\tilde{X}=\left[\begin{array}{cccc}
        1 & 0 &  0 & 0 \\
       0 & 1 &  1 & 0 \\
        0 & 0 &  0 & 1
\end{array}\right].$$
Using the columns of $\tilde{X}=\left[\begin{array}{cccc}
    1 & 1 & 1 & 0\\
    0 & 1 & 1 & 1\\
    1 & 0 & 0 & 0
\end{array}\right]$, and from Eq. (\ref{Eq:Uza2Gate}), we obtain
\begin{align*}
U_4 = \mathrm{CNOT}^{(8,4)}\,
\mathrm{CNOT}^{(7,\{1,4\})}\,
\mathrm{CNOT}^{(6,\{1,4\})}\,
\mathrm{CNOT}^{(5,\{1,9\})},
\end{align*}
where $\mathrm{CNOT}^{(x,*)}$ makes the $x^{\text{th}}$ column of $H_{ez}$ zero based on the positions of the target qubit positions indicated in (*) .
% \begin{align*}
% U_2=\underset{\text{Make 8th column of }H_z\text{ zero}}{\underbrace{\text{CNOT}^{(8,4)}}}\ \ \underset{\text{Make 7th column of }H_z\text{ zero}}{\underbrace{\text{CNOT}^{(7,\{1,4\})}}}\ \ \underset{\text{Make 6th column of }H_z\text{ zero}}{\underbrace{\text{CNOT}^{(6,\{1,4\})}}}\ \ \underset{\text{Make 5th column of }H_z\text{ zero}}{\underbrace{\text{CNOT}^{(5,\{1,9\})}}}.
% \end{align*}
The application of $U_{4}$ transforms the stabilizer matrices as follows:
\begin{align}
    H_{ex}=&\left[\begin{array}{ccc|ccc|ccc||c}
         0 & 0 & 0 & 0 & 0 & 0 &  0 & 0 & 1 & 1\\
       0 & 1 & 0 & 0 & 0 & 0 &  0 & 0 & 1 & 1\\
        0 & 0 & 1 & 0 & 0 & 0 &  0 & 0 & 1 & 1
    \end{array}\right],\nonumber\\
      H_{ez}=&\left[\begin{array}{ccc|ccc|ccc||c}
        0 & 0 & 0 & 0 & 0 & 0 &  0 & 0 & 1 & 1\\
       1 & 0 & 0 & 0 & 0 & 0 &  0 & 0 & 1 & 1\\
       1 & 0 & 0 & 1 & 0 & 0 &  0 & 0 & 1 & 1
    \end{array}\right].
\end{align}
Finally, we apply the Hadamard-gate on the second and third qubits and obtain the complete stabilizer matrix as follows:
    \begin{align}
    H_{\text{s}}=\left[\begin{array}{ccccccccc||c|ccccccccc||c}
        0 & 0 & 0 & 0 & 0 & 0 &  0 & 0 & 1 & 1 & 0 & 0 & 0 & 0 & 0 & 0 & 0 & 0 & 0 & 0\\
        0 & 0 & 0 & 0 & 0 & 0 &  0 & 0 & 1 & 1& 0 & 1 & 0 & 0 & 0 & 0 & 0 & 0 & 0 & 0\\
       0 & 0 & 0 & 0 & 0 & 0 &  0 & 0 & 1 & 1 & 0 & 0 & 1 & 0 & 0 & 0 & 0 & 0 & 0 & 0\\
        \hline
        0 & 0 & 0 & 0 & 0 & 0 & 0 & 0 & 0 & 0 & 0 & 0 & 0 & 0 & 0 & 0 &  0 & 0 & 1 & 1\\
        0 & 0 & 0 & 0 & 0 & 0 & 0 & 0 & 0 & 0 & 1 & 0 & 0 & 0 & 0 & 0 &  0 & 0 & 1 & 1\\
        0 & 0 & 0 & 0 & 0 & 0 & 0 & 0 & 0 & 0 & 1 & 0 & 0 & 1 & 0 & 0 &  0 & 0 & 1 & 1\\
    \end{array}\right].
\end{align} 
The complete encoding operator is $\mathcal{E}=\left(\text{H}^{(2,3)}U_4U_3U_2U_1\right)^{\dagger}$. For bringing the encoder into a systematic form such that the first four qubits are information qubits, a series of SWAP-gates can be applied followed by $\mathcal{E}$ as follows:
$$\text{SWAP}^{(4,8)}\text{SWAP}^{(3,7)}\text{SWAP}^{(2,6)}\text{SWAP}^{(1,5)}.$$
\end{example}
%%%%%%%%%%%%%%%%%%%%%%%%%%%%%%%%%%%%%%%%%%%%%%%%%%%%%%%%%%%%%%%%%%%%%%%%%%%%%%%%%%%%%%%%%%%%%%%%%%%%%%%%%%%%%%%%%%%%%%%%%%%%%%%%%%%%%%%%%%%%%%%%%%%%%%%%%%%%%%%%%%%%%%%%%%%%%%%%%%%%%%%%%%%%%%%%%%%%%%%%%%%%%%%%%%%%%%%%%%%%%%%%%%%%%%%%%%
Figure~\ref{fig:Examp_entang_rate} illustrates the complete encoding schematic in the entanglement-assisted setup, in which a Bell pair $\ket{\Phi}$ is pre-shared between the transmitter and the receiver and assumed to be error free. Four information qubits are encoded into the quantum code $Q = [[9,4,2;1]]_2$ using the encoder $\mathcal{E}$. 
%%%%%%%%%%%%%%%%%%%%%%%%%%%%%%%%%%%%%%%%%%%%%%%%%%%%%%%%%%%%%%%%%%%%%%%%%%%%%%%%%%%%%%%%%%%%%%%%%%%%%%%%%%%%%%%%%%%%%%%%%%%%%%%%%%%%%%%%%%%%%%%%%%%%%%%%%%%%%%%%%%%%%%%%%%%%%%%%%%%%%%%%%%%%%%%%%%%%%%%%%%%%%%%%%%%%%%%%%%%%%%%%%%%%%%%%%%
\begin{figure}
\centering
\begin{subfigure}{\textwidth}
\centering
 \includegraphics[width=0.8\linewidth]{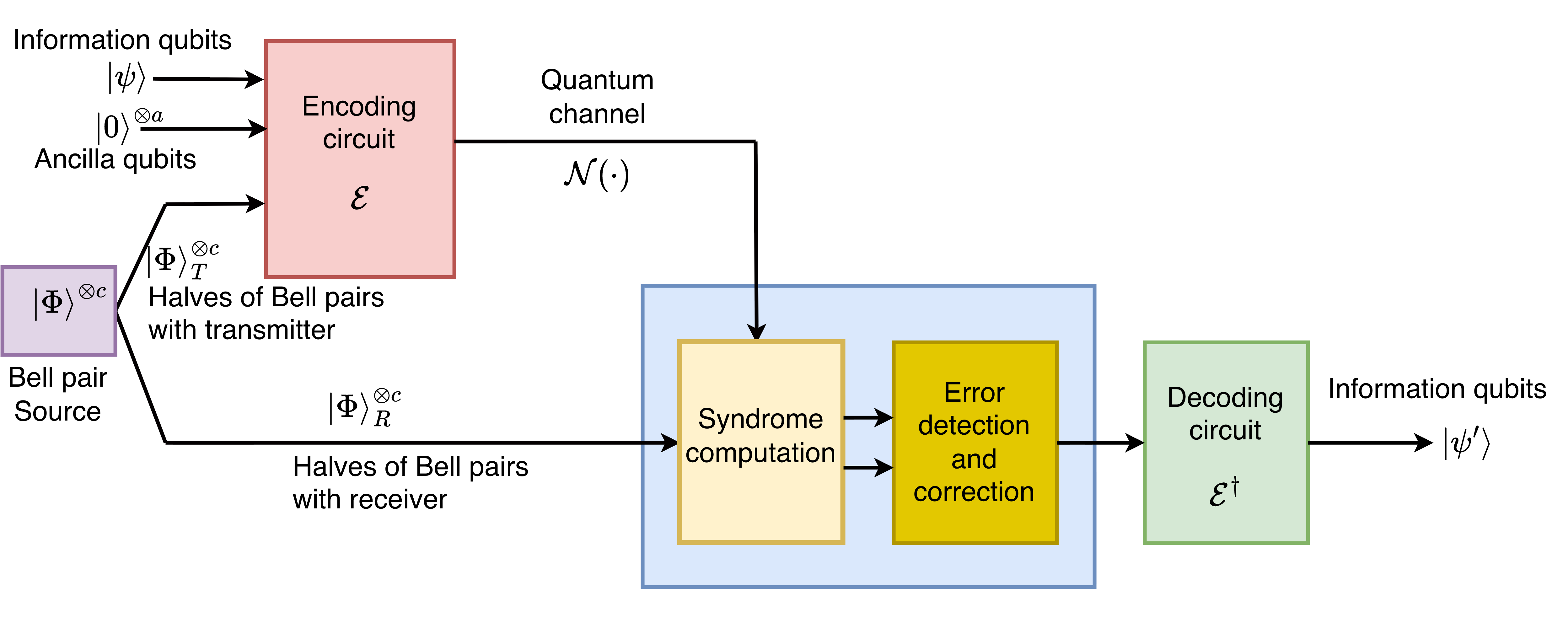}
\caption{One-way communication protocol.}
\label{fig:one_way_comm}
\end{subfigure}

\vspace{0.5cm}

\begin{subfigure}{\textwidth}
\centering
 \includegraphics[width=\linewidth]{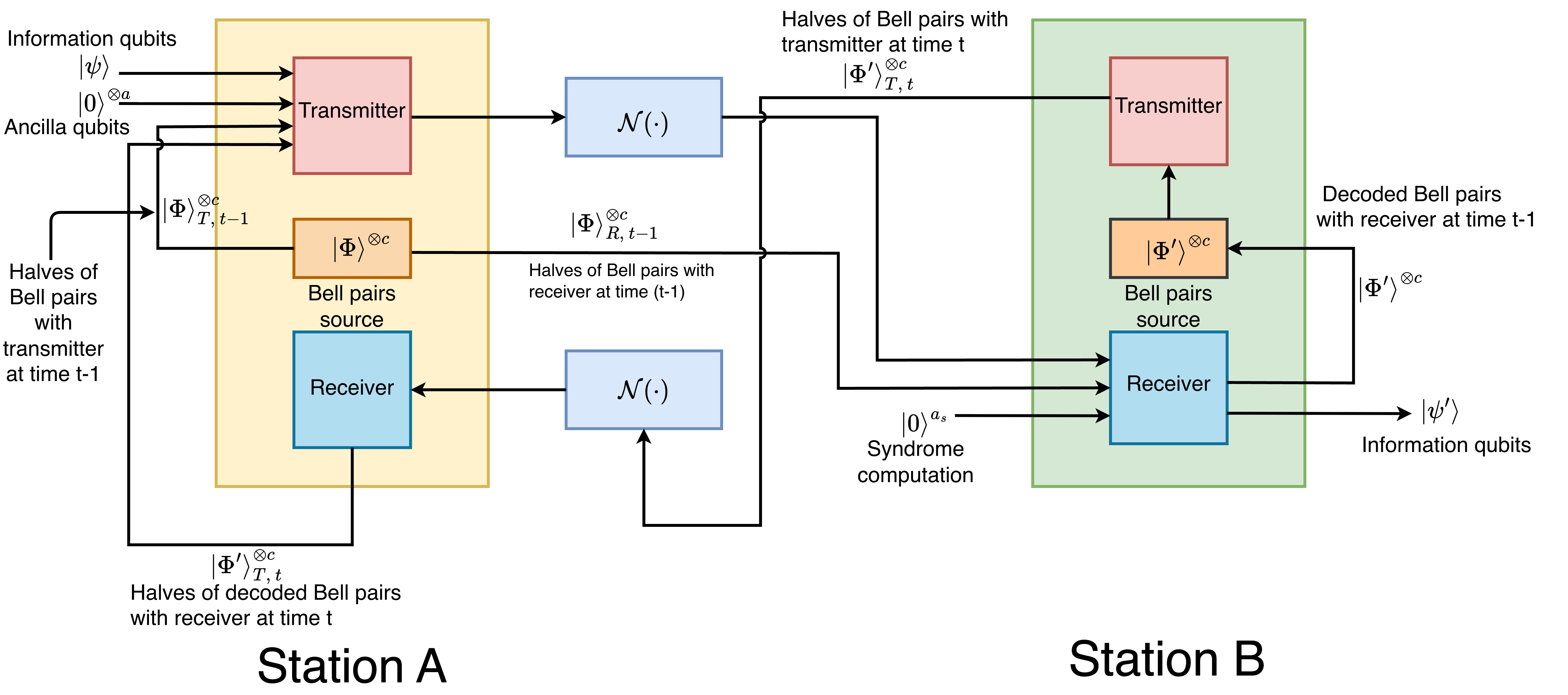}
\caption{Two-way communication protocol.}
\label{fig:two_way_comm}
\end{subfigure}
\caption{(a) Illustration of a one-way communication protocol, in which a new Bell pair is generated for every transmission and consumed upon use. The Bell pair can be generated from a gate-based scheme or from a heralded source. One half of each Bell pair is part of the payload that is encoded and sent to the receiver. The other half of each Bell pair is shared with the receiver and assumed to be error free.
(b) In this scenario, we illustrate a two-way entanglement-assisted communication protocol from stations A and B that operate in catalysis mode. Each station thus acts as a transmitter as well as a receiver. This scenario is practical when the stations can do both computing and communication using a pool of Bell pairs. The transmitter has access to local Bell pair sources and one half of each Bell pair is encoded and sent from the transmitter to the receiver. At the receiver, the decoder uses the other half of the Bell pair assumed to be error free, decodes the information qubits as well as recovering the Bell pair at the receiver for consuming it in the next iteration. Thus, the Bell pairs are constantly cycling between the two stations, in a ping-pong fashion. The reader must note that one has to check for the validity of the decoded Bell pair for successfully using it in a subsequent round of transmission.}
\label{fig:communication_EA_codes}
\end{figure}
%%%%%%%%%%%%%%%%%%%%%%%%%%%%%%%%%%%%%%%%%%%%%%%%%%%%%%%%%%%%%%%%%%%%%%%%%%%%%%%%%%%%%%%%%%%%%%%%%%%%%%%%%%%%%%

\begin{figure}[h]
    \centering
    \includegraphics[width=\linewidth]{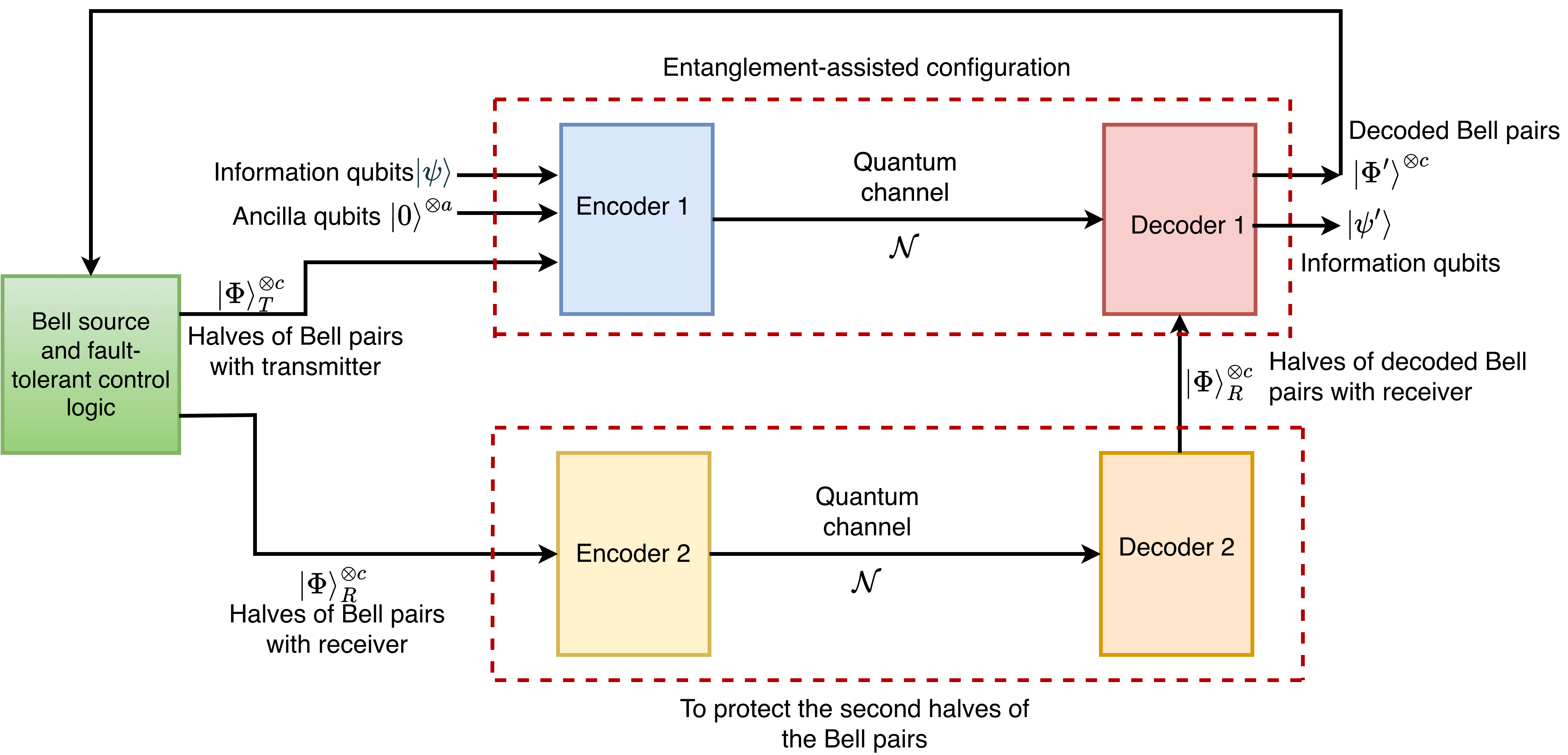}
   \caption{Entanglement-assisted (EA) quantum codes in computation, where halves of shared Bell pairs are protected through an error correction scheme. One half of the Bell pair is part of the payload during the encoding and will be decoded jointly using the other half of the Bell pair that is separately encoded via a second encoder and made available in the entanglement-assisted configuration during decoding.}
    \label{fig:EA-quantum-computing}
\end{figure}

For a given entanglement-assisted quantum code $[[n,k;c]]$, we consider several possible encoding and decoding schemes depending on the specific communication and computational scenarios in which the code is implemented, as follows:
\begin{enumerate}[leftmargin=0pt]
    \item [] (1) \textbf{One-way communication of Bell pairs:} The first $n$ qubits of the resulting codeword state $\ket{\Psi}$ are transmitted to the receiver through a noisy quantum channel $\mathcal{N}(\cdot)$. At the receiver, the stabilizer generators of the code $Q$ are measured on the received $n$ qubits together with the pre-shared Bell-pair qubits. Based on the resulting syndrome, appropriate error correction is applied, followed by the inverse encoding operation $\mathcal{E}^{\dagger}$, thereby successfully recovering the original information qubits, as illustrated in Figure~\ref{fig:one_way_comm}. 
     \item [] (2) \textbf{Two-way communication of Bell pairs (catalytic approach):} The reader must note that the catalytic-rate-based transmission scheme shown in Figure~\ref{fig:two_way_comm} does not require creating new Bell pairs locally and sharing them across the quantum transceiver every time. Instead, once the $n$-qubits are transmitted through the channel and successfully decoded, a portion of that Bell pair successfully decoded at time $t-1$ could be subsequently used for the next transmission at the receiver as part of the payload for encoding the information at time $t$ and relay the portion of that Bell pair to the transmitter. This architecture facilitates a two-way entanglement-assisted coded transmission of quantum information. We note that one half of the EPR pair is maintained error free at the receiver and the other is part of the encoded quantum codeword. In this way, the catalytic scheme is more efficient for the quantum communication scenario. It is obvious that only in case when the Bell pair post decoding at the receiver loses its decoherence, a Bell pair must be locally generated either through gate-based logic or other means, such as entanglement swapping, for the next round of communications.  

     Despite the assumption that one of the EPR pairs at the receiver is error free, the EA-based communications offers significant advantages when soft-decision-based joint decoding is done over the joint Tanner graph, without having to deal with short cycles.
     \item [] (3) \textbf{Entanglement-assisted quantum codes in computing:} In the quantum computing scenario, every qubit undergoes noise. Hence, the EPR pair used in the entanglement-assisted scheme at time $t$ is part of the payload of the previously encoded codeword, i.e., at time $t-1$. Thus, at the expense of the catalytic rate, one can realize coded computing such that none of the qubits are assumed error free as illustrated in Figure~\ref{fig:EA-quantum-computing}. 
\end{enumerate}
The proposed constructions are thus amenable for both quantum communications and computing with a minor sacrifice in the catalytic coding rate in the quantum computing scenario.   
%%%%%%%%%%%%%%%%%%%%%%%%%%%%%%%%%%%%%%%%%%%%%%%%%%%%%%%%%%%%%%%%%%%%%%%%%%%%%%%%%%%%%%%%%%%%%%%%%%%%%%%%%%%%%%%%%%%%%%%%%%%%
%%%%%%%%%%%%%%%%%%%%%%%%%%%%%%%%%%%%%%%%%%%%%%%%%%%%%%%%%%%%%%%%%%%%%%%%%%%%%%%%%%%%%%%%%%%%%%%%%%%%%%%%%%%%%%%%%%%%%%%%%%%%%%%%%%%%%%%%%%%%%%%%%%%%%%%%%%%%%%%%%%%%%%%%%%%%%%%%%%%%%%%%%%%%%%%%%%%%%%%%%%%%%%%%%%%%%%%%%%%%%%%%%%%%%%%%%%
\section{Code Performance Analysis via Simulations}\label{Sec.4}
In this section, we introduce two types of error models: (i) a random error model without burst errors, and (ii) a burst error model with Markovian correlations. We then evaluate the performance of the proposed codes over these models under various decoding configurations.

\textbf{Random Error Model:} Consider a quantum state $\rho$ comprising $\ell$-qubits transmitted over a quantum channel. We assume a depolarizing noise channel model for random errors, denoted by $\pi(p_d)$, where the channel action occurs independently on each qubit. When applied to a single qubit state $\sigma$, the channel introduces a Pauli error with probability $p_d$ as follows:  
$\sigma \mapsto (1 - p_d)\sigma + \frac{p_d}{3} \left( X\sigma X + Y\sigma Y + Z\sigma Z \right),$
where $X$, $Y$ and $Z$ represent the Pauli matrices acting on the state $\sigma$.

\textbf{Burst Error Model:} Consider a quantum state $\rho$ comprising $\ell$-qubits transmitted over a depolarizing quantum channel with Markovian correlations, resulting in burst errors \cite{QBECC,Burstconstruction}. In this model, the Pauli error $E_i$ on the $i^{\text{th}}$ qubit can be an element from the set $\{I, X, Y, Z\}$. Due to Markovian correlations in the channel, an error \(E_i\) occurring on the $i^{\text{th}}$ qubit influences the likelihood of an error \(E_j\) on an adjacent qubit \(j\), where \(j = i+1\). This influence is governed by a correlation parameter \(\eta\), under a first-order Markov process, and is mathematically described by the conditional probability:
\begin{equation*}
   P(E_j | E_i) = (1 - \eta) P(E_j) + \eta \, \delta_{E_i, E_j}, 
\end{equation*}
where \(\delta_{E_i, E_j}\) is an indicator function given by
\begin{equation*}
  \delta_{E_i, E_j} = 
\begin{cases}
1, & \text{if } E_i = E_j, \\
0, & \text{if } E_i \neq E_j.
\end{cases}  
\end{equation*}
For each qubit \(i \in \{1, 2, \dots, \ell\}\), the probability distribution for the error \(E_i\) is specified as \(P(E_i = I) = 1 - p_d\) for no error, and \(P(E_i \in \mathcal{E}) = \frac{p_d}{3}\), where \(\mathcal{E}=\{X, Y, Z\}\). This model captures the correlated nature of errors in the quantum channel, allowing us to analyze the impact of noise propagation on entangled states in a systematic way.
%Since the model is Markov so $P(E_3|E_2,E_1)=P(E_3|E_2)=(1-\eta)P(E_3)+\eta \,\delta_{(E_2,E_3)}$. For the $i^{\mathrm{th}}$ qubit in $\rho$, the probability of error is given by $P(E_i|E_{i-1})=(1-\eta)P(E_i)+\eta \,\delta_{(E_{i-1},E_i)}$.
In general, if $\Phi$ is the Markovian error applied on the $\ell$-qubit state $\rho$, then 
\begin{equation*}  \Phi(\rho)= \displaystyle \sum\limits_{E_1,\ldots,E_\ell}\Big(P(E_1) \prod\limits_{i=2}^{\ell}P(E_{i}|E_{i-1})\Big)  E_{\ell}\cdots E_{1} \rho E^{\dagger}_1\cdots E^{\dagger}_{\ell},  
\end{equation*}
where $E_{i}\in \{I, X, Y, Z\}$ for all $1\leq i\leq \ell$.
% \begin{align*}
	%     \Phi(\rho)=\sum_{E_1,\ldots,E_l}P(E_1) \prod_{i=2}^{l}P(E_{i}|E_{i-1}) \, E_{l}\cdots E_{1} \rho E^{\dagger}_1\cdots E^{\dagger}_{l}.
	% \end{align*}
%\begin{figure}
%	\centering
%	\includegraphics[width=3.3in,height=2.1in]{Plots/PLOT_QLDPC_121_169_plus_5_it_with_121_SS_final.eps}
%	\caption{The logical error rate vs. $p_d$ for $[[121,20,10;1]]_2$ and $[[169,24,12;1]]_2$ EA-QC codes, for $\eta=0.5$. The logical error rate decreases for increasing $p$ with random error as well as for random and burst errors. \textcolor{blue}{The impact of iterations on the $[[121,20,10;1]]_2_2$ code is evident, with the decoder performing poorly after just 5 iterations compared to 100 iterations.}}
%	\label{fig:Depolar_prob}
%\end{figure}

The proposed classical QC-LDPC codes using permutation matrices of order $p$ described in Sections \ref{Sec.2} has a burst erasure correction capability of at least length $p$ \cite{mondal2021efficient}. Consequently, the EA-QC-QLDPC codes constructed in Sections~\ref{Sec.3} and \ref{sec.5}  can correct quantum burst erasures of length $p$ provided that underlying classical codes individually are capable of correcting burst erasures of length at least $p$ \cite{CKM_SGS}. 

We have used four configurations of the decoders, namely: (a) a binary layered sum-product (BLSP) decoder, (b) a quaternary min-sum (QMS) decoder, (c) a quaternary normalized min-sum (QNMS) decoder, and (d) a quaternary block-layered normalized min-sum (QBLNMS) decoder, to evaluate the performance of codes under uncorrelated and correlated $X$- and $Z$-type Pauli errors over the depolarizing channel.
% Using $C$, we can construct a quantum entanglement-assisted stabilizer code $Q=[[n,k,d_q;c]]_2$, detailed in Section \ref{Sec.3}. The quantum code $Q$ is constructed using the CSS construction and the parity-check matrix of $H$. Since the X and Z stabilizer matrices are the non-overlapping block rows of the matrix $H$, so $H_x$ and $H_z$, individually, can correct burst error of length at least $p$. Therefore, the quantum code can correct burst error of length $p$, containing the elements from $\{I,\mathcal{E}\}$.

% \begin{figure}
%     \centering
%     \includegraphics[width=0.5\linewidth]{Figures/Quaternary_121_qubits_code_clean_plot.eps}
%     \caption{}
%     \label{fig:seven_L_code}
% \end{figure}
% \begin{figure}
%     \centering
%     \includegraphics[width=0.5\linewidth]{Figures/Final_itr100_fourty_nine_code.eps}
%     \caption{Caption}
%     \label{fig:eleven_L_code}
% \end{figure}

For the sake of completeness, we now describe the binary and quaternary versions of the decoders along with layering improvements used in the simulations.

\subsection{Binary sum-product (BSP) algorithm}
For an $[[n, k_1 + k_2 - n, d_q]]_2$ quantum LDPC code, we now describe the syndrome-based sum-product algorithm with the $\tanh$-based approach~\cite{Raveendran_2022} over the Tanner graph derived from the following stabilizer matrix 
\begin{equation*}
    H_{S} = \left[\begin{array}{c|c}
	H_x & \mathbf{0}_{(n-k_1)\times n} \\
    \hline
	\mathbf{0}_{(n-k_2)\times n} & H_z
\end{array}\right],
\end{equation*}where \( H_x H^{\mathrm{T}}_z \equiv \mathbf{0} \pmod{2} \). The Tanner graph associated with $H_{S}$ is composed of two separate Tanner graphs: one constructed using \( H_x \) with \( n \) variable nodes and \( n - k_1 \) check nodes, and another using \( H_z \) with \( n \) variable nodes and \( n - k_2 \) check nodes. The sum-product algorithm operates concurrently over both the Tanner graphs.

Let the syndrome $S = [S_x ~ S_z]$ be obtained by measuring observables isomorphic to \( H_x \) and \( H_z \) on a quantum hardware. The $\tanh$-based message-passing algorithm is used to pass messages over each graph, individually. Messages are sent from a variable node to the check node. The check node ascertains if the check is satisfied or not given the incoming messages. Similarly, for each variable node, all the participating check node messages are sent to that variable node. Based on exchanging extrinsic messages, the algorithm converges iteratively. The reader must note that the message passing algorithm in general cannot reach maximum-likelihood performance but reasonably close to that performance. 

For a variable node \( i \) connected to a check node \( j \), the message $\nu_{i,j}$ from the variable node $i$ to the check node $j$ is updated as:
\begin{equation*}
    \nu_{i,j} = L_i + \sum_{k \in \mathcal{N}^{(v)}_i \setminus \{j\}} \mu_{k,i},
\end{equation*}
where \( L_i =\mathrm{log}\left(\frac{1 - p_d}{p_d}\right) \) is the log-likelihood ratio (LLR) associated with each qubit in the depolarizing noise model \( \pi(p_d) \), which is fixed to be the same for all \( n \) qubits, \( \mathcal{N}^{(v)}_i \) represents the neighboring check nodes connected to the variable node \( i \) and $\mu_{k,i}$ represents the message from the check node $k$ to the variable node $i$.

For each check node $j$ and its neighboring variable node $i$, the message $\mu_{j,i}$ from the check node $j$ to the variable node $i$ is updated
\begin{equation*}
     \mu_{j,i} = 2 (-1)^{S_j} \tanh^{-1} \left( \prod_{k \in \mathcal{N}^{(c)}_j \setminus \{i\}} \tanh\left( \frac{\nu_{k,j}}{2} \right) \right),
\end{equation*}
where \( \mathcal{N}^{(c)}_j \) represents the neighboring variable nodes connected to check node \( j \). The sign of each syndrome bit \( S_j \) determines the sign of the messages at the check nodes.

The algorithm continues iterating until the estimated error vector calculated as follows: \( \tilde{e}_{i}^{(a)} = \mathrm{sign} \left( L_i + \sum_{k \in \mathcal{N}^{(v)}_i} \mu_{k,i} \right) \) for $a=x$ or $a=z$ reproduces the syndrome \( S_{a} \), i.e., \( \tilde{S}_x = H_x \tilde{e}_z^{\mathrm{T}} \) and \( \tilde{S}_z = H_z \tilde{e}_x^{\mathrm{T}} \). If \( \tilde{S}_x = S_x \) and \( \tilde{S}_z = S_z \), or the maximum iteration count \( w_{\mathrm{max}} \) is reached, the algorithm terminates.

In practice, a layered implementation of the BSP algorithm can be employed, wherein the check nodes are processed sequentially in layers, and the variable node messages are updated immediately within each layer. This results in faster convergence and improved error-correction performance compared to the conventional flooding schedule. In this work, we adopt the layered BSP (BLSP) decoder as a benchmark for performance comparison with the quaternary decoding schemes.

\subsection{Quaternary sum-product algorithm}
It is important to note that, although $X$ and $Z$ errors are often treated as statistically independent in binary decoding frameworks, a Pauli error $Y$ corresponds to the simultaneous occurrence of both $X$ and $Z$ errors on the same qubit. This induces a form of correlation at the single-qubit level, distinct from spatial correlations across multiple qubits.

This observation naturally motivates the use of quaternary decoding. Quaternary decoders, operating over $\mathbb{F}_{4}$, inherently capture such joint error events by treating $Y$ as a single symbol. In contrast, binary decoders process $X$ and $Z$ components separately and typically assume independence between them, leading to a mismatch with the true error statistics. As a result, quaternary decoders can more accurately model the underlying noise and achieve improved decoding performance compared to their binary counterparts, particularly under depolarizing noise, as demonstrated in Subsection~\ref{subsec:sim_res}. For clarity and completeness, we have included the quaternary sum-product decoder below.

The quaternary sum-product decoder \cite{Quat_dec} operates on a joint Tanner graph 
constructed from the quaternary parity-check matrix $H_{\rho \times n}$, with $n$ 
variable nodes and $\rho$ check nodes, where each edge carries quaternary 
messages over $\{I, X, Z, Y\}$.

Before presenting the complete discussion of the quaternary algorithm, we first introduce the necessary terminology that will be used in its description.

The $(i,j)$-th entry of the quaternary parity-check matrix $H_{\rho \times n}$ is defined as
\begin{equation*}
      H_{i,j} = (H_x + 2H_z)_{i,j} \in \{0,1,2,3\}.
\end{equation*}
Next, we associate a Pauli operator with each entry $H_{i,j}$. Let $H_{i,j}^{(P)}$ denote the Pauli operator corresponding to $H_{i,j}$, defined as
\begin{equation*}
  H_{i,j}^{(P)} =
  \begin{cases}
      I, & \text{if } H_{i,j} = 0, \\
      X, & \text{if } H_{i,j} = 1, \\
      Z, & \text{if } H_{i,j} = 2, \\
      Y, & \text{if } H_{i,j} = 3.
  \end{cases}
\end{equation*}
Now, we are ready to describe the steps involved in quaternary sum-product algorithm.

\textbf{Initialization:} Each variable node $i$ is initialized with a 
quaternary LLR vector reflecting the depolarizing channel:
\begin{equation*}
    l^{(0)}_i = \left[0, \gamma, \gamma, \gamma\right], 
    \quad \gamma = \log\left(\frac{p_{d}}{3(1-p_{d})}\right),
\end{equation*}
where the entries correspond to Pauli operators $\{I, X, Z, Y\}$. The initial message from variable node $i$ to check node $j$, denoted by $M_{i,j}$, is updated as follows:
\begin{equation*}
    M_{i,j} = F_{\mathrm{max}}\!\left(0,\, \gamma\right) 
             - F_{\mathrm{max}}\!\left(\gamma,\, \gamma\right),
\end{equation*}
where $F_{\mathrm{max}}(a,b) = \max(a,b) + \log(1 + e^{-|a-b|})$ is the 
Jacobian logarithm for a numerically stable log-domain computation.

\textbf{Check node update:} To update each check node $j$ with respect to a variable node $i$, we aggregate the messages from all neighboring variable nodes excluding $i$, and update as follows:
\begin{equation*}
    \mu_{i,j} = 2(-1)^{S_j}\tanh^{-1}\!\left(
    \prod_{k \in \mathcal{N}^{(c)}_j \setminus \{i\}} 
    \tanh\!\left(\frac{M_{k,j}}{2}\right)\right),
\end{equation*}
where the syndrome bit $S_j$ flips the sign when the stabilizer is 
violated. The scalar $\mu_{i,j}$ is expanded into a quaternary 
vector, penalizing Pauli operators $P_{k}\in\{I, X, Z, Y\}$ that anticommute with the Pauli element $H_{i,j}^{(P)}$ corresponding to the $(i,j)$-th entry $H_{i,j}$ in the quaternary matrix, as follows:
\begin{equation*}
    L^{(t)}_{i,j} = \begin{cases}
        [0,0,0,0],   & \text{if } H^{(P)}_{i,j} = I,\\
        [0,0,-\mu_{i,j},-\mu_{i,j}],   & \text{if } H^{(P)}_{i,j} = X,\\
         [0,-\mu_{i,j}, -\mu_{i,j},0],
           & \text{if } H^{(P)}_{i,j} = Y,\\
        [0,-\mu_{i,j},0, -\mu_{i,j}],   & \text{if } H^{(P)}_{i,j} = Z.\\
    \end{cases}
\end{equation*}

\textbf{Variable node update and hard decision:} Each variable node $i$ 
updates its outgoing message using extrinsic information:
\begin{equation*}
    l^{(t)}_{i,j} = l^{(0)}_i + \sum_{k \in \mathcal{N}^{(v)}_i \setminus \{j\}} 
    L^{(t)}_{i,k}.
\end{equation*}
A hard decision then selects the most likely Pauli error $E_{i}$ at $i^{\text{th}}$ qubit as follows:
\begin{equation*}
    E_i = \underset{q\,\in\,\{I,X,Z,Y\}}{\mathrm{argmax}}
    \left(l^{(0)}_i + \sum_{k \in \mathcal{N}^{(v)}_i} L^{(t)}_{i,k}\right)(q),
\end{equation*}
which is mapped to its binary representation $\tilde{e}_i$ isomorphic 
to the corresponding Pauli operator.

\textbf{Convergence:} After each iteration, the recomputed syndrome 
$\tilde{S} = H\tilde{e}^{\mathrm{T}} \pmod{2}$, where $\tilde{e}=(\tilde{e}_{1},\tilde{e}_{2},\ldots, \tilde{e}_{n})$, is compared against $S$. 
Decoding succeeds when $\tilde{S} = S$; otherwise message passing 
continues until $w_{\mathrm{max}}$ is reached.

\begin{algorithm}
\caption{Quaternary Block-Layered Normalized Min-Sum Decoding (Part 1)}
\label{algo:block-layered-qnms}
\begin{algorithmic}[1]
    \Procedure{QBLNMS}{$H_{S},\, p_d,\, S,\, w_{\mathrm{max}},\, p,\, \alpha_s,\, \alpha_e,\, \beta_s,\, \beta_e$}
    \State \textbf{Input:} Syndrome vector $S$, stabilizer matrix $H_{S}$, maximum number of iterations $w_{\mathrm{max}}$, depolarizing probability $p_d$, permutation matrix size $p$, normalization parameters $\alpha_s, \alpha_e$, and damping parameters $\beta_s, \beta_e$.
    \State \textbf{Output:} Estimated error vector $\tilde{e}$.
    \State Let $H_{x}$ and $H_{z}$ be the parity-check matrices of the classical LDPC codes. Construct the quaternary parity-check matrix $H_{\rho\times n} = H_x + 2H_z$, where $H_{i,j}\in \{0,1,2,3\}$ and $1\leq i\leq \rho$ denotes a check node index and $1\leq j
    \leq n$ denotes a variable node index. Partition the check nodes into $N = \rho/p$ consecutive blocks $\mathcal{B}_1,\dots,\mathcal{B}_{N}$, each of size $p$.
    \State Define the function
    \(
    F_{\max}(s,t) = \max(s,t) + \log(1 + e^{-|s-t|})
    \)
    and the edge-label-dependent mapping $\phi_h(\boldsymbol{\xi})$ for $h = H_{i,j} \in \{1,2,3\}$:
    \vspace{-0.2cm}
    \begin{equation*}
    \phi_h(\boldsymbol{\xi}) =
    \begin{cases}
        F_{\max}(\xi_1,\xi_2) - F_{\max}(\xi_3,\xi_4), & \text{if }h=1, \\
        F_{\max}(\xi_1,\xi_3) - F_{\max}(\xi_2,\xi_4), & \text{if }h=2, \\
        F_{\max}(\xi_1,\xi_4) - F_{\max}(\xi_2,\xi_3), & \text{if }h=3.
    \end{cases}
    \end{equation*}
    \vspace{-0.2cm}
    \State Initialize the channel LLR vector $\mathbf{L}_\mathrm{ch} = [0, \gamma, \gamma, \gamma]$, where $\gamma = \log\!\bigl(p_{d} / [3(1-p_{d})]\bigr)$.
    \State Initialize the variable-to-check scalar messages $\nu_{i,j}$:
    \vspace{-0.2cm}
    \begin{equation*}
    \nu_{i,j} =
    \begin{cases}
        F_{\max}(0,\gamma) - F_{\max}(\gamma,\gamma), & \text{if } H_{i,j} \in \{1,2,3\}, \\
        +\infty, & \text{if } H_{i,j} = 0.
    \end{cases}
    \end{equation*}
    \vspace{-0.2cm}
    \State Initialize all check-to-variable messages (from $i^\text{th}$ check node to $j^\text{th}$ variable node) to zero: $\boldsymbol{\mu}_{i,j}^{(0,\cdot)} = \mathbf{0}_{1\times4} \text{ for all } i,j$.
    Initialize the extrinsic accumulator to zero:~$\boldsymbol{\Lambda}_j^{(0, 0)} = \mathbf{0}_{1\times4} \text{ for all } j$. For any variable, for example $\nu_{i,j}$, in the superscript tuple $(\cdot,\cdot)$, first coordinate denotes the iteration number and the second coordinate denotes the block layer; this convention is used consistently throughout the algorithm.
    \For{each iteration $w = 1$ to $w_{\mathrm{max}}$}
        \State Select a uniformly random permutation $\pi$ of $\{1,\dots,N\}$. Let the ordered block indices for this iteration be $b_1, b_2, \dots, b_{N}$ where $b_\ell = \pi(\ell)$ and $1\leq \ell\leq N$.
        \State Define the initial accumulator state for this iteration:
        \[
        \boldsymbol{\Lambda}_j^{(w,0)} =
        \begin{cases}
            \boldsymbol{\Lambda}_j^{(0,0)}, & \text{if } w=1, \\
            \boldsymbol{\Lambda}_j^{(w-1,N)}, & \text{if } w > 1.
        \end{cases}
        \]
        \For{each scheduled position $\ell = 1$ to $N$}
            \State Let $b \gets b_\ell$ \quad \Comment{actual block index at scheduling position $\ell$}
            \State Let $\mathcal{R} = \mathcal{B}_{b}$ be the current block-layer of $p$ check nodes.
            \State \textbf{Step 0: Syndrome-driven parameter adaptation}
           
            For each $i^{\text{th}}$ row in $\mathcal{R}$ and $j \in \mathcal{N}^{(v)}_i$, compute the message from $j^\text{th}$ variable node to $i^\text{th}$ check node:
            \vspace{-0.2cm}
            \begin{equation*}
            \nu_{i,j}^{(w,b)} = \phi_h\Bigl(\mathbf{L}_{\mathrm{ch}} + \boldsymbol{\Lambda}_j^{(w,\ell-1)} - \boldsymbol{\mu}_{i,j}^{(w-1,b)}\Bigr), \quad h = H_{i,j}.
            \end{equation*}
           
            Compute the number of unsatisfied checks $n_{\text{u}}$ as follows:
            \vspace{-0.2cm}
            \begin{equation*}
            n_{\text{u}} = \sum_{i \in \mathcal{R}} \mathbf{1}\!\left[(-1)^{S_i} \prod_{k \in \mathcal{N}^{(v)}_i} \operatorname{sign}(\nu_{i,k}^{(w,b)}) < 0 \right], \text{ where } \mathbf{1[\cdot]} \text{ is the indicator function.}
            \end{equation*}
            
            Let $\kappa = n_{\text{u}} / p \in [0,1]$. Set
            \(
            \alpha = \alpha_e + (\alpha_s - \alpha_e)\kappa, \quad
            \beta = \beta_e - (\beta_e - \beta_s)\kappa.
            \)            
            \algstore{qnmsbreak}
            \end{algorithmic}
\end{algorithm}
\begin{algorithm}
\ContinuedFloat
\caption{Quaternary Block-Layered Normalized Min-Sum Decoding (Part 2)}
\begin{algorithmic}[1]
\algrestore{qnmsbreak}
  \State \textbf{Step 1: Variable-to-check message computation}
           
            For each $i^{\text{th}}$ row in $\mathcal{R}$ and $j \in \mathcal{N}^{(v)}_{i}$ we compute:
            \vspace{-0.2cm}
            \begin{equation*}
            \nu_{i,j}^{(w,b)} = \phi_h\Bigl(\mathbf{L}_{\mathrm{ch}} + \boldsymbol{\Lambda}_j^{(w,\ell-1)} - \boldsymbol{\mu}_{i,j}^{(w-1,b)}\Bigr), \quad \text{where }h = H_{i,j}.
            \end{equation*}
            \State \textbf{Step 2: Check-to-variable message computation}
           
             For each $i^{\text{th}}$ row in $\mathcal{R}$ and $j \in \mathcal{N}^{(v)}_{i}$:
            \begin{itemize}
                \item Let $m_1 \leq m_2$ be the two smallest absolute values among $\{|\nu_{i,k}^{(w,b)}|\}_{k \in \mathcal{N}^{(v)}_{i}}$.
                \item Compute the scalar check-node output:
                \vspace{-0.2cm}
                \begin{equation*}
                x = \alpha \cdot (-1)^{S_i} \prod_{k \in \mathcal{N}^{(v)}_{i} \setminus \{j\}} \operatorname{sign}(\nu_{i,k}^{(w,b)}) \cdot
                \begin{cases}
                    1, & \text{if } m_1 = 0, \\
                    m_2, & \text{if } |\nu_{i,j}^{(w,b)}| = m_1, \\
                    m_1, & \text{otherwise}.
                \end{cases}
                \end{equation*}
                \vspace{-0.2cm}
                \item Compute the quaternary vector message $\boldsymbol{\mu}_{i,j}^{\text{(q)}}$:
                \vspace{-0.2cm}
                \begin{equation*}
                \boldsymbol{\mu}_{i,j}^{\text{(q)}} =
                \begin{cases}
                    [0,0,-x,-x], &\text{if } H_{i,j}=1, \\
                    [0,-x,0,-x], &\text{if } H_{i,j}=2, \\
                    [0,-x,-x,0], &\text{if } H_{i,j}=3.
                \end{cases}
                \end{equation*}
            \end{itemize}
            \State \textbf{Step 3: Damped update and accumulator commitment}
           
             For each $i^{\text{th}}$ row in $\mathcal{R}$ and $j \in \mathcal{N}^{(v)}_{i}$, we update:
            $\boldsymbol{\mu}_{i,j}^{(w,b)} = \beta \,\boldsymbol{\mu}_{i,j}^{\text{(q)}} + (1-\beta)\,\boldsymbol{\mu}_{i,j}^{(w-1,b)}.$
           
            Accumulate:
            $\Delta\boldsymbol{\Lambda}_j^{(w,b)} \leftarrow \Delta\boldsymbol{\Lambda}_j^{(w,b)} + \bigl(\boldsymbol{\mu}_{i,j}^{(w,b)} - \boldsymbol{\mu}_{i,j}^{(w-1,b)}\bigr).$
           
            Commit the updates:
            \(
            \boldsymbol{\Lambda}_j^{(w,\ell)} \leftarrow \boldsymbol{\Lambda}_j^{(w,\ell-1)} + \Delta\boldsymbol{\Lambda}_j^{(w,b)} \quad \text{ for all }\, j.
            \)
        \EndFor
        \State \textbf{Hard decision and syndrome check}
       
        For each variable node $j=1,\dots,n$, we evaluate:
        \vspace{-0.2cm}
        \begin{equation*}
            \tilde{e}(j) = \underset{q\,\in\,\{I,X,Z,Y\}}{\mathrm{argmax}}\;\bigl(\mathbf{L}_{\mathrm{ch}} + \boldsymbol{\Lambda}_j^{(w,N)}\bigr)(q).
        \end{equation*}
        Map the vector $\tilde{e}$ to binary vectors $(\tilde{e}_x, \tilde{e}_z)$ and compute the estimated syndrome $\tilde{S} = H_{S}\,[\tilde{e}_z \mid \tilde{e}_x]^\top \pmod{2}$ \footnotemark. 
    
        \If{$\tilde{S} = S$ or $w = w_{\mathrm{max}}$}
            \State \Return $\tilde{e} = [\tilde{e}_x \mid \tilde{e}_z]$
        \EndIf
    \EndFor
    \EndProcedure
\end{algorithmic}
\rule{\linewidth}{0.4pt}
\footnotesize{ $^{4}$ The reader should note that if the hard-decision outputs of a block layer show no sign changes relative to the preceding layers processed earlier in the same iteration, the decoder may exit early to save clock cycles; skipping this check risks a miscorrection.}
\end{algorithm}
%%%%%%%%%%%%%%%%%%%%%%%%%%%%%%%%%%%%%%%%%%%%%%%%%%%%%%%%%%%%%
\subsection{Quaternary min-sum algorithm with normalization, layering, and damping}
The quaternary min-sum algorithm is a low-complexity approximation to the sum-product algorithm, where the $\tanh$-based check node update is replaced by a min function operating solely on the magnitudes and signs of the incoming messages. Although this approximation introduces a slight performance degradation, it significantly reduces computational complexity by eliminating the $\tanh$ and $\tanh^{-1}$ operations. The variable node update and hard decision steps remain structurally identical to the sum-product algorithm, and the quaternary message structure encoding Pauli commutation and anti-commutation relations is preserved.

The min-sum approximation is known to overestimate the magnitude of check node messages, since the min operation is an upper bound on the true sum-product update. A normalization factor $\alpha \in (0,1]$ is applied to the outgoing check node message to compensate for this overestimation,
\begin{equation*}
    x \leftarrow \alpha \cdot x,
\end{equation*}
effectively scaling down the check node contributions before they are mapped to the quaternary message vector $\boldsymbol{\mu}_{i,j}$. Without this correction, inflated messages bias the posterior belief accumulator $\boldsymbol{\Lambda}_j$, degrading error-correction performance. In practice, $\alpha \in [0.6, 0.9]$ yields consistent gains over the unscaled min-sum decoder, and $\alpha$ is treated as a tunable hyperparameter optimized for the code and the channel under consideration.

Layered decoding updates one group of check nodes at a time and immediately propagates the updated messages to subsequent groups within the same iteration, unlike flooding schedules that wait for a full iteration. This typically accelerates the information flow across the Tanner graph, providing coding gains, reducing the number of iterations required for convergence, and thereby lowering the decoding latency. For QC-QLDPC codes, this advantage is reinforced by having a structured parity-check matrix, where each group corresponds to a block-row composed of circulant permutation matrices, and interconnections are related by cyclic shifts. This regularity enables efficient message routing using a barrel shifter, eliminating the need for complex interconnects. Consequently, the decoder can be realized using a compact and regular hardware architecture with reusable processing units, where only the shift values vary across groups. Such low-latency implementations with the needed parallelism are particularly important in quantum error correction, where decoding must be completed within the physical limits of qubit decoherence times.

We also make use of a stabilization mechanism in the layered setting via message damping. Since the updated beliefs are immediately visible to subsequent check nodes within the same iteration, any aggressive updates can cause oscillations, thereby hindering convergence due to the graphical structure of the code. Damping addresses this issue by blending the newly computed message with the existing one,
\begin{equation*}
    \boldsymbol{\mu}^{(w,b)}_{i,j} = \beta\,\boldsymbol{\mu}^{\mathrm{(q)}}_{i,j} + (1-\beta)\,\boldsymbol{\mu}^{(w-1,b)}_{i,j},
\end{equation*}
where $\beta \in (0,1]$ is a damping factor that controls how much the new message influences the update. The term $\boldsymbol{\mu}_{i,j}^{(w,b)}$ denotes the blended message at iteration $w$ for block-layer $b$, $\boldsymbol{\mu}_{i,j}^{(w-1,b)}$ is the message from the previous iteration, and $\boldsymbol{\mu}_{i,j}^{(\mathrm{q})}$ is the newly computed quaternary message vector. Setting $\beta = 1$ recovers the undamped decoder, while smaller values suppress oscillations at the cost of slower convergence. In practice, $\beta \in [0.7, 0.98]$ provides a good balance, and $\beta$ is treated as a tunable hyperparameter alongside $\alpha$. These practical strategies help in overcoming the convergence issues inherent to the nonlinear dynamics of the decoding algorithm.

\begin{figure}[t]
    \centering    \includegraphics[width=0.8\textwidth]{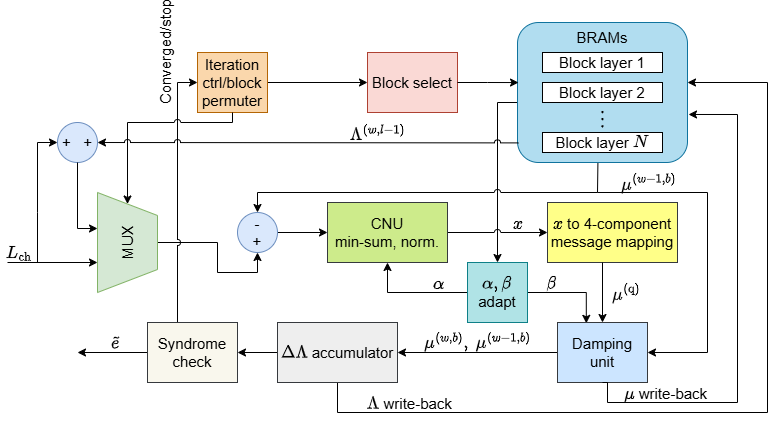}
    \caption{Figure shows the high-level hardware architecture of the QBLNMS decoder (Algorithm~\ref{algo:block-layered-qnms}). The BRAMs store the CN-to-VN messages $\boldsymbol{\mu}^{(w-1,b)}$ and the extrinsic accumulator $\boldsymbol{\Lambda}$, organized by block layer. At the start of each block, the block select unit reads the current block's messages from the BRAMs under the control of the iteration controller and block permuter. The adder combines the channel LLR $L_{\mathrm{ch}}$ with the block reference accumulator $\boldsymbol{\Lambda}^{(w,l-1)}$, and the subtractor removes the prior CN-to-VN contribution $\boldsymbol{\mu}^{(w-1,b)}$ to form the extrinsic belief, which is passed to the CNU. The CNU computes the normalized min-sum scalar $x$, using the normalization factor $\alpha$ supplied by the $\alpha$/$\beta$ adaptation unit, which determines $\alpha$ and $\beta$ based on the fraction of unsatisfied parity checks in the current block. The scalar $x$ is expanded into a four-component CN-to-VN message via the symplectic mapping, which is then blended with $\boldsymbol{\mu}^{(w-1,b)}$ in the damping unit using the factor $\beta$ to produce the damped message $\boldsymbol{\mu}$. The resulting message change is accumulated in the $\Delta\boldsymbol{\Lambda}$ accumulator, and upon completion of the block, the updated $\boldsymbol{\Lambda}$ and $\boldsymbol{\mu}$ are written back to the BRAMs. After each full iteration, the syndrome check unit forms the hard decision and verifies the estimated syndrome $\tilde{e}$, signaling the controller to stop upon convergence.}
    \label{fig:lay_dec_archi}
\end{figure}

In general, a layered normalized min-sum decoder relies on fixed scaling and damping parameters $\alpha$ and $\beta$ that are tuned once before decoding begins. This works well on an average but is blind to how the decoder is actually performing on a given syndrome instance, i.e., the same parameter values are applied regardless of whether the decoder is far from convergence or nearly there. The key idea of syndrome-driven adaptation is to use the decoder's own internal state as a feedback signal. Before processing each block of check nodes, we estimate $\kappa$ from the soft messages as the fraction of unsatisfied parity checks within that block. This serves as a measure of local convergence. Blocks with a high fraction of unsatisfied checks require more aggressive updates, while blocks that are close to convergence benefit from more conservative, stabilizing updates. Accordingly, the parameters $\alpha$ and $\beta$ are adapted as simple functions of $\kappa$. When $\kappa$ is large, $\alpha$ is increased (reducing normalization) and $\beta$ is decreased to allow stronger updates. When $\kappa$ is small, $\alpha$ is reduced and $\beta$ is increased to damp oscillations and stabilize the solution. The parameters are constrained within predefined ranges, with $\alpha \in [\alpha_s, \alpha_e]$ and $\beta \in [\beta_s, \beta_e]$, where $\alpha_s$ and $\alpha_e$ denote the maximum and minimum allowable values of the normalization factor, respectively, while $\beta_s$ and $\beta_e$ denote the minimum and maximum allowable values of the damping factor, respectively. This adaptation is performed independently for each block at every iteration, allowing different regions of the Tanner graph to evolve at different rates depending on their local state. As a result, the decoder naturally transitions from aggressive correction in poorly converged regions to fine-grained stabilization near convergence, with only minimal additional computational overhead.

Based on the above discussion, in our simulations, we use the quaternary block-layered normalized min-sum (QBLNMS) approximation of Algorithm~\ref{algo:block-layered-qnms}. The high-level architecture of the corresponding decoder is as shown in Figure~\ref{fig:lay_dec_archi}.

\subsection{Simulation Results and Discussions}
\label{subsec:sim_res}
In this section, we evaluate the performance of the constructed codes under various decoding schemes and identify the optimal choice as the quaternary block-layered normalized min-sum decoder. Furthermore, we compare the proposed codes with several prior works and observe significant performance gains.

\begin{figure}[ht]
    \centering
    % ---------- Subfigure 1 ----------
    \begin{subfigure}[h]{0.45\textwidth}
        \centering
        \includegraphics[width=\textwidth]{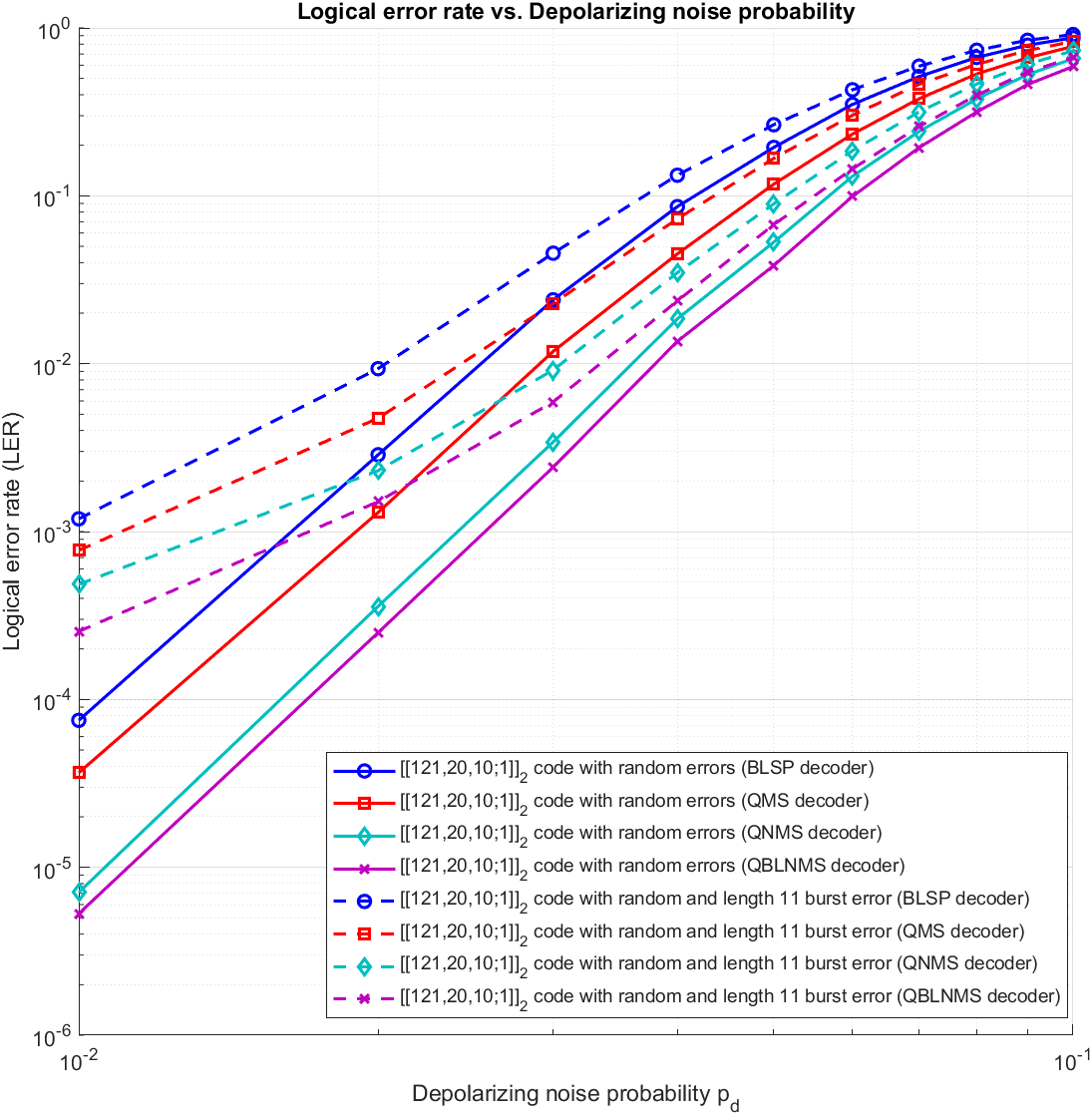}
        \caption{Performance of the $[[121,20,10;1]]_2$ code.}
        \label{fig:eleven_L_code}
    \end{subfigure}
    \hfill % adds horizontal space between figures
    % ---------- Subfigure 2 ----------
    \begin{subfigure}[h]{0.45\textwidth}
        \centering
    \includegraphics[width=\textwidth]{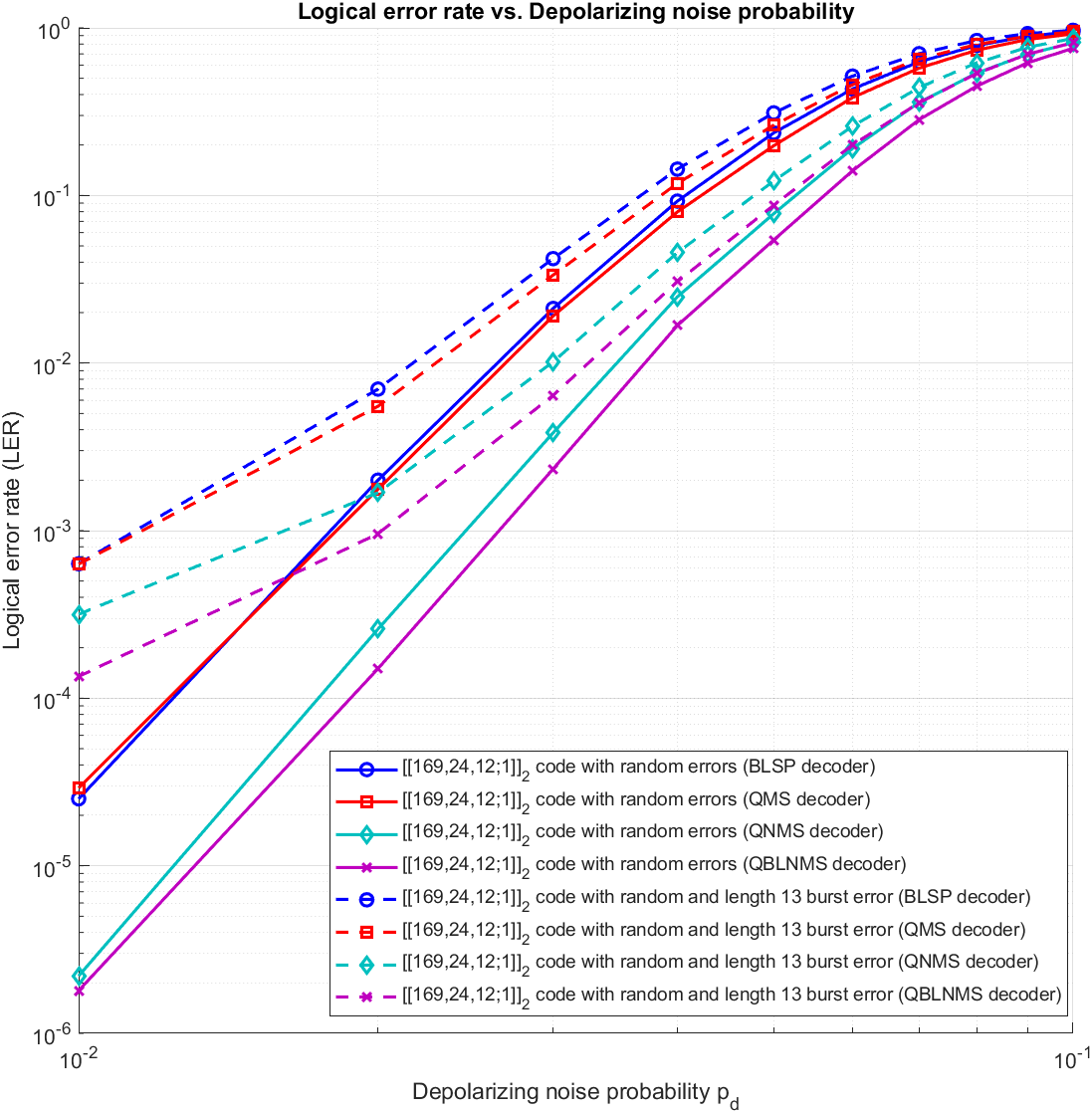}
        \caption{Performance of the $[[169,24,12;1]]_2$ code.}
        \label{fig:thirteen_L_code}
    \end{subfigure}
       \caption{Performance curves for the $[[121,20,10;1]]_2$ and $[[169,24,12;1]]_2$ EA-QC-QLDPC codes are shown under depolarizing noise and depolarizing noise with Markovian burst errors, where the logical error rate is plotted against the depolarizing probability $p_d$. For $p = 11$ case (Subfigure (a)), the quaternary min-sum (QMS) decoder outperforms the binary layered sum-product (BLSP) decoder. For $p = 13$ case (Subfigure (b)), both the decoders exhibit comparable performance, as the gain from layered decoding in BLSP compensates for the advantage of quaternary processing. The normalized variant (QNMS) provides further performance improvement. The best performance in both cases is achieved by the block-layered normalized decoder (QBLNMS). Additionally, Subfigures (a) and (b) illustrate the performance improvement obtained with increasing size of the underlying circulant permutation matrices from $p = 11$ to $p = 13$.}
        \label{fig:three graphs}
\end{figure}

%%%%%%%%%%%%%%%%%%%%%%%%%%%%%%%%%%%%%%%%%%%%%%%%%%%%%%%%%%%%%%%%%%%%%%%%%%%%%%%%%%%%%%%%%%%%%%%%%%%%%%%%%%%%%%%%%%%%%%%%%%%%%%%%%%%%%%%%%%%%%%%%%%%%%%%%%%%%%%%%%%%%%%%%%%%%%%%%%%%%%%%%%%%%%%%%%%%%%%%%%%%%%%%%%%%%%%%%%%%%%%%%%%%%%%%%%%%%%%%%%%%%%%%%%%%%%%%%%%%%%%%%

Before presenting the performance analysis of the proposed codes, we first introduce the following definition.
\begin{definition}[Logical Error Rate (LER)]
Let $m_E$ denote the total number of error instances. The logical error rate (LER) of a quantum code is defined as
\begin{equation*}
    \mathrm{LER} = \frac{n_E}{m_E},
\end{equation*}
where $n_E$ represents the number of instances resulting in a logical failure. An error instance is said to result in a logical failure if either of the following conditions holds:
\begin{enumerate}[leftmargin=*]
    \item $E_2^\dagger E_1 \in C(S)\setminus S$, where $E_1$ and $E_2$ are the actual and estimated error operators after decoding, respectively, $S$ is the stabilizer group, and  $C(S)$ is the centralizer of $S$.
    \item The decoder reaches the maximum number of allowed iterations without successful convergence.
\end{enumerate}
\end{definition}
For clarity, we present and discuss the simulation results based on the following points, using 100 iterations per simulation\footnote{MATLAB codes available at: https://github.com/Karthik-Bharadwaj-GM/QC-QLDPC-Decoders.git}.

\paragraph{Performance analysis of proposed codes under different decoding schemes:}
In Figures (\ref{fig:eleven_L_code}) and (\ref{fig:thirteen_L_code}), for correlation coefficient $\eta=0.5$, the logical error rate versus depolarizing probability $p_d$ is shown for BLSP, QMS, QNMS, and QBLNMS decoders under the random and Markovian error models. Monte Carlo simulations are used to evaluate the $[[121,20,10;1]]_2$ and $[[169,24,12;1]]_2$ EA-QC-QLDPC codes over both random depolarizing noise and burst errors. Binary decoding is performed on separate Tanner graphs, while quaternary decoding is carried out on the joint Tanner graph.

From Figures~(\ref{fig:eleven_L_code}) and~(\ref{fig:thirteen_L_code}) under random errors, the QMS decoder outperforms the BLSP decoder for the $p=11$ case, while for the $p=13$ case both decoders show comparable performance, indicating that the gain from layering in the binary decoder compensates for the advantage of quaternary processing; this is because as the code length increases, the ability to update variable nodes after every layer provides more immediate gains than the quaternary decoder, which must wait for a full iteration to update its variable nodes. The QNMS decoder provides improved performance over both BLSP and QMS in both cases, whereas the QBLNMS decoder achieves the best overall performance. In particular, QBLNMS offers an improvement of approximately one order of magnitude over BLSP for a maximum of $w_{\mathrm{max}}=100$ decoding iterations.

Under Markovian burst errors, the quaternary normalized decoders continue to outperform the binary decoder, as they more effectively capture the correlated nature of the channel. In this setting, the QBLNMS decoder achieves an improvement of slightly less than an order of magnitude over the BSP decoder.

Additionally, it is also observed that the performance improves with increasing permutation size $p$ (from $p=11$ to $p=13$), even when the burst length scales linearly with $p$ (refer to Figures~(\ref{fig:eleven_L_code}) and (\ref{fig:thirteen_L_code})).
%%%%%%%%%%%%%%%%%%%%%%%%%%%%%%%%%%%%%%%%%%%%%%%%%%%%%%%%%%
 \begin{figure}
    \centering
    \includegraphics[width=0.8\linewidth]{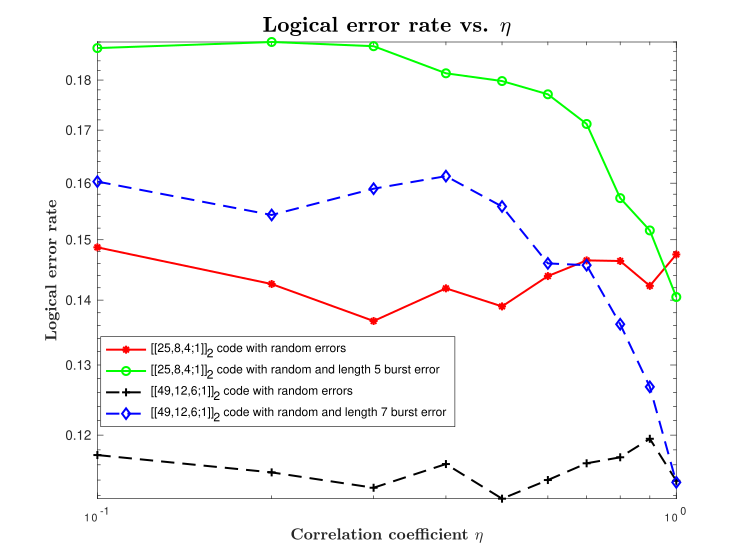}
    \caption{The LER vs. $\eta$ for $p_d=0.03$ is shown for the EA-QC-QLDPC codes $[[25,8,4;1]]_2$ and $[[49,12,6;1]]_2$. For both the codes, we assess the LER against (a) the random depolarizing model and (b) the random depolarizing along with the Markovian noise bursts. We observe the convergence of the LER for both the channels for these codes when the $\eta\rightarrow 1$ since the code can correct all bursts of length $p$.}
    \label{fig:Corr_Coeff}
\end{figure}
%%%%%%%%%%%%%%
\begin{figure}
    \centering
    \includegraphics[width=0.7\textwidth]{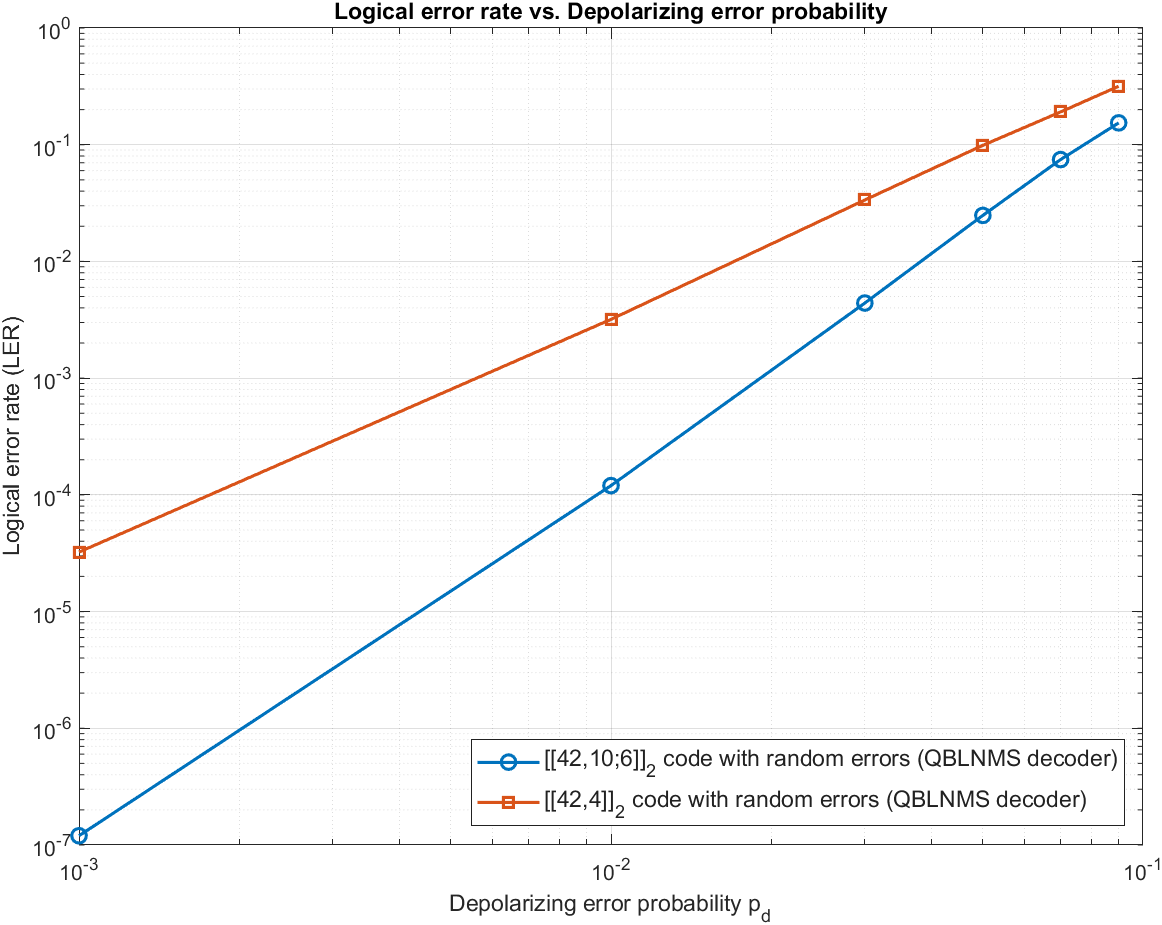}
    \caption{Performance comparison of the EA-QC-QLDPC code $[[42,10;6]]_2$ from Example~\ref{ex:hagiwara_comparison} with the $[[42,4]]_2$ CSS quantum code reported in~\cite{hagiwara2007quantum}. The decoder employed is the QBLNMS decoder, and the result clearly illustrates the superior performance of the proposed EA-CSS code, which can be attributed to the elimination of 4-cycles in the unassisted portion of the joint Tanner graph.}
    \label{fig:Comp_Hagiwara}
\end{figure}
%%%%%%%%%%%%%%%%%%%%%%%%%%%%%%%%%%%%%%%%%%%%%%%%%%%%%%%%%%%%%%%%%%
\begin{figure}
    \centering
    \includegraphics[width=0.7\textwidth]{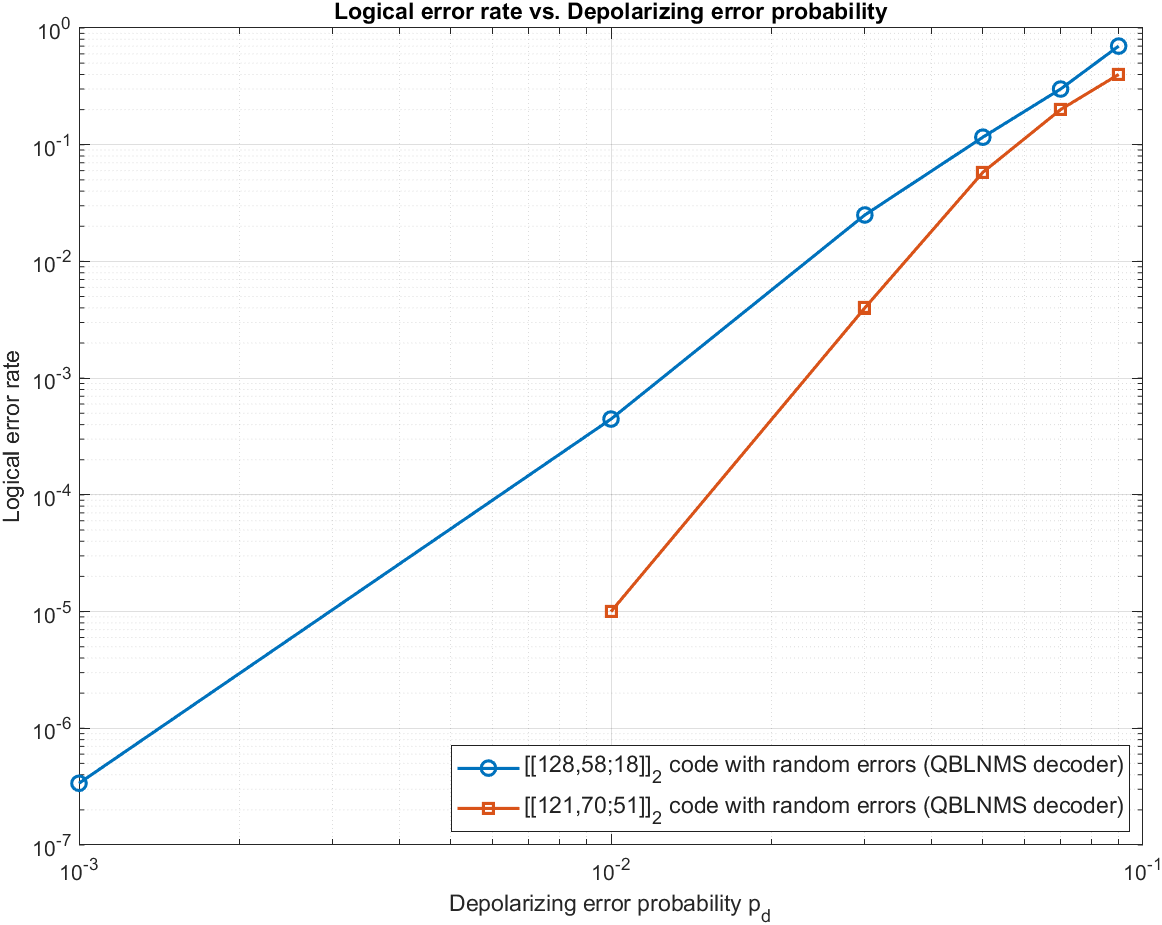}
    \caption{The proposed EA-QC-QLDPC code $[[121,70;51]]_2$ is compared with the $[[128,58;18]]_2$ EA quantum code from~\cite{hsieh2009entanglement} using the QBLNMS decoder. The proposed code achieves a higher rate of $0.578$ compared to $0.453$ for the reference code, and delivers better error correction performance despite having a slightly smaller block length.}
    \label{fig:comp_min_hsieh}
\end{figure}
%%%%%%%%%%%%%%%%%%%%%%%%%%%%%%%%%%%%%
\paragraph{Role of correlation coefficient $\eta$:} Figure \ref{fig:Corr_Coeff} shows the logical error rate vs. $\eta$. For simplicity, we have considered the two codes $[[25,8,4;1]]_2$ and $[[49,12,6;1]]_2$. From Figure~\ref{fig:Corr_Coeff}, we observe that the logical error rate of these codes becomes similar for the random errors case as well as for a mixture of random and burst errors over higher $\eta$. So, we could state that the codes are capable of correcting nearly all burst errors of length $p$ when $\eta \simeq 1$.
%%%%%%%%%%%%%%%%%%%%%%%%%%%%%%%%%%%%%%%%%%%%%%%%%%%%%%%%%%%%%%%%%%%%%%%%%%%%%%%%%%%%%%%%%%%%%%%%%%%%%%%%%%%%%%%%%%%%%%%%%%%%%%%%%%%%%%%%%%%%%%%%%%%%%%%%%%%%%%%%%%%%%%%%%%%%%%%%%%%%%%%%%%%%%%%%%%%%%%%%%%%%%%%%
%%%%%%%%%%%%%%%%%%%%%%%%%%%%%%%%%%%%%%%%%%%%%%%%%%%%%%%%%%%%%%%%%%%%%%%%%%%%%%%%%%%%%%%%%%%%%%%%%%%%%%%%%%%%%%%%%%%%%%%%%%%%%%%%%%%%%%%%%%%%%%
\paragraph{Effects of short cycles in joint Tanner graphs:} In Figure~\ref{fig:Comp_Hagiwara}, we compare the logical error rate performance of the $[[42,4]]_2$ code from~\cite{hagiwara2007quantum} with proposed code $[[42,10;6]]_2$ constructed using Theorem~\ref{theorem3.1}, under the QBLNMS decoder. At a noise probability of $10^{-3}$, our code achieves more than two orders of magnitude improvement in the logical error rate. This significant improvement is mainly due to the proposed code construction, which eliminates 4-cycles from the \textit{entanglement-unassisted portion of the joint Tanner graph} and thereby enhances the decoding performance compared to the reference code. It is also important to note that both codes have the same column weight; hence, the additional gain from layered decoding is limited in this case as the benefit of layering is closely related to the column weight of the parity-check matrix.

Additionally, in Figure~\ref{fig:Comp_Hagiwara}, it is important to note that although our EA quantum code $[[42,10;6]]_2$ requires $6$ pre-shared entangled qubits at the receiver side, compared to $[[42,4]]_2$, the decoding is performed only on the $42$ physical qubits transmitted by the sender. Therefore, both codes transmit the same number of qubits from the sender’s side; however, our code encodes significantly more logical qubits, resulting in a higher effective coding rate. Therefore, the performance gain reported in this figure is achieved alongside a higher code rate, making our code doubly advantageous.

\begin{figure}
    \centering
    \includegraphics[width=0.6\linewidth]{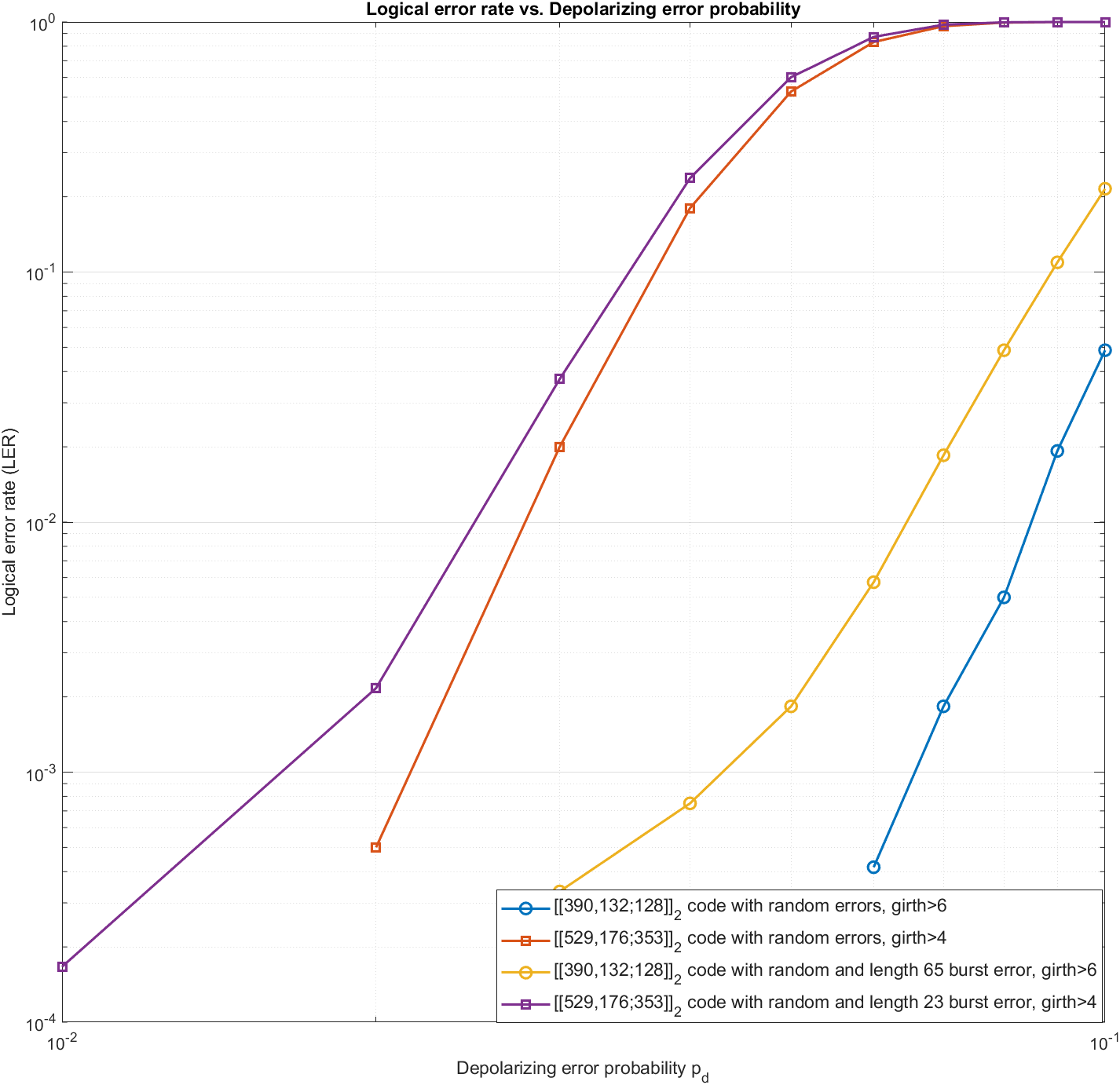}
    \caption{The $[[390,132;128]]_2$ EA-QC-QLDPC code with girth greater than 6 outperforms the $[[529,176;353]]_2$ EA-QC-QLDPC code with girth greater than 4 under the QBLNMS decoder over both the random depolarizing and Markovian burst error channels, despite having a smaller block length. This confirms that higher girth leads to better error correction performance. Additionally, the $[[390,132;128]]_2$ code has a burst error correction capability of 65, compared to only 23 for the $[[529,176;353]]_2$ code due to the size of the permutation matrices within those codes, making it significantly more robust against burst errors.}
    \label{fig:SIx_cycle}
\end{figure}
  \begin{figure}
    \centering
    \includegraphics[width=0.7\textwidth]{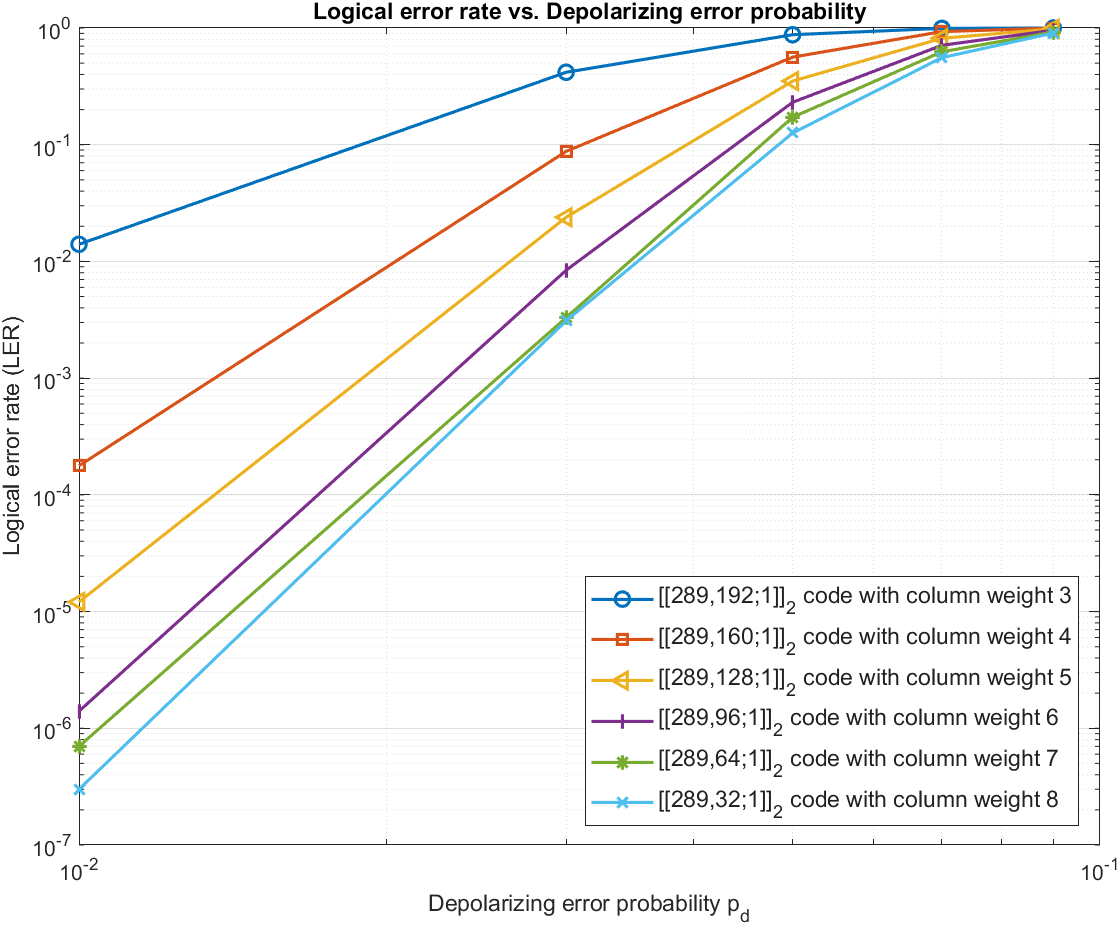}
    \caption{The plot shows the performance of  EA-QC-QLDPC codes constructed from Theorem \ref{theorem3} (for $p = 17$) under depolarizing noise for various column weights varying  from $3$ to $8$. The decoder employed is the QBLNMS decoder, and it is evident that increasing the column weights in the $H_x$ and $H_z$ parity-check matrices leads to enhanced error-correction performance of the quantum code.}
    \label{fig:weight_p_17}
\end{figure}
\paragraph{Impact of column weight:} In Figure~\ref{fig:comp_min_hsieh}, we compare the logical error rate performance of the $[[128,58;18]]_2$ code from~\cite{hsieh2009entanglement} with our $[[121,70;51]]_2$ EA-QC-QLDPC code constructed using Theorem~\ref{theorem4}. With the QMS decoder, our code already achieves approximately one order of magnitude improvement in logical error rate at a noise probability of~$10^{-3}$. The performance gap widens further when the QBLNMS decoder is used, owing to the higher column weight of our code (weight 5) compared to the reference code (weight 3), since layered decoding provides greater gains for higher column-weight codes. Specifically, with the QBLNMS decoder, our code achieves roughly one and a half orders of magnitude improvement at $p_d = 10^{-2}$. At $p_d = 10^{-3}$, we simulated $10^9$ depolarizing error instances for our code and observed zero logical errors, while the reference code registers a logical error rate of approximately $2\times 10^{-7}$ at the same noise level, confirming a coding gain of at least \textit{three orders} of magnitude.

Moreover, in Figure~\ref{fig:comp_min_hsieh}, it is worth emphasizing that although our EA quantum code $[[121,70;51]]_2$ requires a larger number of pre-shared entangled qubits at the receiver side compared to $[[128,58;18]]_2$, the decoding process operates solely on the $n$ physical qubits transmitted by the sender. In this comparison, our code transmits fewer qubits from the sender’s side while encoding a greater number of logical qubits, thereby achieving a higher effective coding rate. Consequently, the performance improvement observed in this figure is attained alongside a higher code rate, making our code \textit{doubly} advantageous.

\paragraph{Impact of girth on performance:}  Figure~\ref{fig:SIx_cycle} compares the logical error rate performance of the proposed code  $[[529,176;353]]_2$ which has girth greater than 4, with proposed code $[[390,132;128]]_2$ which has girth greater than~6. Despite having a smaller block length\footnote{Here, block length refers to the number of physical qubits transmitted through the noisy quantum channel, excluding the qubits pre-shared with the receiver.}, the $[[390,132;128]]_2$ code significantly outperforms the $[[529,176;353]]_2$ code under the QBLNMS decoder. This clearly demonstrates that increasing the girth of the Tanner graph leads to substantial improvements in error correction performance.
%%%%%%%%%%%%%%%%%%%%%%%%%%%%%%%%%%%%%%%%%%%%%%%%%%%%%%%%%%%%%%%%%%%%%%%%%%%%%%%%%%%%%%%%%%%%%%%%%%%%%%%%%%%%%%%%%%%%%%%%%%%%%%%%%%%%%%%%%%%%%%%%%%%%%%%%%%%%%%%%%%%%%%%%%%%%%%%%%%%%%%%%%%%%%%%%%%%%%%%%%%%%%%%%%%%%%%%
\paragraph{Effect of column weight on code performance:}   Figure \ref{fig:weight_p_17} illustrates the performance of the EA-QC-QLDPC code $[[289,k;1]]_2$ as the column weight is varied from 3 to~8, resulting in a reduction in the number of logical qubits from $k = 192$ to $k = 32$. The decoder employed is the QBLNMS decoder. Increasing the column weight leads to a higher variable node degree, which enables more effective use of layered scheduling, resulting in better error correction performance. Also, specific to the proposed code constructions, increasing the column weight improves the minimum distance and hence enhances error correction performance under depolarizing noise. This behavior reflects the fundamental trade-off between code rate and error correction capability, where higher column weights yield stronger codes at the expense of reduced code rate.

\section{Conclusions}\label{Sec.5}
We introduced several families of EA-QC-QLDPC codes derived by tiling permutation matrices of prime and composite orders, requiring only a minimum number of ebits, useful in entanglement-assisted quantum coded communications. When the construction is based on two distinct classical codes, we ensure that the unassisted portion of the joint Tanner graph is devoid of short cycles. % Further, while constructing entanglement-assisted quantum codes using a single classical code, the underlying classical code is devoid of short cycles as well. 
We present two other families of EA-QC-QLDPC codes, derived from a single classical QC-LDPC code, whose Tanner graph has a girth~>~6. We have further designed an efficient encoder for the EA-QC-QLDPC codes with computational complexity of $\mathcal{O}(n(\rho_{1}+\rho_{2})-\rho_1^2)$. The gate complexity can be possibly slightly improved using greedy algorithms based on circuit equivalences in the design pipeline.

We presented simulation results for various classes of EA-QC-QLDPC codes under the depolarizing channel with random errors and with Markovian burst errors along with random errors. The performance of the proposed  codes is evaluated using both binary and quaternary decoders on separate, as well as the joint Tanner graphs, with quaternary decoders showing performance improvements. We also proposed improved decoding algorithms — the quaternary min-sum and its normalized and layered variants (QNMS and QBLNMS), which deliver better decoding performance than the binary sum-product algorithm while being more hardware-friendly compared to the quaternary sum-product algorithm. The proposed EA-QC-QLDPC codes can correct a constant number of burst errors. Furthermore, several comparisons are made: (a) the proposed EA-QC-QLDPC code $[[42,10;6]]_2$ versus the code $[[42,4]]_2$ in \cite{hagiwara2007quantum}$,$ illustrating the superior performance of the EA-CSS code due to the elimination of 4-cycles in the unassisted portion of the joint Tanner graph; (b) the EA-QC-QLDPC code $[[121,70;51]]_2$ from Theorem~\ref{theorem4} versus the $[[128,58;18]]_2$ EA quantum code in~\cite{hsieh2009entanglement}; and (c) the EA-QC-QLDPC codes $[[390,132;128]]_2$ with girth greater than 6 and $[[529,176;353]]_2$ with girth greater than 4, highlighting the advantages of larger girth despite shorter code length. The proposed code constructions are scalable across various coding rates and data payloads, along with the desired parallelism for high-throughput circuits. With a complete characterization of several code families along with encoders and decoders, we hope that this work will serve as a blueprint for using such quantum codes in practice. 

Last, we would like to point out several structural advantages based on the geometry of EA-QC-QLDPC codes against other EA quantum codes derived from combinatorial means by minimizing the hull dimension of the constituent codes. The stabilizer generators for the EA-QC-QLDPC codes has predetermined weights along with known properties on the commutativity or anti-commutativity of the stabilizer generators, which is not the case for arbitrary EA quantum codes. Based on this, we can conveniently insert the EPR pairs during the code design and derive a systematic form of the encoding circuit, useful for the synthesis pipeline. 

\section*{Acknowledgment}
P. Kumar is supported by the Institute Postdoctoral Fellowship from Indian Institute of Science. A. Sharma and K. Bharadwaj are supported by PhD Fellowships from the Ministry of Education, Govt. of India. S. S. Garani is supported through the grant award CRG/2022/003469 from Anusandhan National Research Foundation, Govt. of India.
\bibliographystyle{quantum}
\bibliography{references}

@article{Quat_dec,
  author = {Ostrev, Dimiter and Orsucci, Davide and  Lázaro, Francisco and Matuz, Balazs},
  title = {Classical product code constructions for quantum Calderbank-Shor-Steane codes},
  journal = {Quantum},
  volume = {8},
  pages = {1420},
  year = {2024},
  doi={https://doi.org/10.22331/q-2024-07-22-1420}
}

@article{calderbank1997quantum,
  title={Quantum error correction and orthogonal geometry},
  author={Calderbank, A Robert and Rains, Eric M and Shor, Peter W and Sloane, Neil JA},
  journal={Phys. Rev. Lett.},
  volume={78},
  number={3},
  pages={405},
  year={1997},
  publisher={APS},
doi={https://doi.org/10.1103/PhysRevLett.78.405}
}

@ARTICLE{CKM_SGS,
  author={Matcha, Chaitanya Kumar and Roy, Shounak and Bahrami, Mohsen and Vasic, Bane and Srinivasa, Shayan Garani},
  journal={IEEE Trans. Magn.}, 
  title={{2-D LDPC Codes and Joint Detection and Decoding for Two-Dimensional Magnetic Recording}}, 
  year={2018},
  volume={54},
  number={2},
  pages={1-11},
doi={https://doi.org/10.1109/TMAG.2017.2735181}
}

@book{gottesman1997stabilizer,
  title={Stabilizer codes and quantum error correction},
  author={Gottesman, Daniel},
  year={1997},
  publisher={California Institute of Technology},
doi={https://arxiv.org/pdf/quant-ph/9705052}
}

@article{gallager1963low,
  title={Low-density parity-check codes},
  author={Gallager, Robert G},
  journal={Ph.D. dissertation, Cambridge University},
  year={1963},
doi={http://hdl.handle.net/1721.1/11804}
}

@article{fossorier2004quasicyclic,
  title={Quasicyclic low-density parity-check codes from circulant permutation matrices},
  author={Fossorier, Marc PC},
  journal={IEEE {T}rans. {I}nf. {T}heory},
  volume={50},
  number={8},
  pages={1788--1793},
  year={2004},
  publisher={IEEE},
doi={https://doi.org/10.1109/TIT.2004.831841}
}

@ARTICLE{Wireless,
  author={Richardson, Tom and Kudekar, Shrinivas},
  journal={IEEE Commun. Mag.}, 
  title={Design of Low-Density Parity Check Codes for 5{G} New Radio}, 
  year={2018},
  volume={56},
  number={3},
  pages={28-34},
  doi={https://doi.org/10.1109/MCOM.2018.1700839}}

@article{Shayan_chann,
  title={Channel Engineering in Magnetic Recording: From Theory to Practice},
  author={Shayan Srinivasa Garani and Bane Vasić},
  journal={IEEE BITS: The Information Theory Magazine~},
  volume={},
  number={},
  pages={1--36},
  year={2023},
  publisher={IEEE},
  doi={10.1109/MBITS.2023.3336213}
}

@article{mondal2021efficient,
  title={Efficient parallel decoding architecture for cluster erasure correcting {2-D LDPC} codes for {2-D} data storage},
  author={Mondal, Arijit and Garani, Shayan Srinivasa},
  journal={IEEE Trans. Magn.},
  volume={57},
  number={12},
  pages={1--16},
  year={2021},
  publisher={IEEE},
doi={https://doi.org/10.1109/TMAG.2021.3119723}
}

@article{ketkar2006nonbinary,
  title={Nonbinary stabilizer codes over finite fields},
  author={Ketkar, Avanti and Klappenecker, Andreas and Kumar, Santosh and Sarvepalli, Pradeep Kiran},
  journal={IEEE Trans.  Inf.  Theory},
  volume={52},
  number={11},
  pages={4892--4914},
  year={2006},
  publisher={IEEE},
doi={https://doi.org/10.1109/TIT.2006.883612}
}

@article{brun2006correcting,
  title={Correcting quantum errors with entanglement},
  author={Brun, Todd and Devetak, Igor and Hsieh, Min-Hsiu},
  journal={science},
  volume={314},
  number={5798},
  pages={436--439},
  year={2006},
  publisher={American Association for the Advancement of Science},
doi={https://doi.org/10.1126/science.1131563}
}

@article{chen2017entanglement,
  title={Entanglement-assisted quantum {MDS} codes constructed from negacyclic codes},
  author={Chen, Jianzhang and Huang, Yuanyuan and Feng, Chunhui and Chen, Riqing},
  journal={Quantum Inf. Process.},
  volume={16},
  pages={1--22},
  year={2017},
  publisher={Springer},
doi={https://doi.org/10.1007/s11128-017-1750-4}
}

@article{guenda2018constructions,
  title={Constructions of good entanglement-assisted quantum error correcting codes},
  author={Guenda, Kenza and Jitman, Somphong and Gulliver, T Aaron},
  journal={Des. Codes Cryptogr.},
  volume={86},
  pages={121--136},
  year={2018},
  publisher={Springer},
doi={https://doi.org/10.1007/s10623-017-0330-z}
}

@article{liu2018application,
  title={Application of constacyclic codes to entanglement-assisted quantum maximum distance separable codes},
  author={Liu, Yang and Li, Ruihu and Lv, Liangdong and Ma, Yuena},
  journal={Quantum Inf. Process.},
  volume={17},
  number={8},
  pages={210},
  year={2018},
  publisher={Springer},
doi={https://doi.org/10.1007/s11128-018-1978-7}
}

@article{qian2018mds,
  title={On {MDS} linear complementary dual codes and entanglement-assisted quantum codes},
  author={Qian, Jianfa and Zhang, Lina},
  journal={Des. Codes Cryptogr.},
  volume={86},
  pages={1565--1572},
  year={2018},
  publisher={Springer},
doi={https://doi.org/10.1007/s10623-017-0413-x}
}

@article{galindo2019entanglement,
  title={Entanglement-assisted quantum error-correcting codes over arbitrary finite fields},
  author={Galindo, Carlos and Hernando, Fernando and Matsumoto, Ryutaroh and Ruano, Diego},
  journal={Quantum Inf. Process.},
  volume={18},
  number={4},
  pages={116},
  year={2019},
  publisher={Springer},
doi={https://doi.org/10.1007/s11128-019-2234-5}
}

@article{luo2016non,
  title={Non-binary entanglement-assisted quantum stabilizer codes},
  author={Luo, Lan and Ma, Zhi and Wei, Zhengchao and Leng, Riguang},
  journal={Sci. China Inf. Sci},
  volume={60},
  number={4},
  year={2016},
doi={https://doi.org/10.1007/s11432-015-0932-y}
}

@article{nadkarni2021encoding,
  title={Encoding of nonbinary entanglement-unassisted and assisted stabilizer codes},
  author={Nadkarni, Priya J and Garani, Shayan Srinivasa},
  journal={IEEE Trans. Quantum Eng.},
  volume={2},
  pages={1--22},
  year={2021},
  publisher={IEEE},
doi={https://doi.org/10.1109/TQE.2021.3050848}
}

@article{nadkarni2021entanglement,
  title={{Entanglement-assisted Reed--Solomon codes over qudits: theory and architecture}},
  author={Nadkarni, Priya J and Garani, Shayan Srinivasa},
  journal={Quantum Inf. Process.},
  volume={20},
  pages={1--68},
  year={2021},
  publisher={Springer},
doi={https://doi.org/10.1007/s11128-021-03028-w}
}

@article{hsieh2011high,
  title={High performance entanglement-assisted quantum {LDPC} codes need little entanglement},
  author={Hsieh, Min-Hsiu and Yen, Wen-Tai and Hsu, Li-Yi},
  journal={IEEE Trans.  Inf.  Theory},
  volume={57},
  number={3},
  pages={1761--1769},
  year={2011},
  publisher={IEEE},
doi={https://doi.org/10.1109/TIT.2011.2104590}
}

@inproceedings{hagiwara2007quantum,
  title={Quantum quasi-cyclic {LDPC} codes},
  author={Hagiwara, Manabu and Imai, Hideki},
  booktitle={IEEE Int. Symp. Inf.},
  pages={806--810},
  year={2007},
doi={https://doi.org/10.1109/ISIT.2007.4557323}
}

@article{panteleev2021quantum,
  title={Quantum {LDPC} codes with almost linear minimum distance},
  author={Panteleev, Pavel and Kalachev, Gleb},
  journal={IEEE Trans.  Inf.  Theory},
  volume={68},
  number={1},
  pages={213--229},
  year={2021},
  publisher={IEEE},
doi={https://doi.org/10.1109/TIT.2021.3119384}
}

@inproceedings{miao2024joint,
  title={{A joint code and belief propagation decoder design for quantum LDPC codes}},
  author={Miao, Sisi and Mandelbaum, Jonathan and J{\"a}kel, Holger and Schmalen, Laurent},
  booktitle={Proc. IEEE Int. Symp. Inf. Theory (ISIT)},
  pages={2263--2268},
  year={2024},
doi={}
}

@INPROCEEDINGS{EA_ITW,
  author={Kumar, Pavan and Sharma, Abhi Kumar and Garani, Shayan Srinivasa},
  booktitle={Proc. IEEE Inf. Theory Workshop (ITW)}, 
  title={{Entanglement-Assisted Quasi-Cyclic LDPC Codes}}, 
  year={2024},
  volume={},
  number={},
  pages={205-210},
  doi={10.1109/ITW61385.2024.10806978}}

@article{hsieh2009entanglement,
  title={Entanglement-assisted quantum quasicyclic low-density parity-check codes},
  author={Hsieh, Min-Hsiu and Brun, Todd A and Devetak, Igor},
  journal={Phys. Rev. A},
  volume={79},
  number={3},
  pages={032340},
  year={2009},
  publisher={APS},
doi={https://doi.org/10.1103/PhysRevA.79.032340?_gl=1*1selfky*_gcl_au*NTY4NTkxMjc5LjE3MzM5MTg5NDU.*_ga*MTI3MTM4MjYwMS4xNzMzOTE4OTQ0*_ga_ZS5V2B2DR1*MTczNjYwNjExMS4xMi4xLjE3MzY2MDc3ODYuNjAuMC44Njc0OTA1MjY.}
}

@inproceedings{zhang2012new,
  title={New quasi-cyclic {LDPC} codes with girth at least eight based on Sidon sequences},
  author={Zhang, Guohua and Sun, Rong and Wang, Xinmei},
  booktitle={IEE Int. Symp. Turbo Codes Iterative Inf. Process. ISTC},
  pages={31--35},
  year={2012},
doi={https://doi.org/10.1109/ISTC.2012.6325193}
}

@article{wilde2008optimal,
  title={Optimal entanglement formulas for entanglement-assisted quantum coding},
  author={Wilde, Mark M and Brun, Todd A},
  journal={Phys. Rev. A},
  volume={77},
  number={6},
  pages={064302},
  year={2008},
  publisher={APS},
  doi={https://doi.org/10.1103/PhysRevA.77.064302}
}

@INPROCEEDINGS{QBECC,
  author={Trinca, Cibele Cristina and De Albuquerque, Clarice Dias and Junior, Reginaldo Palazzo and Interlando, J. Carmelo and De Andrade, Antonio Aparecido and Watanabe, Ricardo Augusto},
  booktitle={GLOBECOM}, 
  title={New {Quantum Burst-Error Correcting Codes from Interleaving Technique}}, 
  year={2022},
  volume={},
  number={},
  pages={5243-5248},
  doi={https://doi.org/10.1109/GLOBECOM48099.2022.10000761}}

@inproceedings{Burstconstruction,
  title={Construction and performance of quantum burst error correction codes for correlated errors},
  author={Fan, Jihao and Hsieh, Min-Hsiu and Chen, Hanwu and Chen, He and Li, Yonghui},
  booktitle={IEEE Int. Symp. Inf. Theory (ISIT)},
  pages={2336--2340},
  year={2018},
  doi={https://doi.org/10.1109/ISIT.2018.8437493}
}

@INPROCEEDINGS {Raveendran_2022,
author = {Raveendran, Nithin and Rengaswamy, Narayanan and Pradhan, Asit Kumar and Vasi{\'c}, Bane},
booktitle = {IEEE Int. Conf. Quantum Comput. Eng. (QCE)},
title = {{Soft Syndrome Decoding of Quantum {LDPC} Codes for Joint Correction of Data and Syndrome Errors}},
year = {2022},
volume = {},
issn = {},
pages = {275-281},
doi={https://doi.org/10.1109/QCE53715.2022.00047}
}

@article{calderbank1998quantum,
  title={Quantum error correction via codes over {GF (4)}},
  author={Calderbank, A Robert and Rains, Eric M and Shor, Peter M and Sloane, Neil JA},
  journal={IEEE Trans.  Inf.  Theory},
  volume={44},
  number={4},
  pages={1369--1387},
  year={1998},
  publisher={IEEE},
doi={https://doi.org/10.1109/18.681315}
}

@ARTICLE{Shayan_1,
  author={Garani, Shayan Srinivasa and Dolecek, Lara and Barry, John and Sala, Frederic and Vasić, Bane},
  journal={Proceedings of the IEEE}, 
  title={Signal Processing and Coding Techniques for 2-{D} Magnetic Recording: An Overview}, 
  year={2018},
  volume={106},
  number={2},
  pages={286-318},  
doi={https://doi.org/10.1109/JPROC.2018.2795961}
}

@article{terhal2015quantum,
  title={Quantum error correction for quantum memories},
  author={Terhal, Barbara M},
  journal={Rev. Mod. Phys.},
  volume={87},
  number={2},
  pages={307--346},
  year={2015},
  publisher={APS},
doi={https://doi.org/10.1103/RevModPhys.87.307?_gl=1*30vnzc*_gcl_au*NTY4NTkxMjc5LjE3MzM5MTg5NDU.*_ga*MTI3MTM4MjYwMS4xNzMzOTE4OTQ0*_ga_ZS5V2B2DR1*MTczNjYwNjExMS4xMi4xLjE3MzY2MDg2NDIuNjAuMC44Njc0OTA1MjY.}
}

@article{breuckmann2021quantum,
  title={Quantum low-density parity-check codes},
  author={Breuckmann, Nikolas P and Eberhardt, Jens Niklas},
  journal={PRX Quantum},
  volume={2},
  number={4},
  pages={040101},
  year={2021},
  publisher={APS},
doi={https://doi.org/10.1103/PRXQuantum.2.040101}
}

@article{lieb2022number,
  title={A Number Theoretic Approach to Cycles in {LDPC} Codes},
  author={Lieb, Julia and Tinani, Simran},
  journal={IFAC-Pap.},
  volume={55},
  number={30},
  pages={67--72},
  year={2022},
  publisher={Elsevier},
  doi={https://doi.org/10.1016/j.ifacol.2022.11.030}
}

@article{mackay2004sparse,
  title={Sparse-graph codes for quantum error correction},
  author={MacKay, David JC and Mitchison, Graeme and McFadden, Paul L},
  journal={IEEE Trans.  Inf.  Theory.},
  volume={50},
  number={10},
  pages={2315--2330},
  year={2004},
  publisher={IEEE},
doi={https://doi.org/10.1109/TIT.2004.834737}
}

@article{kitaev1997quantum,
  title={Quantum computations: algorithms and error correction},
  author={Kitaev, A Yu},
  journal={Russ. Math. Surv.},
  volume={52},
  number={6},
  pages={1191},
  year={1997},
  publisher={IOP Publishing},
doi={https://doi.org/10.1070/RM1997v052n06ABEH002155}
}

@article{calderbank1996good,
  title={Good quantum error-correcting codes exist},
  author={Calderbank, A Robert and Shor, Peter W},
  journal={Phys. Rev. A.},
  volume={54},
  number={2},
  pages={1098},
  year={1996},
  publisher={APS},
doi={https://doi.org/10.1103/PhysRevA.54.1098?_gl=1*1uiinvc*_gcl_au*NTY4NTkxMjc5LjE3MzM5MTg5NDU.*_ga*MTI3MTM4MjYwMS4xNzMzOTE4OTQ0*_ga_ZS5V2B2DR1*MTczNjYwNjExMS4xMi4xLjE3MzY2MDkyMTQuNjAuMC44Njc0OTA1MjY.}
}

@article{ashikhmin2001asymptotically,
  title={Asymptotically good quantum codes},
  author={Ashikhmin, Alexei and Litsyn, Simon and Tsfasman, Michael A},
  journal={Phys. Rev. A.},
  volume={63},
  number={3},
  pages={032311},
  year={2001},
  publisher={APS},
doi={https://doi.org/10.1103/PhysRevA.63.032311?_gl=1*1r2udg5*_gcl_au*NTY4NTkxMjc5LjE3MzM5MTg5NDU.*_ga*MTI3MTM4MjYwMS4xNzMzOTE4OTQ0*_ga_ZS5V2B2DR1*MTczNjYwNjExMS4xMi4xLjE3MzY2MDkyNDAuMzQuMC44Njc0OTA1MjY.}
}

@article{gallager1962low,
  title={Low-density parity-check codes},
  author={Gallager, Robert},
  journal={IEEE Trans.  Inf.  Theory.},
  volume={8},
  number={1},
  pages={21--28},
  year={1962},
  publisher={IEEE},
doi={https://doi.org/10.1109/TIT.1962.1057683}
}

@article{gottesman2013fault,
  author  = {Daniel Gottesman},
  title   = {Fault-tolerant Quantum Computation with Constant Overhead},
  journal = {Quantum Inf. Comput.},
  volume  = {14},
  number  = {15--16},
  pages   = {1338--1372},
  year    = {2014},
  doi     = {https://doi.org/10.48550/arXiv.1310.2984}
}

@article{knill1998resilient,
  title={Resilient quantum computation: error models and thresholds},
  author={Knill, Emanuel and Laflamme, Raymond and Zurek, Wojciech H},
  journal={Proc. R. Soc. A: Math. Phys. Eng. Sci.},
  volume={454},
  number={1969},
  pages={365--384},
  year={1998},
  publisher={The Royal Society},
doi={https://doi.org/10.1098/rspa.1998.0166}
}

@inproceedings{aharonov1997fault,
  title={Fault-tolerant quantum computation with constant error},
  author={Aharonov, Dorit and Ben-Or, Michael},
  booktitle={Proc. Annu. ACM Symp. Theory Comput.},
  pages={176--188},
  year={1997},
doi={https://dl.acm.org/doi/pdf/10.1145/258533.258579}
}

@book{richardson2008modern,
  title={Modern coding theory},
  author={Richardson, Tom and Urbanke, Ruediger},
  year={2008},
  publisher={Cambridge University Press},
doi={}
}

@inproceedings{raveendran2017stochastic,
  title={Stochastic resonance decoding for quantum {LDPC} codes},
  author={Raveendran, Nithin and Nadkarni, Priya J and Garani, Shayan Srinivasa and Vasi{\'c}, Bane},
  booktitle={IEEE Int. Conf. Commun. (ICC)},
  pages={1--6},
  year={2017},
doi={https://doi.org/10.1109/ICC.2017.7996747}
}

@article{wu2023high,
  title={{High-Performance QC-LDPC Code Co-Processing Approach and VLSI Architecture for Wi-Fi 6}},
  author={Wu, Yujun and Wu, Bin and Zhou, Xiaoping},
  journal={Electronics},
  volume={12},
  number={5},
  pages={1210},
  year={2023},
  publisher={MDPI},
doi={https://doi.org/10.3390/electronics12051210}
}

@article{chen2018performance,
  title={{Performance analysis of practical QC-LDPC codes: From DVB-S2 to ATSC 3.0}},
  author={Chen, Shuang and Peng, Kewu and Song, Jian and Zhang, Yushu},
  journal={IEEE Trans. Broadcast},
  volume={65},
  number={1},
  pages={172--178},
  year={2018},
  publisher={IEEE},
doi={https://doi.org/10.1109/TBC.2018.2881364}
}

@book{lidar2013quantum,
  author    = {Daniel A. Lidar and Todd A. Brun},
  title     = {Quantum Error Correction},
  publisher = {Cambridge University Press},
  year      = {2013},
doi={}
}

@article{suzuki2022quantum,
  title={{Quantum error mitigation as a universal error reduction technique: Applications from the NISQ to the fault-tolerant quantum computing eras}},
  author={Suzuki, Yasunari and Endo, Suguru and Fujii, Keisuke and Tokunaga, Yuuki},
  journal={PRX Quantum},
  volume={3},
  number={1},
  pages={010345},
  year={2022},
  publisher={APS},
doi={}
}

@article{zhao2022realization,
  title={Realization of an error-correcting surface code with superconducting qubits},
  author={Zhao, Youwei and Ye, Yangsen and Huang, He-Liang and Zhang, Yiming and Wu, Dachao and Guan, Huijie and Zhu, Qingling and Wei, Zuolin and He, Tan and Cao, Sirui and others},
  journal={Phys. Rev. Lett.},
  volume={129},
  number={3},
  pages={030501},
  year={2022},
  publisher={APS},
doi={}
}

@article{pecorari2025high,
  title={{High-rate quantum LDPC codes for long-range-connected neutral atom registers}},
  author={Pecorari, Laura and Jandura, Sven and Brennen, Gavin K and Pupillo, Guido},
  journal={Nature Communications},
  volume={16},
  number={1},
  pages={1111},
  year={2025},
  publisher={Nature Publishing Group UK London},
doi={}
}
%%%%%%%%%%%%%%%%%%%%%%%%%%%%%%%%%%%%%%%%%%%%%%%%%%%%%%%%%%%%%%%%%%%%%%%%%%%%%%%%%%%%%%%%%%%%%%%%%%%%%%%%%%%%%%%%%%%%%%%%%%%%%%%%%%%%%%%%%%%%%%%%%%%%%%%%%%%%%%%%%%%%%%%%%%%%%%%%%%%%%%%%%%%%%%%%%%%%%%%%%%%%%%%%%%%%%%%%%%%%%%%%%%%%%%%%%%%%%%%%%%%%%%%%%%%%%%%%%%%%%%%%%%%%%%%%%%%%%%%%%%%%%%%%%%%%%%%%%%%%%%%%%%%%%%%%%%%%%%%%%%%%%%%%%%%%%%%%%%%%%%%%%%%%%%%%%%%%%%%%%%%%%%%%%%%%%%%%%%%%%%%%%%%%%%%%%%%%%%%%%%%%%%%%%%%%%%%%%%%%%%%%%%%%%%%%%%%%%%%%%%%%%%%%%%%%%%%%%%%%%%%%%%%%%%%%%%%%%%%%%%%%%%%%%%%%%%%%%%%%%%%%%%%%%%%%%%%%%%%%%%%%%%%%%%%%%%%%%%%%%%%%%%%%%%%%%%%%%%%%%%%%%%%%%%%%%%%%%%%%%%%%%%%%%%%%%%%%%%%%%%%%%%%%%%%%%%%%%%%%%%%%%%%%%%%%%%%%%%%%%%%%%%%%%%%%%%%%%%%%%%%%%%%%%%%%%%%%%%%%%%%%%%%%%%%%%%%%%%%%%%%%%%%%%%%%%%%%%%%%%%%%%%%%%%%%%%%%%%%%%%%%%%%%%%%%%%%%%%%%%%%%%%%%%%%%%%%%%%%%%%%%%%%%%%%%%%%%%%%%%%%%%%%%%%%%%%%%%%%%%%%%%%%%%%%%%%%%%%%%%%%%%%%%%%%%%%%%%%%%%%%%%%%%%%%%%%%%%%%%%%%%%%%%%%%%%%%%%%%%%%%%%%%%%%%%%%%%%%%%%%%%%%%%%%%%%%%%%%%%%%%%%%%%%%%%%%%%%%%%%%%%%%%%%%%%%%%%%%
\newpage
\appendix
\section{The Proof of Theorem \ref{2k-cycletheorem}}\label{2k-cycletheoremproof}
\textbf{Statement:} A cycle of length $2k$ exists in the Tanner graph of a QC-LDPC code with parity-check matrix $H$, defined in \eqref{modelmatrix}, if and only if there exists a closed path $(i_0, j_0), (i_0, j_1), (i_1, j_1),$ $ (i_1, j_2), \ldots, (i_{k-1}, j_{k-1}), (i_{k-1}, j_0)$ in $H$ such that
\begin{equation*}
P^{a_{i_{0},j_{0}}} (P^{a_{i_{0},j_{1}}})^{-1} P^{a_{i_{1},j_{1}}} (P^{a_{i_{1},j_{2}}})^{-1} \cdots P^{a_{i_{k-1},j_{k-1}}} (P^{a_{i_{k-1},j_{0}}})^{-1}=I~(\text{identity matrix}).
\end{equation*}

\begin{proof} In \cite{fossorier2004quasicyclic}, the proof of the theorem is omitted. Consequently, it is imperative to present the complete proof. Therefore, we shall provide a detailed proof, utilizing certain hints provided in \cite{lieb2022number}.

For any $0\leq s\leq n-1$, define a right circulant permutation $R^{s}: \mathbb{Z}_{n}\to \mathbb{Z}_{n}$ by $R^{s}(i)= (i+s) \pmod{n}.$
Let $a_{i,j}\in\mathbb{Z}_{n}$, then the right circulant matrix $P^{(a_{i,j})}$ can be defined by a right circulant permutation $R^{a_{i,j}}: \mathbb{Z}_{n}\rightarrow\mathbb{Z}_{n}$ as follows; $P^{(a_{i,j})}(u,v)=1$ if and only if $v=R^{a_{i,j}}(u)=u+a_{i,j}\pmod{n}$, where $(u,v)$ stands for $(u,v)$-th entry in the right circulant matrix $P^{(a_{i,j})}$. For two positive integers $m,t\geq1$, we can define the parity-check matrix $H$ in terms of right circulant matrices and permutations, respectively as follows:
\begin{equation*}
H=\begin{bmatrix}	P^{a_{0,0}}&P^{a_{0,1}}&\cdots&P^{a_{0,t}}\\
P^{a_{1,0}}&P^{a_{1,1}}&\cdots&P^{a_{1,t}}\\
\vdots&\vdots&\ddots&\vdots\\	P^{a_{m,0}}&P^{a_{m,1}}&\cdots&P^{a_{m,t}}
\end{bmatrix},\text{ or }H=\begin{bmatrix}	R^{a_{0,0}}&R^{a_{0,1}}&\cdots&R^{a_{0,t}}\\	R^{a_{1,0}}&R^{a_{1,1}}&\cdots&R^{a_{1,t}}\\
\vdots&\vdots&\ddots&\vdots\\	R^{a_{m,0}}&R^{a_{m,1}}&\cdots&R^{a_{m,t}}
\end{bmatrix}.
\end{equation*}
 % For each pair $(i,j)$, where $0\leq i\leq m$ and $0\leq j\leq t$, we can write the entries of $P^{a_{i,j}}$ as $h_{in+i', jn+j'}$,  $0\leq i',j'\leq n-1$. Then $h_{in+i',  jn+j'}=1$ $\iff$ $j'=R^{a_{i,j}}(i')$.
  Consider the closed path of indices as follows:
	\begin{equation*}
	(i_{0},j_{0}),(i_{0},j_{1});(i_{1},j_{1}),(i_{1},j_{2});\ldots,(i_{k-2},j_{k-2}),(i_{k-2},j_{k-1});(i_{k-1},j_{k-1}),(i_{k-1},j_{0}),
	\end{equation*}
    where $i_\ell \neq i_{\ell+1}, j_\ell \neq j_{\ell+1}$, for $\ell = 0, 1, 2,\ldots, k-2$, and  $i_{k-1} \neq i_0, j_{k-1} \neq j_0 $.
To check the existence of cycle of $2k$-length in the matrix $H$ corresponding to the indices $(i_{0},j_{0}),$ $(i_{0},j_{1});$ $(i_{1},j_{1}),$ $(i_{1},j_{2})$ $;\ldots;(i_{k-2},j_{k-2}),$ $(i_{k-2},j_{k-1});$ $(i_{k-1},j_{k-1}),$ $(i_{k-1},j_{0})$, we need the following steps:

\textbf{Step 1:} Choose any nonzero entry at some position $(r_{0},c_{0})$, where $0\leq r_{0},c_{0}\leq n-1$, in the circulant matrix $P^{(a_{i_{0},j_{0}})}$ and join it horizontally to a nonzero entry at some position $(r_{0},c_{1})$ in the circulant matrix $P^{a_{i_{0},j_{1}}}$. By definition of $R^{a_{i,j}}$ associated with  $P^{a_{i,j}}$, we have 
\begin{align*}
    P^{a_{i_{0},j_{0}}}(r_{0},c_{0}) = P^{a_{i_{0},j_{1}}}(r_{0},c_{1}) = 1&
     \iff c_{0}=R^{a_{i_{0},j_{0}}}(r_{0}) \text{ and }c_{1}=R^{a_{i_{0},j_{1}}}(r_{0})\\
   &\iff   c_{1} = R^{a_{i_{0},j_{1}}}(R^{a_{i_{0},j_{0}}})^{-1}(c_{0}).
\end{align*}   
    \textbf{Step 2:} Join, horizontally, the nonzero entry at some position $(r_{1},c_{1})$  in the circulant matrix $P^{a_{i_{1},j_{1}}}$ to a nonzero entry at some position $(r_{1},c_{2})$ in the circulant matrix $P^{a_{i_{1},j_{2}}}$. Then, by definition of $R^{a_{i,j}}$ associated with  $P^{(a_{i,j})}$, we acquire
    \begin{equation*}
        P^{a_{i_{1},j_{1}}}(r_{1},c_{1}) = P^{a_{i_{1},j_{2}}}(r_{1},c_{2}) = 1 \iff c_{2} = R^{a_{i_{1},j_{2}}}(R^{a_{i_{1},j_{1}}})^{-1}(c_{1}).
    \end{equation*}
    To avoid redundancy, we, now, consider the final step.
    
    \textbf{Step k:} Connect, horizontally, the nonzero entry at some position $(r_{k-1}, c_{k-1})$ in the circulant matrix $P^{a_{i_{k-1}, j_{k-1}}}$ to a nonzero entry at some position $(r_{k-1}, c_{k})$ in the circulant matrix $P^{a_{i_{k-1}, j_{0}}}$.  By definition of the permutation $P^{a_{i,j}}$ associated with $R^{(a_{i,j})}$, we obtain 
\begin{equation*}
    P^{a_{i_{k-1},j_{k-1}}}(r_{k-1},c_{k-1})= P^{a_{i_{k-1},j_{0}}}(r_{k-1},c_{k}) = 1 \iff c_{k} = R^{a_{i_{k-1},j_{0}}}(R^{a_{i_{k-1},j_{k-1}}})^{-1}(c_{k-1}).
\end{equation*}
From the above steps, one can easily conclude that
\begin{equation}\label{comp.mapp.}
	c_{k}=R^{a_{i_{k-1},j_{0}}}(R^{a_{i_{k-1},j_{k-1}}})^{-1}\cdots R^{a_{i_{1},j_{2}}}(R^{a_{i_{1},j_{1}}})^{-1} R^{a_{i_{0},j_{1}}}(R^{a_{i_{0},j_{0}}})^{-1}(c_{0}).
\end{equation}
Since, each $R^{a_{i,j}}$, where $0\leq i\leq m$ and $0\leq j\leq t$, corresponds to some exponent of right circulant permutation matrix $P$, their composition 
\begin{equation*}
  R^{a_{i_{k-1},j_{0}}}(R^{a_{i_{k-1},j_{k-1}}})^{-1}\cdots R^{a_{i_{1},j_{2}}}(R^{a_{i_{1},j_{1}}})^{-1} R^{a_{i_{0},j_{1}}}(R^{a_{i_{0},j_{0}}})^{-1}(c_{0})
\end{equation*} as mappings corresponds to some power of right circulant permutation matrix $P$. Consequently, from equation \eqref{comp.mapp.}, a cycle of length $2k$ will exist in the Tanner graph of the parity-check matrix $H$ if and only if $c_{k} = c_{0}$, which is only possible under the following condition:
\begin{equation}\label{eq.cycle.comp}
  R^{a_{i_{k-1},j_{0}}}(R^{a_{i_{k-1},j_{k-1}}})^{-1}\cdots R^{a_{i_{1},j_{2}}}(R^{a_{i_{1},j_{1}}})^{-1} R^{a_{i_{0},j_{1}}}(R^{a_{i_{0},j_{0}}})^{-1}(c_{0})=id_{\mathbb{Z}_{n}},
\end{equation}
where $id_{\mathbb{Z}_{n}}$ is the identity map on $\mathbb{Z}_{n}$. The Eq~\eqref{eq.cycle.comp} holds if and only if
\begin{align*}
P^{a_{i_{k-1},j_{0}}}(P^{a_{i_{k-1},j_{k-1}}})^{-1}\cdots P^{a_{i_{1},j_{2}}}(P^{a_{i_{1},j_{1}}})^{-1} P^{a_{i_{0},j_{1}}}(P^{a_{i_{0},j_{0}}})^{-1}&=I~(\text{identity matrix})\\
\iff P^{a_{i_{0},j_{0}}} (P^{a_{i_{0},j_{1}}})^{-1} P^{a_{i_{1},j_{1}}} (P^{a_{i_{1},j_{2}}})^{-1} \cdots P^{a_{i_{k-1},j_{k-1}}} (P^{a_{i_{k-1},j_{0}}})^{-1}&=I~(\text{identity matrix}),
\end{align*}
since for any invertible matrix $A$, $A=I$ if and only if $A^{-1}=I$.
Hence the result established.
\end{proof}
%%%%%%%%%%%%%%%%%%%%%%%%%%%%%%%%%%%%%%%%%%%%%%%%%%%%%%%%%%%%%%%%%%%%%%%%%%%%%%%%%%%%%%%%%%%%%%%%%%%%%%%%%%%%%%%%%%%%%%%%%%%%%%%%%%%%%%%%%%%%%%%%%%%%%%%%%%%%%%%%%%%%%%%%%%%%%%%%%%%%%%%%%%%%%%%%%%%%%%%%%%%%%%%%%%%%%%%%%%%%%%%%%%%%%%%%%%%%%%%%%%%%%%%%%%%%%%%%%%%%%%%%%%%%%%%%%%%%%%%%%%%%%%%%%%%%%%%%%%%%%%%%%%%%%%%%%%%%%%%%%%%%%%%%%%%%%%%%%%%%%%%%%%%%%%%%%%%%%%%%%%%%%%%%%%%%%%%%%%%%%%%%%%%%%%%%%%%%%%%%%%%%%%%%%%%%%%%%%%%%%%%%%%%%%%%%%%%%%%%%%%%%%%%%%%%%%%%%%%%%%%%%%%%%%%%%%%%%%%%%%%
\section{The Proof of Theorem \ref{columnweight-4-even}}
\textbf{Statement: }\label{appendix.a}
	Let $S=\{a_{1},a_{2},\ldots,a_{t}\}$, where $2\leq a_{1}<a_{2}\cdots<a_{t}$ and for any $p<q<r$, $a_{q}-a_{p}=a_{r}-a_{q}=1$ does not hold. Let $P$ be a right circulant permutation matrix of order $(w^{a_{t}+1}-1)$, where\footnote{The parameter \(a_t\) is selected so that the catalytic rate remains positive.} $a_{t}\geq 12$ is an integer and $w\geq2$ is an even integer, and let
	\begin{equation}\label{apxmodel}
		M=\begin{bmatrix}
		   0 & 0&\cdots&0\\
			w^{a_{1}} & w^{a_{2}}&\cdots&w^{a_{t}}\\   
			w^{a_{1}-1} & w^{a_{2}-1}&\cdots&w^{a_{t}-1}\\   
			w^{a_{1}-2} & w^{a_{2}-2}&\cdots&w^{a_{t}-2} 
		\end{bmatrix}
	\end{equation} be a degree matrix associated with the parity-check matrix $H$. Then there exists an EA quantum QC-LDPC code with parameters $[[n,2k-n+c;c]]_2$, where $n=(w^{a_{t}+1}-1)t$, $k\geq (w^{a_{t}+1}-1)(t-4)+3$ and $c\leq 4(w^{a_{t}+1}-1)-3$. In addition, the Tanner graph of the parity-check matrix $H$ is of girth greater than 6.

\begin{proof} The length of the code is obvious to determine, and the lower bound on the dimension $k$ can easily be obtained using Theorem \ref{theorem2}. We will show that the Tanner graph of $H$ is free from 4-cycles. Verifying that each nonzero row contains different elements from $\mathbb{Z}_{(w^{a_{t}+1}-1)}$ is easy. Therefore, by Proposition \ref{4-cyclepropo}, the presence of a 4-cycle is not possible for one zero row and one nonzero row. Consider any two non-zero rows from the matrix $M$.  Then there exists a 4-cycle in the Tanner graph of $H$ if and only if there exist $p,q\in\{1,2,\ldots,t\}$, where $p<q$,  such that
\begin{equation}\label{eq.4-cycle-weight-4}
      w^{a_{p}-\ell_{1}}-w^{a_{p}-\ell_{2}}=w^{a_{q}-\ell_{1}}-w^{a_{q}-\ell_{2}}\quad\textrm{mod }(w^{a_{t}+1}-1),
\end{equation}
where $\ell_{1}\in\{0,1\},\ell_{2}\in\{1,2\}$ and  $\ell_{1}<\ell_{2}$. Equation~\eqref{eq.4-cycle-weight-4} holds if and only if
	\begin{align*} 
		(w^{(\ell_{2}-\ell_{1})}-1)w^{a_{p}-\ell_{2}}&=(w^{(\ell_{2}-\ell_{1})}-1)w^{a_{q}-\ell_{2}}\quad\textrm{mod }(w^{a_{t}+1}-1)\\
		\iff  (w^{(\ell_{2}-\ell_{1})}-1)&=(w^{(\ell_{2}-\ell_{1})}-1)w^{(a_{q}-a_{p})}\quad(\textrm{since }\textrm{gcd}(w^{a_{p}-\ell_{2}},w^{a_{t}+1}-1)=1),
	\end{align*}
which leads to a contradiction since the number on the left is odd whereas the number on the right is even.

Next, we will prove that the graphical representation of $H$ does not have cycles of length 6. We divide the proof into two cases. In the first case, we have one zero row and two nonzero rows, and in the second case, we have all nonzero rows.

Consider the following three rows and three  arbitrary columns from the matrix $M$ defined in \eqref{apxmodel}:
	\begin{equation}\label{1-portion_of_H}
		\left[\begin{array}{ccccccc}
			\cdots&0&\cdots & 0& \cdots&0&\cdots\\
			\cdots&w^{a_{p}-\ell_{1}}&\cdots& w^{a_{q}-\ell_{1}}&\cdots&w^{a_{r}-\ell_{1}}&\cdots\\   
			\cdots&w^{a_{p}-\ell_{2}}&\cdots& w^{a_{q}-\ell_{2}}&\cdots&w^{a_{r}-\ell_{2}}&\cdots\\ 
		\end{array}\right],
	\end{equation}
	where $a_{1}\leq a_{p}<a_{q}<a_{r}\leq a_{t}$ and $0\leq\ell_{1}<\ell_{2}\leq 2$. We investigate the existence of all possible 6-cycles, as shown in Figure \ref{fig:all-6-cycles}, within the segment of the Tanner graph that corresponds to a portion of the degree matrix defined by Eq.~\eqref{1-portion_of_H}.
	\begin{enumerate}[leftmargin=0pt,itemsep=0pt]
		\item []\textbf{Type-I 6-cycle} exists if and only if the following holds:
		\begin{align*}
			w^{a_{p}-\ell_{2}}&= w^{a_{r}-\ell_{2}}-w^{a_{r}-\ell_{1}}+w^{a_{q}-\ell_{1}}\quad\textrm{mod }(w^{a_{t}+1}-1)\\
			\implies w^{a_{r}-\ell_{1}}-w^{a_{q}-\ell_{1}}&= w^{a_{r}-\ell_{2}}-w^{a_{p}-\ell_{2}}\quad\textrm{mod }(w^{a_{t}+1}-1)\\
			\implies w^{a_{q}-\ell_{1}}(w^{a_{r}-a_{q}}-1)&= w^{a_{p}-\ell_{2}}(w^{a_{r}-a_{p}}-1)\quad\textrm{mod }(w^{a_{t}+1}-1)\\
			\implies w^{a_{q}-a_{p}+(\ell_{2}-\ell_{1})}(w^{a_{r}-a_{q}}-1)&= (w^{a_{r}-a_{p}}-1)\quad\textrm{mod }(w^{a_{t}+1}-1),
		\end{align*} since gcd$(w^{a_{p}-\ell_{2}},w^{a_{t}+1}-1)=1$. This is a contradiction, as the number on the left is even, while the number on the right is odd.
		\item[] \textbf{Type-II 6-cycle} exists if and only if 
    $w^{a_{p}-\ell_{2}} = w^{a_{q}-\ell_{2}} - w^{a_{q}-\ell_{1}} + w^{a_{r}-\ell_{1}} \textrm{ mod }(w^{a_{t}+1} - 1)$ if and only if   $w^{a_{p}-\ell_{2}} + w^{a_{q}-\ell_{1}} = w^{a_{q}-\ell_{2}} + w^{a_{r}-\ell_{1}}$ $  \textrm{ mod }(w^{a_{t}+1} - 1)$, which leads to a contradiction, as the left-hand side is smaller than the right-hand side.
		\item[] \textbf{Type-III 6-cycle}  exists if and only if the following holds:
		\begin{align*}
			w^{a_{r}-\ell_{2}}&= w^{a_{p}-\ell_{2}}-w^{a_{p}-\ell_{1}}+w^{a_{q}-\ell_{1}}\quad\textrm{mod }(w^{a_{t}+1}-1)\\
			\implies w^{a_{r}-\ell_{2}}-w^{a_{p}-\ell_{2}}&= w^{a_{q}-\ell_{1}}-w^{a_{p}-\ell_{1}}\quad\textrm{mod }(w^{a_{t}+1}-1)\\
			\implies w^{a_{p}-\ell_{2}}(w^{a_{r}-a_{p}}-1)&= w^{a_{p}-\ell_{1}}(w^{a_{q}-a_{p}}-1)\quad\textrm{mod }(w^{a_{t}+1}-1)\\
			\implies(w^{a_{r}-a_{p}}-1)&= w^{\ell_{2}-\ell_{1}}(w^{a_{q}-a_{p}}-1)\quad\textrm{mod }(w^{a_{t}+1}-1),
		\end{align*}
		 since gcd$(w^{a_{p}-\ell_{2}},w^{a_{t}+1}-1)=1$, which is a contradiction since the number on the left is odd, whereas the number on the right is even.
		\item[] \textbf{Type-IV 6-cycle } exists if and only if $
    w^{a_{r}-\ell_{2}} = w^{a_{q}-\ell_{2}} - w^{a_{q}-\ell_{1}} + w^{a_{p}-\ell_{1}}  \textrm{ mod }(w^{a_{t}+1} - 1)$, which implies
    $w^{a_{r}-\ell_{2}} + w^{a_{q}-\ell_{1}} = w^{a_{q}-\ell_{2}} + w^{a_{p}-\ell_{1}} $ $\textrm{ mod }(w^{a_{t}+1} - 1)$
and leads to a contradiction, as the left-hand side is greater than the right-hand side.
		\item[] \textbf{Type-V 6-cycle}  exists if and only if the following condition holds:
		\begin{align*}
			w^{a_{q}-\ell_{2}}&= w^{a_{p}-\ell_{2}}-w^{a_{p}-\ell_{1}}+w^{a_{r}-\ell_{1}}\quad\textrm{mod }(w^{a_{t}+1}-1)\\
			\implies w^{a_{q}-\ell_{2}}-w^{a_{p}-\ell_{2}}&= w^{a_{r}-\ell_{1}}-w^{a_{p}-\ell_{1}}\quad\textrm{mod }(w^{a_{t}+1}-1)\\
			\implies w^{a_{p}-\ell_{2}}(w^{a_{q}-a_{p}}-1)&= w^{a_{p}-\ell_{1}}(w^{a_{r}-a_{p}}-1)\quad\textrm{mod }(w^{a_{t}+1}-1)\\
			\implies (w^{a_{q}-a_{p}}-1)&= w^{\ell_{2}-\ell_{1}}(w^{a_{r}-a_{p}}-1)\quad\textrm{mod }(w^{a_{t}+1}-1),   
		\end{align*}
		since gcd$(w^{a_{p}-\ell_{2}},w^{a_{t}+1}-1)=1$, which leads to a contradiction since the number on the left is odd,  whereas the number on the right is even.
		\item[] \textbf{Type-VI 6-cycle} exists if and only if the following holds:
		\begin{align*}
			w^{a_{q}-\ell_{2}}&= w^{a_{r}-\ell_{2}}-w^{a_{r}-\ell_{1}}+w^{a_{p}-\ell_{1}}\quad\textrm{mod }(w^{a_{t}+1}-1)\\
			\implies w^{a_{r}-\ell_{2}}-w^{a_{q}-\ell_{2}}&= w^{a_{r}-\ell_{1}}-w^{a_{p}-\ell_{1}}\quad\textrm{mod }(w^{a_{t}+1}-1)\\
			\implies w^{a_{q}-\ell_{2}}(w^{a_{r}-a_{q}}-1)&= w^{a_{p}-\ell_{1}}(w^{a_{r}-a_{p}}-1)\quad\textrm{mod }(w^{a_{t}+1}-1). 		
		\end{align*}
        If $a_{p}-\ell_{1}\leq a_{q}-\ell_{2}$, then we have
        \begin{equation*}
            w^{a_{q}-a_{p}-(\ell_{2}-\ell_{1})}(w^{a_{r}-a_{q}}-1)= (w^{a_{r}-a_{p}}-1)\quad\textrm{mod }(w^{a_{t}+1}-1),
        \end{equation*}
		since gcd$(w^{a_{p}-\ell_{1}},w^{a_{t}+1}-1)=1$. This is a contradiction since the number on the left is even, whereas the number on the right is odd, unless $a_{q}-a_{p}=\ell_{2}-\ell_{1}$. If $a_{q}-a_{p}=\ell_{2}-\ell_{1}$, then $(w^{a_{r}-a_{q}}-1)= (w^{a_{r}-a_{p}}-1)\quad\textrm{mod }(w^{a_{t}+1}-1)$ if and only if $a_{p}=a_{q}$, which is a contradiction to the fact that $a_{p}<a_{q}.$ Similarly, we can examine the case if $a_{p}-\ell_{1}\geq a_{q}-\ell_{2}$.
	\end{enumerate} %%%%%%%%%%%%%%%%%%%%%%%%%%%%%%%%%%%%%%%%%%%%%%%%%%%%%%%%%%%%%%%%%%%%%%%%%%%%%%%%%%%%%%%%%%%%%%%%%%%%%%%%%%%%%%%%%%%%%%%%%%%%%%%%%%%%%%%%%%%%%%%%%%%%%%%%%%%%%%%%%%%%%%%%%%%%%%%%%%%%%%%%
	Now, consider the following three rows and three columns from the matrix $M$ in \eqref{apxmodel}:
	\begin{equation}\label{2-portion_of_H}
		\left[\begin{array}{ccccccc}
			\cdots&w^{a_{p}}&\cdots & w^{a_{q}}& \cdots&w^{a_{r}}&\cdots\\
			\cdots&w^{a_{p}-1}&\cdots& w^{a_{q}-1}&\cdots&w^{a_{r}-1}&\cdots\\   
			\cdots&w^{a_{p}-2}&\cdots & w^{a_{q}-2}&\cdots&w^{a_{r}-2}&\cdots
		\end{array}\right],
	\end{equation}
	where $a_{1}\leq a_{p}<a_{q}<a_{r}\leq a_{t}$.  We explore the presence of all possible 6-cycles, as depicted in Figure \ref{fig:all-6-cycles}, in the segment of the Tanner graph that corresponds to a portion of the degree matrix defined by Eq.~\eqref{2-portion_of_H}.

	\begin{enumerate}[leftmargin=0pt,itemsep=0pt]
		\item[] \textbf{Type-I 6-cycle} exists if and only if the following condition holds:
		\begin{align*}
			w^{a_{p}-2}-w^{a_{p}}&= w^{a_{r}-2}-w^{a_{r}-1}+w^{a_{q}-1}-w^{a_{q}}\quad\textrm{mod }(w^{a_{t}+1}-1)\\
		\implies	(1-w^{2})w^{a_{p}-2}&= w^{a_{r}-2}(1-w)+(1-w)w^{a_{q}-1}\quad\textrm{mod }(w^{a_{t}+1}-1)\\
		\implies	w^{a_{p}-2}+w^{a_{p}-1}&= w^{a_{r}-2}+w^{a_{q}-1}\quad\textrm{mod }\left(\frac{w^{a_{t}+1}-1}{w-1}\right),
		\end{align*}
		which is a contradiction since the number on the left is less than the number  on the right.
		\item[] \textbf{Type-II 6-cycle} exists if and only if the following condition is satisfied:
		\begin{align*}
			w^{a_{p}-2}-w^{a_{p}}&= w^{a_{q}-2}-w^{a_{q}-1}+w^{a_{r}-1}-w^{a_{r}}\quad\textrm{mod }(w^{a_{t}+1}-1)\\
		\implies	(1-w^{2})w^{a_{p}-2}&= w^{a_{q}-2}(1-w)+(1-w)w^{a_{r}-1}\quad\textrm{mod }(w^{a_{t}+1}-1)\\
		\implies	w^{a_{p}-2}+w^{a_{p}-1}&= w^{a_{r}-2}+w^{a_{q}-1}\quad\textrm{mod }\left(\frac{w^{a_{t}+1}-1}{w-1}\right),
		\end{align*}
which is a contradiction, as the number on the left is less than the number on the right.
		\item[] \textbf{Type-III 6-cycle}  exists if and only if the following condition is satisfied:
		\begin{align*}
			w^{a_{r}-2}-w^{a_{r}}&= w^{a_{p}-2}-w^{a_{p}-1}+w^{a_{q}-1}-w^{a_{q}}\quad\textrm{mod }(w^{a_{t}+1}-1)\\
		\implies	(1-w^{2})w^{a_{r}-2}&= w^{a_{p}-2}(1-w)+w^{a_{q}-1}(1-w)\quad\textrm{mod }(w^{a_{t}+1}-1)\\
		\implies	w^{a_{r}-2}+w^{a_{r}-1}&= w^{a_{p}-2}+w^{a_{q}-1}\quad\textrm{mod }\left(\frac{w^{a_{t}+1}-1}{w-1}\right), 
		\end{align*}
		which is a contradiction since the number on the left is greater than the number on the right.
		\item[] \textbf{Type-IV 6-cycle} exists if and only if the following condition is satisfied:
		\begin{align*}
			w^{a_{r}-2}-w^{a_{r}}&= w^{a_{q}-2}-w^{a_{q}-1}+w^{a_{p}-1}-w^{a_{p}}\quad\textrm{mod }(w^{a_{t}+1}-1)\\
			\implies (1-w^{2})w^{a_{r}-2}&= w^{a_{q}-2}(1-w)+w^{a_{p}-1}(1-w)\quad\textrm{mod }(w^{a_{t}+1}-1)\\
			\implies w^{a_{r}-2}+w^{a_{r}-1}&= w^{a_{q}-2}+w^{a_{p}-1}\quad\textrm{mod }\left(\frac{w^{a_{t}+1}-1}{w-1}\right).
		\end{align*}
This leads to a contradiction, as the number on the right is less than the number on the left.
		\item[] \textbf{Type-V 6-cycle} exists if and only if the following condition holds:
		\begin{align*}
			w^{a_{q}-2}-w^{a_{q}}&= w^{a_{p}-2}-w^{a_{p}-1}+w^{a_{r}-1}-w^{a_{r}}\quad\textrm{mod }(w^{a_{t}+1}-1)\\
		\implies	(1-w^{2})w^{a_{q}-2}&= w^{a_{p}-2}(1-w)+(1-w)w^{a_{r}-1}\quad\textrm{mod }(w^{a_{t}+1}-1)\\
			\implies w^{a_{q}-2}+w^{a_{q}-1}&= w^{a_{p}-2}+w^{a_{r}-1}\quad\textrm{mod }\left(\frac{w^{a_{t}+1}-1}{w-1}\right),
		\end{align*}
		which is a contradiction since the number on the left is less than the number on the right.
		\item[] \textbf{Type-VI 6-cycle} exists if and only if the following condition is satisfied:
		\begin{align*}
			w^{a_{q}-2}-w^{a_{q}}&= w^{a_{r}-2}-w^{a_{r}-1}+w^{a_{p}-1}-w^{a_{p}}\quad\textrm{mod }(w^{a_{t}+1}-1)\\
			\implies (1-w^{2})w^{a_{q}-2}&= w^{a_{r}-2}(1-w)+(1-w)w^{a_{p}-1}\quad\textrm{mod }(w^{a_{t}+1}-1)\\
			\implies w^{a_{q}-2}+w^{a_{q}-1}&= w^{a_{r}-2}+w^{a_{p}-1}\quad\textrm{mod }\left(\frac{w^{a_{t}+1}-1}{w-1}\right)\\  
            \implies   (w^{a_{q}-a_{p}}-1)w^{a_{p}-1}&=(w^{a_{r}-a_{q}}-1)w^{a_{q}-2}\quad\textrm{mod }\left(\frac{w^{a_{t}+1}-1}{w-1}\right)\\
			\implies (w^{a_{q}-a_{p}}-1)&=(w^{a_{r}-a_{q}}-1)w^{a_{q}-a_{p}-1}\quad\textrm{mod }\left(\frac{w^{a_{t}+1}-1}{w-1}\right).
		\end{align*}
		This leads to a contradiction, as the number on the left is odd while the number on the right is even, unless $a_{q} - a_{p} = 1$. If $a_{q} - a_{p} = 1$, then the condition $(w^{a_{q} - a_{p}} - 1) = (w^{a_{r} - a_{q}} - 1) \mod (w^{a_{t} + 1} - 1)$
holds if and only if \( a_{q} - a_{p} = a_{r} - a_{q} = 1 \), which leads to a contradiction with the original assumptions.
	\end{enumerate} 
	The minimum number of required ebits is given by $c=\mathrm{rank}(HH^{T}),$
	where
	\begin{align*}
		HH^{T}&=\left[\begin{array}{cccc}
			I & I&\cdots&I\\
			P^{(w^{a_{1}})} & P^{(w^{a_{2}})}&\cdots&P^{(w^{a_{t}})}\\   
			P^{(w^{a_{1}-1})} & P^{(w^{a_{2}-1})}&\cdots&P^{(w^{a_{t}-1})}\\  
			P^{(w^{a_{1}-2})} & P^{(w^{a_{2}-2})}&\cdots&P^{(w^{a_{t}-2})}\\   
		\end{array}\right]\left[\begin{array}{cccc}
			I & I&\cdots&I\\
			P^{(w^{a_{1}})} & P^{(w^{a_{2}})}&\cdots&P^{(w^{a_{t}})}\\   
			P^{(w^{a_{1}-1})} & P^{(w^{a_{2}-1})}&\cdots&P^{(w^{a_{t}-1})}\\  
			P^{(w^{a_{1}-2})} & P^{(w^{a_{2}-2})}&\cdots&P^{(w^{a_{t}-2})}\\   
		\end{array}\right]^{T}\\
		&=\left[\begin{array}{cccc}
			tI & \sum\limits_{i=1}^{t}P^{-w^{a_{i}}}&\sum\limits_{i=1}^{t}P^{-w^{a_{i}-1}}&\sum\limits_{i=1}^{t}P^{-w^{a_{i}-2}}\\
			\sum\limits_{i=1}^{t}P^{w^{a_{i}}} & tI&\sum\limits_{i=1}^{t}P^{(w-1)w^{a_{i}-1}}&\sum\limits_{i=1}^{t}P^{(w^{2}-1)w^{a_{i}-2}}\\
			\sum\limits_{i=1}^{t}P^{w^{a_{i}-1}} & \sum\limits_{i=1}^{t}P^{-(w-1)w^{a_{i}-1}}&tI&\sum\limits_{i=1}^{t}P^{(w-1)w^{a_{i}-2}}\\
			\sum\limits_{i=1}^{t}P^{(w^{a_{i}-2})} &     \sum\limits_{i=1}^{t}P^{-(w^{2}-1)w^{a_{i}-2}}&\sum\limits_{i=1}^{t}P^{-(w-1)w^{a_{i}-2}}&tI
		\end{array} \right].
	\end{align*}
	It is easy to verify that the rank$(HH^{T})\leq 4(w^{a_{t}+1}-1)-3$  since the vector sum of each block row is either all-ones vector or zero-vector depending on whether $t$ is even or odd, completing the proof.
\end{proof}
\section{The proof of Theorem \ref{columnweight-4-general}}\label{appendix.b}
\textbf{Statement: }
    Let $S=\{a_{1},a_{2},a_{3},\ldots,a_{t}\}$, where $2\leq a_{1}<a_{2}<a_{3}\cdots<a_{t}$ and for any $1\leq p<q\leq t$, $a_{q}-a_{p}\geq 2$. For two integers\footnote{The parameter \(a_t\) is selected so that the catalytic rate remains positive.} $a_{t}\geq 16$ and $w\geq2$, let $P$ be a right circulant permutation matrix of order $(w^{a_{t}+1}-1)$, and let
	\begin{equation}\label{apxmodel2}
		M=\begin{bmatrix}
		    0 & 0&\cdots&0\\
			w^{a_{1}} & w^{a_{2}}&\cdots&w^{a_{t}}\\   
			w^{a_{1}-1} & w^{a_{2}-1}&\cdots&w^{a_{t}-1}\\   
			w^{a_{1}-2} & w^{a_{2}-2}&\cdots&w^{a_{t}-2}
		\end{bmatrix}
	\end{equation} be a degree matrix associated with the parity-check matrix $H$. Then there exists an entanglement assisted quantum QC-LDPC code with parameters $[[n,2k-n+c;c]]_2$, where $n=(w^{a_{t}+1}-1)t$, $k\geq (w^{a_{t}+1}-1)(t-4)+3$ and $c\leq 4(w^{a_{t}+1}-1)-3$. In addition, the Tanner graph of the parity-check matrix $H$ is of girth greater than 6.   

\begin{proof} The length of the code is straightforward to determine, and the lower bound on the dimension $k$ can be easily derived using Theorem \ref{theorem2}. Following similar arguments used in proving Theorem \ref{columnweight-4-even}, one can easily show that the Tanner graph of $H$ is free from 4-cycles. 
	
Next, we will prove that the graphical representation of $H$ does not contain cycles of length 6. To streamline the proof and avoid repetition, we will consider two cases. In the first case, we have one zero row and two nonzero rows, while in the second case, all rows are nonzero.
	Consider the following three rows and three  arbitrary columns from the matrix $M$ defined in \eqref{apxmodel2}:
	\begin{equation}\label{1-portion_of_H1}
		\left[\begin{array}{ccccccc}
			\cdots&0&\cdots & 0& \cdots&0&\cdots\\
			\cdots&w^{a_{p}-\ell_{1}}&\cdots& w^{a_{q}-\ell_{1}}&\cdots&w^{a_{r}-\ell_{1}}&\cdots\\   
			\cdots&w^{a_{p}-\ell_{2}}&\cdots& w^{a_{q}-\ell_{2}}&\cdots&w^{a_{r}-\ell_{2}}&\cdots\\ 
		\end{array}\right],
	\end{equation}
	where $a_{1}\leq a_{p}<a_{q}<a_{r}\leq a_{t}$ and $0\leq \ell_{1}<\ell_{2}\leq2$. We investigate the existence of all possible 6-cycles, as shown in Figure \ref{fig:all-6-cycles}, within the segment of the Tanner graph that corresponds to a portion of the degree matrix defined by Eq.~\eqref{1-portion_of_H1}.
	\begin{enumerate}[leftmargin=0pt,itemsep=0pt]
		\item[] \textbf{Type-I 6-cycle} exists if and only if $w^{a_{p}-\ell_{2}}= w^{a_{r}-\ell_{2}}-w^{a_{r}-\ell_{1}}+w^{a_{q}-\ell_{1}}$ under modulo $(w^{a_{t}+1}-1)$, which implies
		\begin{align*}			
			w^{a_{r}-\ell_{1}}-w^{a_{q}-\ell_{1}}&= w^{a_{r}-\ell_{2}}-w^{a_{p}-\ell_{2}}\quad\textrm{mod }(w^{a_{t}+1}-1)\\
            \implies (w-1)w^{a_{r}-\ell_{1}-1}+(w^{a_{r}-\ell_{1}-1}-w^{a_{q}-\ell_{1}})&= w^{a_{r}-\ell_{2}}-w^{a_{p}-\ell_{2}}\quad\textrm{mod }(w^{a_{t}+1}-1).
		\end{align*}
		This is a contradiction, as the number on the left is greater than the number on the right.
		\item[] \textbf{Type-II 6-cycle} exists if and only if 
    $w^{a_{p}-\ell_{2}} = w^{a_{q}-\ell_{2}} - w^{a_{q}-\ell_{1}} + w^{a_{r}-\ell_{1}} \textrm{ mod }(w^{a_{t}+1} - 1)$ if and only if   $w^{a_{p}-\ell_{2}} + w^{a_{q}-\ell_{1}} = w^{a_{q}-\ell_{2}} + w^{a_{r}-\ell_{1}}$ $ \textrm{ mod }(w^{a_{t}+1} - 1)$
which leads to a contradiction, as the left-hand side is smaller than the right-hand side.

		\item[] \textbf{Type-III 6-cycle} exists if and only if $w^{a_{r}-\ell_{2}}= w^{a_{p}-\ell_{2}}-w^{a_{p}-\ell_{1}}+w^{a_{q}-\ell_{1}}$ under modulo $(w^{a_{t}+1}-1)$, which implies
		\begin{align*}			
			w^{a_{r}-\ell_{2}}-w^{a_{p}-\ell_{2}}&= w^{a_{q}-\ell_{1}}-w^{a_{p}-\ell_{1}}\quad\textrm{mod }(w^{a_{t}+1}-1)\\
            \implies (w-1)w^{a_{r}-\ell_{2}-1}+(w^{a_{r}-\ell_{2}-1}-w^{a_{p}-1})&= w^{a_{q}-\ell_{1}}-w^{a_{p}-\ell_{1}}\quad\textrm{mod }(w^{a_{t}+1}-1),
		\end{align*}
		leads to a contradiction since the number on the right is less than the number on the left.
		\item[] \textbf{Type-IV 6-cycle} exists if and only if $
    w^{a_{r}-\ell_{2}} = w^{a_{q}-\ell_{2}} - w^{a_{q}-\ell_{1}} + w^{a_{p}-\ell_{1}} \textrm{  mod }(w^{a_{t}+1} - 1)$, which implies
    $w^{a_{r}-\ell_{2}} + w^{a_{q}-\ell_{1}} = w^{a_{q}-\ell_{2}} + w^{a_{p}-\ell_{1}}$ $ \textrm{ mod }(w^{a_{t}+1} - 1)$
and leads to a contradiction, as the left-hand side is greater than the right-hand side.
		\item[] \textbf{Type-V 6-cycle} exists if and only if the following condition holds:
		\begin{align*}
			w^{a_{q}-\ell_{2}}&= w^{a_{p}-\ell_{2}}-w^{a_{p}-\ell_{1}}+w^{a_{r}-\ell_{1}}\quad\textrm{mod }(w^{a_{t}+1}-1)\\
			\implies w^{a_{q}-\ell_{2}}-w^{a_{p}-\ell_{2}}&= w^{a_{r}-\ell_{1}}-w^{a_{p}-\ell_{1}}\quad\textrm{mod }(w^{a_{t}+1}-1)\\
          \implies w^{a_{q}-\ell_{2}}-w^{a_{p}-\ell_{2}}&= (w-1)w^{a_{r}-\ell_{1}-1}+(w^{a_{r}-\ell_{1}-1}-w^{a_{p}-\ell_{1}})\quad\textrm{mod }(w^{a_{t}+1}-1),   
		\end{align*}
		which is a contradiction since the number on the left is less than the number on the right.
		\item[] \textbf{Type-VI 6-cycle} exists if and only if the following holds:
		\begin{align*}
			w^{a_{q}-\ell_{2}}&= w^{a_{r}-\ell_{2}}-w^{a_{r}-\ell_{1}}+w^{a_{p}-\ell_{1}}\quad\textrm{mod }(w^{a_{t}+1}-1)\\
			\implies w^{a_{r}-\ell_{2}}-w^{a_{q}-\ell_{2}}&= w^{a_{r}-\ell_{1}}-w^{a_{p}-\ell_{1}}\quad\textrm{mod }(w^{a_{t}+1}-1)\\
            \implies w^{a_{r}-\ell_{2}}-w^{a_{q}-\ell_{2}}&= (w-1)w^{a_{r}-\ell_{1}-1}+(w^{a_{r}-\ell_{1}-1}-w^{a_{p}-\ell_{1}})\quad\textrm{mod }(w^{a_{t}+1}-1),
		\end{align*}
		which is a contradiction since the number on the right is greater than the number the number on the left.
	\end{enumerate} %%%%%%%%%%%%%%%%%%%%%%%%%%%%%%%%%%%%%%%%%%%%%%%%%%%%%%%%%%%%%%%%%%%%%%%%%%%%%%%%%%%%%%%%%%%%%%%%%%%%%%%%%%%%%%%%%%%%%%%%%%%%%%%%%%%%%%%%%%%%%%%%%%%%%%%%%%%%%%%%%%%%%%%%%%%%%%%%%%%%%%%%
	Consider the following three rows and three columns from the matrix $M$ in \eqref{apxmodel2}:
	\begin{equation}\label{2-portion_of_H1}
		\left[\begin{array}{ccccccc}
			\cdots&w^{a_{p}}&\cdots & w^{a_{q}}& \cdots&w^{a_{r}}&\cdots\\
			\cdots&w^{a_{p}-1}&\cdots& w^{a_{q}-1}&\cdots&w^{a_{r}-1}&\cdots\\   
			\cdots&w^{a_{p}-2}&\cdots & w^{a_{q}-2}&\cdots&w^{a_{r}-2}&\cdots
		\end{array}\right],
	\end{equation}
	where $a_{1}\leq a_{p}<a_{q}<a_{r}\leq a_{t}$.  We investigate the existence of all possible 6-cycles, as shown in Figure \ref{fig:all-6-cycles}, within the segment of the Tanner graph that corresponds to a portion of the degree matrix defined by Eq.~\eqref{2-portion_of_H1}.
	\begin{enumerate}[leftmargin=0pt, itemsep=0pt]
		\item[]  There exists \textbf{type-I 6-cycle} if and only if the following condition holds:
		\begin{align*}
			w^{a_{p}-2}-w^{a_{p}}&= w^{a_{r}-2}-w^{a_{r}-1}+w^{a_{q}-1}-w^{a_{q}}\quad\textrm{mod }(w^{a_{t}+1}-1)\\
		\implies	(1-w^{2})w^{a_{p}-2}&= w^{a_{r}-2}(1-w)+(1-w)w^{a_{q}-1}\quad\textrm{mod }(w^{a_{t}+1}-1)\\
		\implies	w^{a_{p}-2}+w^{a_{p}-1}&= w^{a_{r}-2}+w^{a_{q}-1}\quad\textrm{mod }\left(\frac{w^{a_{t}+1}-1}{w-1}\right),
		\end{align*}
		which is a contradiction since the number on the left is less than the number  on the right.
		\item [] \textbf{Type-II 6-cycle} exists if and only if the following condition is satisfied:
		\begin{align*}
			w^{a_{p}-2}-w^{a_{p}}&= w^{a_{q}-2}-w^{a_{q}-1}+w^{a_{r}-1}-w^{a_{r}}\quad\textrm{mod }(w^{a_{t}+1}-1)\\
		\implies	(1-w^{2})w^{a_{p}-2}&= w^{a_{q}-2}(1-w)+(1-w)w^{a_{r}-1}\quad\textrm{mod }(w^{a_{t}+1}-1)\\
		\implies	w^{a_{p}-2}+w^{a_{p}-1}&= w^{a_{r}-2}+w^{a_{q}-1}\quad\textrm{mod }\left(\frac{w^{a_{t}+1}-1}{w-1}\right),
		\end{align*}
leading to a contradiction, as the number on the left is less than the number on the right.
		\item[] \textbf{Type-III 6-cycle} exists if and only if the following condition is satisfied:
		\begin{align*}
			w^{a_{r}-2}-w^{a_{r}}&= w^{a_{p}-2}-w^{a_{p}-1}+w^{a_{q}-1}-w^{a_{q}}\quad\textrm{mod }(w^{a_{t}+1}-1)\\
		\implies	(1-w^{2})w^{a_{r}-2}&= w^{a_{p}-2}(1-w)+w^{a_{q}-1}(1-w)\quad\textrm{mod }(w^{a_{t}+1}-1)\\
		\implies	w^{a_{r}-2}+w^{a_{r}-1}&= w^{a_{p}-2}+w^{a_{q}-1}\quad\textrm{mod }\left(\frac{w^{a_{t}+1}-1}{w-1}\right), 
		\end{align*}
		which is a contradiction since the number on the right is less than the number on the left.
		\item[] \textbf{Type-IV 6-cycle} exists if and only if the following condition is satisfied:
		\begin{align*}
			w^{a_{r}-2}-w^{a_{r}}&= w^{a_{q}-2}-w^{a_{q}-1}+w^{a_{p}-1}-w^{a_{p}}\quad\textrm{mod }(w^{a_{t}+1}-1)\\
			\implies (1-w^{2})w^{a_{r}-2}&= w^{a_{q}-2}(1-w)+w^{a_{p}-1}(1-w)\quad\textrm{mod }(w^{a_{t}+1}-1)\\
			\implies w^{a_{r}-2}+w^{a_{r}-1}&= w^{a_{q}-2}+w^{a_{p}-1}\quad\textrm{mod }\left(\frac{w^{a_{t}+1}-1}{w-1}\right),
		\end{align*}
leading to a contradiction, as the number on the left is greater than the number on the right.
		\item[]  There exists \textbf{type-V 6-cycle} if and only if the following condition holds:
		\begin{align*}
			w^{a_{q}-2}-w^{a_{q}}&= w^{a_{p}-2}-w^{a_{p}-1}+w^{a_{r}-1}-w^{a_{r}}\quad\textrm{mod }(w^{a_{t}+1}-1)\\
		\implies	(1-w^{2})w^{a_{q}-2}&= w^{a_{p}-2}(1-w)+(1-w)w^{a_{r}-1}\quad\textrm{mod }(w^{a_{t}+1}-1)\\
			\implies w^{a_{q}-2}+w^{a_{q}-1}&= w^{a_{p}-2}+w^{a_{r}-1}\quad\textrm{mod }\left(\frac{w^{a_{t}+1}-1}{w-1}\right),
		\end{align*}
		which is a contradiction since the number  on the left is less than the number on the right.
		\item[] \textbf{Type-VI 6-cycle} exists if and only if the following condition is satisfied:
		\begin{align*}
			w^{a_{q}-2}-w^{a_{q}}&= w^{a_{r}-2}-w^{a_{r}-1}+w^{a_{p}-1}-w^{a_{p}}\quad\textrm{mod }(w^{a_{t}+1}-1)\\
			\implies (1-w^{2})w^{a_{q}-2}&= w^{a_{r}-2}(1-w)+(1-w)w^{a_{p}-1}\quad\textrm{mod }(w^{a_{t}+1}-1)\\
			\implies w^{a_{q}-2}+w^{a_{q}-1}&= w^{a_{r}-2}+w^{a_{p}-1}\quad\textrm{mod }\left(\frac{w^{a_{t}+1}-1}{w-1}\right)\\ 
            \implies w^{a_{q}-2}+w^{a_{q}-1}&= (w-1)w^{a_{r}-3}+(w^{a_{r}-3}+w^{a_{p}-1})\quad\textrm{mod }\left(\frac{w^{a_{t}+1}-1}{w-1}\right).  
		\end{align*}
		This leads to a contradiction, as the number on the left is less the number on the right.
	\end{enumerate} 
	The minimum number of required maximally entangled bits is given by	$c=\mathrm{rank}(HH^{T})$,	where
	\begin{align*}
		HH^{T}&=\left[\begin{array}{cccc}
			I & I&\cdots&I\\
			P^{(w^{a_{1}})} & P^{(w^{a_{2}})}&\cdots&P^{(w^{a_{t}})}\\   
			P^{(w^{a_{1}-1})} & P^{(w^{a_{2}-1})}&\cdots&P^{(w^{a_{t}-1})}\\  
			P^{(w^{a_{1}-2})} & P^{(w^{a_{2}-2})}&\cdots&P^{(w^{a_{t}-2})}\\   
		\end{array}\right]\left[\begin{array}{cccc}
			I & I&\cdots&I\\
			P^{(w^{a_{1}})} & P^{(w^{a_{2}})}&\cdots&P^{(w^{a_{t}})}\\   
			P^{(w^{a_{1}-1})} & P^{(w^{a_{2}-1})}&\cdots&P^{(w^{a_{t}-1})}\\  
			P^{(w^{a_{1}-2})} & P^{(w^{a_{2}-2})}&\cdots&P^{(w^{a_{t}-2})}\\   
		\end{array}\right]^{T}\\
		&=\left[\begin{array}{cccc}
			tI & \sum\limits_{i=1}^{t}P^{-w^{a_{i}}}&\sum\limits_{i=1}^{t}P^{-w^{a_{i}-1}}&\sum\limits_{i=1}^{t}P^{-w^{a_{i}-2}}\\
			\sum\limits_{i=1}^{t}P^{w^{a_{i}}} & tI&\sum\limits_{i=1}^{t}P^{(w-1)w^{a_{i}-1}}&\sum\limits_{i=1}^{t}P^{(w^{2}-1)w^{a_{i}-2}}\\
			\sum\limits_{i=1}^{t}P^{w^{a_{i}-1}} & \sum\limits_{i=1}^{t}P^{-(w-1)w^{a_{i}-1}}&tI&\sum\limits_{i=1}^{t}P^{(w-1)w^{a_{i}-2}}\\
			\sum\limits_{i=1}^{t}P^{(w^{a_{i}-2})} &     \sum\limits_{i=1}^{t}P^{-(w^{2}-1)w^{a_{i}-2}}&\sum\limits_{i=1}^{t}P^{-(w-1)w^{a_{i}-2}}&tI
		\end{array} \right].
	\end{align*}
	It is easy to verify that the rank$(HH^{T})\leq 4(w^{a_{t}+1}-1)-3$  since the vector sum of each block row is either all-ones vector or zero-vector depending on whether $t$ is even or odd, completing the proof.
\end{proof}
\end{document}